\UseRawInputEncoding
\PassOptionsToPackage{english}{babel}
\documentclass[
preprint,
aps,
groupedaddress,jmp,amsmath,amssymb,longbibliography]{revtex4-2}

\usepackage[english]{babel}
\usepackage{amsmath}
\usepackage{lineno,hyperref}
\usepackage{caption}
\usepackage{subcaption}
\usepackage{xcolor}

 \usepackage{dcolumn}
 \newcolumntype{d}{D{.}{.}{-1}}
  \makeglossary
 \usepackage{fancyvrb}
  \fvset{fontsize=\footnotesize,xleftmargin=2em}
 \usepackage{hyperref}
  \usepackage{graphicx, color, psfrag, float}
  \usepackage{dcolumn}
\usepackage{bm}
\usepackage{epstopdf}
\usepackage{booktabs}
\newcolumntype{L}{>{\centering\arraybackslash}m{3cm}}
\modulolinenumbers[5]

\begin{document}

\title[]{Wake-foil Interactions and Energy Harvesting Efficiency in Tandem Oscillating Foils}

\author{Bernardo Luiz R. Ribeiro}
\affiliation{Department of Engineering Physics, University of Wisconsin-Madison, Madison, WI}
\author{Yunxing Su}
\affiliation{Center for Fluid Mechanics, School of Engineering, Brown University, Providence, RI}
\author{Quentin Guillaumin}
\affiliation{Center for Fluid Mechanics, School of Engineering, Brown University, Providence, RI}
\author{Kenneth S. Breuer}
\affiliation{Center for Fluid Mechanics, School of Engineering, Brown University, Providence, RI}
\author{Jennifer A. Franck}
\affiliation{{Department of Engineering Physics, University of Wisconsin-Madison, Madison, WI  }}

\begin{abstract}

Oscillating foils in synchronized pitch/heave motions can be used to harvest hydrokinetic energy. By understanding the wake structure
and its correlation with the foil kinematics, predictive models for how
foils can operate in array configurations can be developed. To establish a relationship between foil
kinematics and wake characteristics, a wide range of kinematics is explored in a two-foil tandem
configuration with inter-foil spacing from 4-9 chord lengths separation and multiple inter-foil phases. Using data from experiments and simulations, an in-depth wake
analysis is performed and the mean velocity and the
turbulent kinetic energy are quantified in the wake. With this energy quantification, the trailing foil efficiency is modified to account for the mean flow in addition
to the energy transported by the coherent leading edge vortices (LEVs) shed from the leading
foil. With the mean wake velocity, a predictive wake model is able to distinguish three regimes through analyzing trailing foil efficiency profiles and the strength of the primary LEV shed from the leading foil. Dividing the wake into regimes is an insightful way to narrow the range of foil kinematics and configurations and improve the energy harvesting in a two-tandem foil-array.

\end{abstract}

\maketitle

\section{Introduction} \label{s:intro}

Oscillating foils in a combined heave/pitch motion are an effective way to extract hydrokinetic energy \cite{Xiao2014,Young2014} and offer benefits of shallow water operation, scalability, and low cut-in speeds. The efficiency and kinematics of a single foil in freestream conditions are well understood both numerically \cite{Jones2003,kindum2008,zhu2009,Wu2014,wu2015} and experimentally \cite{mckdel,Karakas2016,Kim2017,Su2019conf,Su2019}, however foils in coordinated array configurations have not received as much attention. As observed in other wind and hydrokinetic turbines, the downstream foils of arrays suffer from reduced freestream velocity and wake disturbances. Thus, a major challenge within the industry is to accurately predict and model how the wake of one turbine affects the others downstream. Progress has been made in wind farm layouts \cite{gocmen2016,shakoor2016}, however due to the oscillatory, rather than rotational, kinematics the wake structure of oscillating foils is vastly different from that of traditional horizontal-axis turbines. Moreover, the wake structure is a strong yet nonlinear function of the precise kinematic motion of the foil. Using a two-foil in-line array, this paper aims to quantify the wake structure in terms of the leading foil kinematics and subsequently model the downstream’s foil performance.

\begin{figure}[htbp]
\centering
\includegraphics[width=0.3\textwidth]{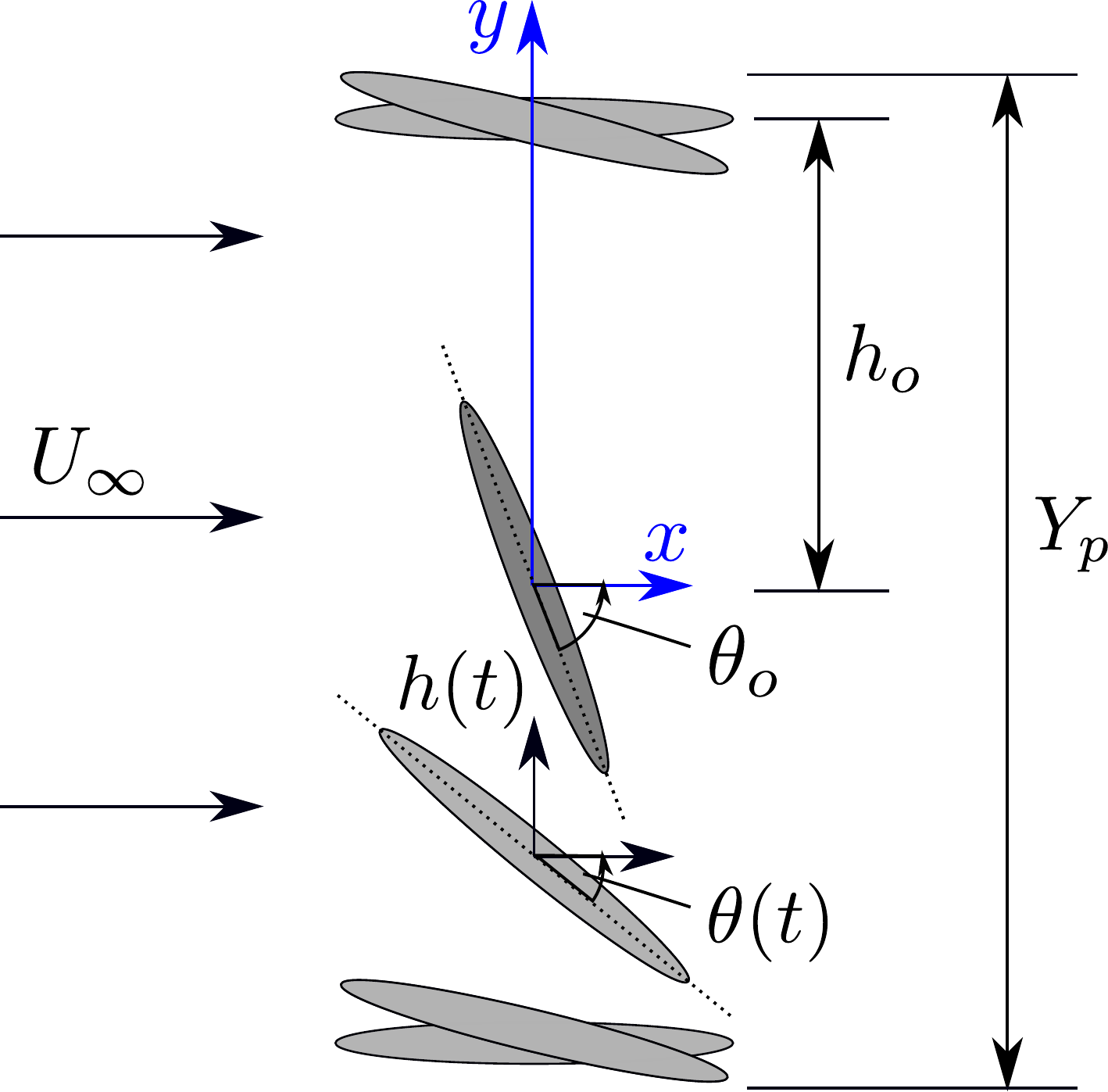}
\caption{Foil kinematics: pitch amplitude $\theta_o$, heave amplitude $h_o$, reduced oscillating frequency $fc/U_{\infty}$, swept area $Y_p$.}
\label{f:single}
\end{figure}

McKinney and DeLaurier \cite{mckdel} introduced the first modern demonstration of oscillating foils for energy harvesting. To extract hydrokinetic energy, oscillating foils heave and pitch periodically with the pitch leading the heave by $90$ degrees. When operated at optimal frequencies, this motion creates a heave stroke at high angle of attack, subsequently generating a strong lift force, which is converted to linear power. Depending on the precise kinematics, power can also be generated through torque on the foil. The kinematics strongly affect the foil's performance, as discussed thoroughly in review articles by Young et al. \cite{Young2014} and Xiao and Zhu \cite{Xiao2014}. The kinematic motion with respect to the freestream flow is summarized in Figure \ref{f:single}, with $h(t)$ denoting the heave position and $\theta(t)$ the pitch angle. The frequency of oscillation, $f$, non-dimensionalized by chord length $c$ and freestream velocity $U_{\infty}$, is in the range $fc/U_\infty=0.10-0.15$, lower than that found in propulsive oscillating foils. The heave and pitch kinematics are typically sinusoidal and characterized by high amplitudes, where the pitch amplitude $\theta_o$ ranges from $55-85$ degrees, and the heave amplitude $h_o$ is between $0.5-2.0$ chord lengths. 

A distinguishing flow feature of the high-pitch and high-heave kinematics is the trail of coherent vortices formed in the wake. Throughout most of the energy harvesting kinematic parameter space one or more leading edge vortices (LEVs) are shed on each half-stroke, sometimes accompanied by a trailing edge vortex (TEV) \cite{kindum2008, zhu2009}. 
The structured pattern of these relatively two-dimensional coherent vortices differs widely from the three-dimensional wake of horizontal-axis turbines \cite{Young2014, Xiao2014, boudreau2018}.
The formation and wake dynamics of the shed vortices are strongly correlated with the specific kinematic parameters. However, for the optimal kinematics it has been shown that the boundary layer separates very close to the leading edge of the foil. Due to the leading edge separation the dynamics and energy harvesting performance are relatively insensitive to Reynolds number \cite{RibeiroFranck2020} and blade geometry \cite{Kim2017}. Evidence has also indicated that although the wake structure is tied to the kinematics, it is relatively independent of Reynolds number and inflow conditions \cite{RibeiroFranck2020,ekaterinaris1994}, unlike horizontal-axis turbines whose wakes are very sensitive to Reynolds number and the turbulent characteristics of the oncoming flow \cite{mctavish2013,boudreau2018}.

When foils are arranged in a tandem array the layout introduces two configuration parameters, the inter-foil distance, $S_x$, and inter-foil phase, $\psi$. Inter-foil phase describes the phase difference between leading and trailing foils which are both undergoing the same kinematics. Through experiments in a water tunnel at $Re=20,000$ and using fixed foil kinematics of $fc/U_\infty=0.80; \; h_o/c=1.05; \; \theta_o=73^{\circ}$, Platzer et al. \cite{platzer2009} discovered that any slight change in the configuration parameters will affect the trailing foil’s performance. Using the same flow conditions and kinematics as Platzer et al., Ashraf et al. \cite{Ashraf2011} performed numerical simulations using two mesh zones, a dynamic mesh close to the foil and a stationary zone with a sliding interface in-between zones. They varied $S_x$ from $2c$ to $6c$ demonstrating that the trailing foil efficiency increased when the inter-foil phase was nonzero. Subsequent simulations by Broering and Lian \cite{broering2012a} and Broering et al. \cite{broering2012b} at $Re=10,000$ observed that both inter-foil phase and spacing had similar effects on power generation, however these were performed at lower pitch and heave amplitudes ($h_o/c=0.50; \theta_o=20^{\circ}$), likely avoiding the flow separation and LEV generation characteristic of higher amplitudes. Karakas and Fenercioglu \cite{Karakas2017} experimentally investigated the inter-foil phase effects on power generation of a two-foil fixed set of kinematics and obtained an optimal inter-foil phase of $135^{\circ}$ from the observation of wake-foil interactions through a inter-foil phase variation of $-180^{\circ}-+180^{\circ}$. Numerical work from Ma et al. \cite{ma2019} has coupled the motion of the leading and trailing foils in passive tandem arrays and noted how the inter-foil distance impact the system dynamics. Experimental work from Ramananarivo et al. \cite{ramananarivo2016} and Newbolt et al. \cite{newbolt2019} showed that when two foils are arranged in a tandem array and are randomly perturbed, the flow interaction between the trailing foil and the wake generated by the leading foil promoted group cohesion, similar to those found in fish schooling and bird flocking. Their results show that foil kinematics and configuration parameters can be used to control locomotion within wakes.

To combine the two configuration parameters into a single variable, Kinsey and Dumas \cite{kinseytandem2012} introduced a global phase parameter, $\Phi$,  

\begin{equation}
\label{e:dumas}
\Phi = 2\pi \frac{S_x}{U_\infty T} + \psi.
\end{equation}

\noindent The global phase parameter is defined as the summation of the inter-foil phase and a non-dimensional term comprised of the inter-foil distance, freestream velocity, and the oscillation period, $T$. Kinsey and Dumas performed numerical simulations at $Re=500,000$ varying the frequency of oscillation and inter-foil spacing for two distinct phases, $\psi=-90^{\circ},-180^{\circ}$. From this data they determined a global phase of $90^{\circ}$ typically leads to an increase in trailing foil performance. However, they emphasize that cases with the same global phase do not share the same wake-foil interactions, unless they all have same kinematics and configurations parameters. Xu and Xu \cite{xuWake2017} also investigated the global phase model using potential flow theory and through a variation of inter-foil distance ($S_x=3.4c-6.7c$) and inter-foil phase ($\psi=-180^{\circ}- +180^{\circ}$), they discovered an optimal global phase of $160^{\circ}$ that increased trailing foil performance. The optimal phase was different than that found by Kinsey and Dumas \cite{kinseytandem2012}, which could be explained by the single set of kinematics Xu and Xu investigated compared to the variation in foil kinematics performed by Kinsey and Dumas. Similar to Broering and Lian  \cite{broering2012a} and Broering et al. \cite{broering2012b}, Xu and Xu also noticed that a similar trailing foil performance can be achieved by either changing inter-foil phase or inter-foil distance.

Understanding the wake dynamics is critical for developing a physics-based model of how the kinematics and array configuration parameters influence energy harvesting performance. The path of individual vortices within the wake is highly dependent on the leading foil kinematics \cite{RibeiroFranck2020} and affects how or if the trailing foil interacts with the vortices originating from the leading foil. Through experiments and simulations performed in a flow at $Re=30,000$, Rival et al. \cite{rival2011} discovered two types of wake-foil interactions. The first type occurs when the LEV from the leading foil induces a leading edge suction region generating a thrust force on the trailing foil, hence decreasing power generation. The second type of interaction occurs when the trailing edge vortex (TEV) from the leading foil induces flow separation on the trailing foil’s upper surface, increasing power generation. Their conclusions are obtained using a two-foil system placed in close proximity to one another ($S_x=2c$) with small heave and pitch amplitudes ($h_o/c=0.50; \; \theta_o=8^{\circ}$). Through two-dimensional simulations at $Re=44,000$ and fixed kinematics and configuration parameters of $h_o/c=1.00; \; \theta_o=70^{\circ}; \; \psi=180^{\circ}$ and $S_x=5.4c$, Xu et al. \cite{xuLEV2016} demonstrate that the array efficiency linearly increases with increasing frequency, reaching a maximum at $fc/U_\infty=0.14$. As the frequency surpasses $fc/U_\infty=0.14$ the trailing foil performance is heavily influenced by the TEV from the leading foil. 

Using actuator disk theory, the optimal efficiency of a single turbine is up to $59.3\%$, as attributed to Betz \cite{betz}. This is expanded to inline turbine arrays by Newman, who demonstrates an optimal system efficiency of $64\%$ \cite{newman1986}, meaning that the sum of the power extracted by two devices is 64\% of the freestream kinetic energy. These traditional actuator disk models have been recently revisited by considering both the steady and unsteady components within the flow. Dabiri \cite{dabiri2020} has developed a theoretical framework that surpasses Betz's limit by relaxing the steady flow assumption, noting that this approach may be particularly useful in oscillating foil arrays to increase performance. From his framework, a theoretical time-averaged power coefficient of $76.4\%$ was obtained, which is significantly higher than Betz's limit. Inspired by Dabiri's framework, Young et al. \cite{young2020} analyzed the mean and unsteady flow around oscillating foils using a control volume analysis to compute efficiency in single and tandem arrangements. The unsteady terms that arose from the formation and shedding of vortices would entrain additional energy and momentum into the wake, which could be used to increase energy and hence efficiency on downstream foils. Young et al. validated their methodology through a numerical analysis on optimal kinematics obtained from Kinsey and Dumas \cite{kinseytandem2012} and achieved similar efficiencies with the control volume analysis as obtained via force computations. They estimated a theoretical maximum efficiency of $77.7\%$ for the tandem foil arrangement, emphasizing the effect of the unsteady terms not previously incorporated by Newman's limit. 

Through these recent investigations there is strong evidence that the structured wake, including the unsteadiness imposed by the coherent vortices in the wake, can be utilized beneficially in oscillating foil arrays. However, the generation of the vortices and wake dynamics are primarily governed by the kinematics, which in turn have a large parameter space within energy harvesting regime. Understanding how the leading foil kinematics change the wake structure and dynamics will lead to better wake-foil models, and development of optimal array configurations for energy harvesting. This can lay the foundation for models and control laws that govern the optimization of kinematics between foils within an array, and may also have implications in the bio-inspired propulsion field in terms of interactions between groups of swimmers or fliers.

This paper focus on quantifying the effects of foil kinematics within the wake structure and dynamics, and using this information to model the trailing foil’s performance. Simulations are performed with two-dimensional direct numerical simulation (DNS), along with experiments and  Particle Image Velocimetry (PIV). A wide range of kinematics with varying reduced frequencies, heave and pitch amplitudes, inter-foil phase and spacing will be used in a two-foil (tandem) array configuration in order to quantify the effects of foil kinematics in the wake. The vortex structure is used to characterize the wake into three main regimes, that correlate strongly with foil kinematics, and are described by the steady and unsteady components within the flow. Furthermore, a modification to the global phase model is proposed that can predict trailing foil performance over a wide range of kinematics, and a modified efficiency is introduced that incorporates available energy from steady and unsteady regions within the wake.

\section{Methods}

\subsection{Tandem-foil array parameters and performance metrics} \label{ss:parameters}

The kinematic motion of the foil is described below in lab-fixed coordinates as

\begin{equation}
h(t)=h_{o}\cos(2\pi f t + \psi) 
\label{eq:heave}
\end{equation}

\noindent and 

\begin{equation}
\theta(t)=\theta_{o}\cos(2\pi f t + \pi/2 + \psi)
\label{eq:pitch}
\end{equation}

\noindent where $h(t)$ and $\theta(t)$ are the prescribed heave and pitch kinematics, respectively, with a pitching motion about the center-chord. Using a fore-aft symmetric elliptical cross-section with a $c/2$ pivot location enables the design to be used in tidal flows that have regular flow reversal. Both the leading and trailing foil have the same kinematics except for an inter-foil phase ($\psi$) which is zero for the leading foil, and varies between $-180^{\circ}$ and $180^{\circ}$ for the trailing foil. Modifying pitch and heave simultaneously generates a time-varying relative angle of attack with respect to the freestream flow, which is given by

\begin{equation}
\alpha_{rel}(t) = \tan^{-1} (-\dot{h}(t)/U_{\infty}) + \theta(t),
\label{eq:alpharel}
\end{equation} 

\noindent with $\dot{h}(t)$ representing the time derivative of the heave motion and $\alpha_{rel}(t)$ is in radians.

A representative relative angle of attack is evaluated when the foil is at maximum $\theta$ and maximum heave velocity, which occurs at one quarter of the cycle period $T$, or

\begin{equation}
\alpha_{T/4} =\alpha_{rel}(t = 0.25T).
\label{eq:at4}
\end{equation}

In order to evaluate the performance of different kinematic conditions, the foil's efficiency is defined as

\begin{equation}
\eta = \frac{\overline{P}}{\frac{1}{2}\rho U_{\infty}^3 Y_p}
\label{eq:eta}
\end{equation}

\noindent which is the ratio of the average power extracted, $\overline{P}$, to the power available in the oncoming flow window defined by the swept area $Y_p$. Power generation on an oscillating foil is defined as

\begin{equation}
P(t)=F_y \dot{h} + M_z \dot{\theta}
\label{eq:power}
\end{equation}

which is comprised on a translational component from the vertical force $F_y$, and an angular component from the spanwise moment $M_z$. All quantities reported are non-dimensionalized by the freestream velocity, $U_\infty$, and the chord length of a single foil, $c$. To remove small cycle-to-cycle variations, the efficiency, forces, and flow fields are all phase-averaged over the last two cycles of simulation and the experiments are phase-averaged over $10$ cycles.

The array configuration parameters of inter-foil phase, $\psi$, and inter-foil distance, $S_x$, are varied between computations and experiments in order to cover a wider parameter space, but contain overlap for validation purposes.

The computations are performed with three heave amplitudes in the range of $h_o=0.75-1.25$, three pitch amplitudes between $\theta_o=55^{\circ}-75^{\circ}$, and eight reduced frequencies in the range of $fc/U_{\infty}=0.10-0.17$ for a total of 21 unique kinematics. These kinematics are performed at an inter-foil distance of $S_x=6$ for 12 distinct inter-foil phases. To further explore separation distance, the kinematics of $fc/U_\infty=0.12; h_o=1.00; \theta_o=65^{\circ}$ is also explored for separation distances of $S_x=4$ and $5$ at 12 inter-foil phases. There is a total of 276 unique simulations. 

The experiments are performed with a fixed heave/pitch amplitudes of $h_o=1.0$ and $\theta_o=65^{\circ}$, with six reduced frequencies from $fc/U_{\infty}=0.10-0.15$ for a total of the six unique kinematics. Experiments are performed at a single inter-foil distance of $S_x=6$ with 36 different inter-foil phases. To explore separation distance, the same set of foil kinematics as the simulations is performed at $S_x=8$ and $9$, for 36 inter-foil phases.  In experiments the inter-foil phase is varied from $-180^{\circ}$ to $180^{\circ}$ in increments of $10^{\circ}$ for each set of kinematics, whereas the simulations explore the same range of inter-foil phase sampled at every $30^{\circ}$.

Table \ref{t:tablekin} summarizes the kinematics and configuration of all cases, including the $\alpha_{T/4}$ values.

\begin{table}[htbp]
\setlength{\tabcolsep}{8pt}
\centering
\caption{Kinematics and array configurations with their respective $\alpha_{T/4}$ values. S: Simulations; E: Experiments; PIV: Particle Image Velocimetry. Markers used in the results section are shown.}
\begin{tabular}{ccccc}
\hline
\textbf{Marker} & \textbf{Kinematics} & \textbf{$\alpha_{T/4}$} &\textbf{$S_x$} & \textbf{Data} \\ \hline
\begin{minipage}{0.03\textwidth}
 \includegraphics[width=\linewidth]{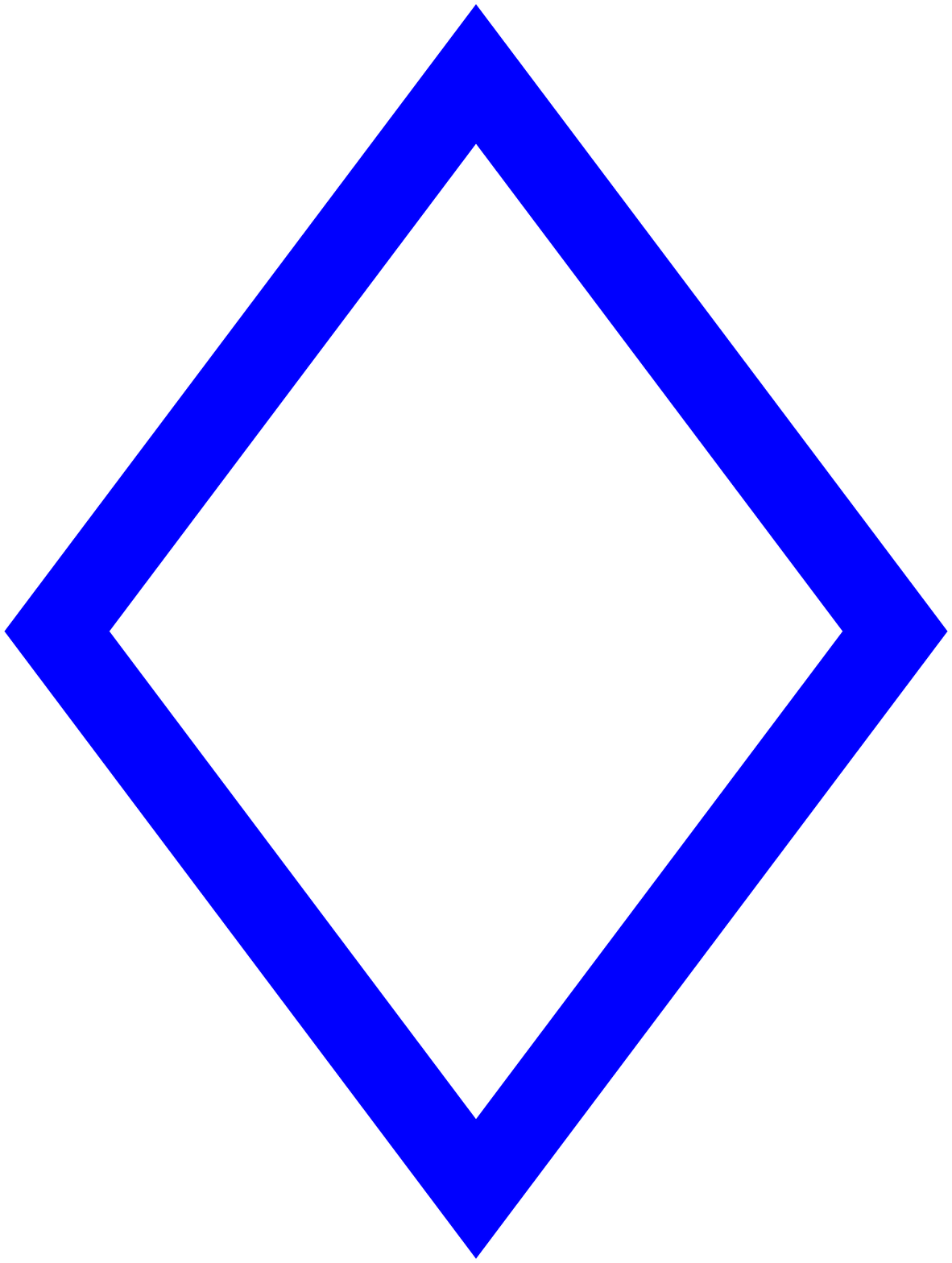}
\end{minipage} & $fc/U_\infty = 0.12 ; h_o = 1.50 ; \theta_o = 55^{\circ}$ & 0.11 & 6 & S\\
\begin{minipage}{0.03\textwidth}
 \includegraphics[width=\linewidth]{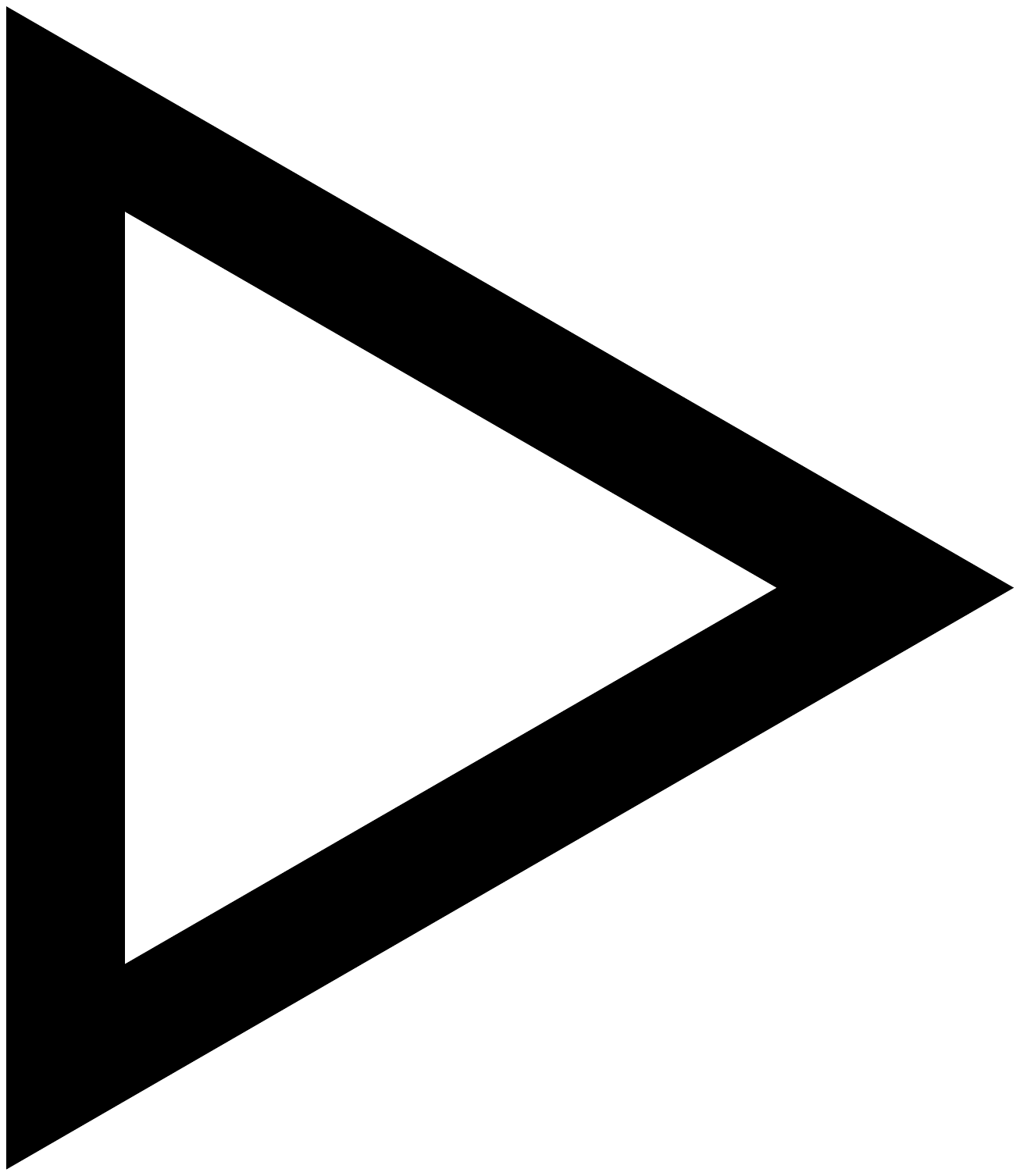}
\end{minipage} & $fc/U_\infty = 0.17 ; h_o = 1.00 ; \theta_o = 55^{\circ}$ & 0.14 & 6 & S\\
\begin{minipage}{0.03\textwidth}
 \includegraphics[width=\linewidth]{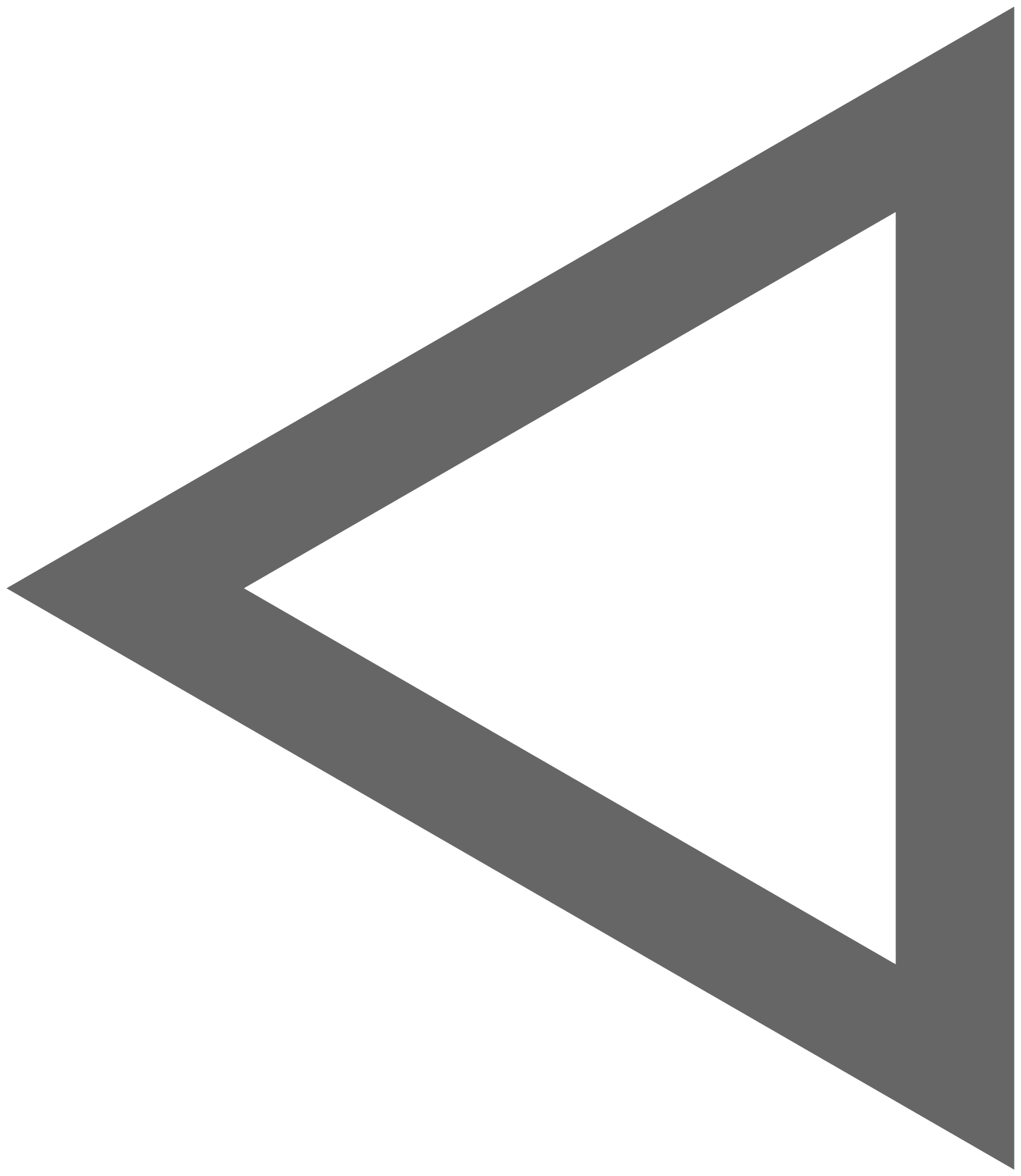}
\end{minipage} & $fc/U_\infty = 0.16 ; h_o = 1.00 ; \theta_o = 55^{\circ}$ & 0.17 & 6 & S\\
\begin{minipage}{0.03\textwidth}
 \includegraphics[width=\linewidth]{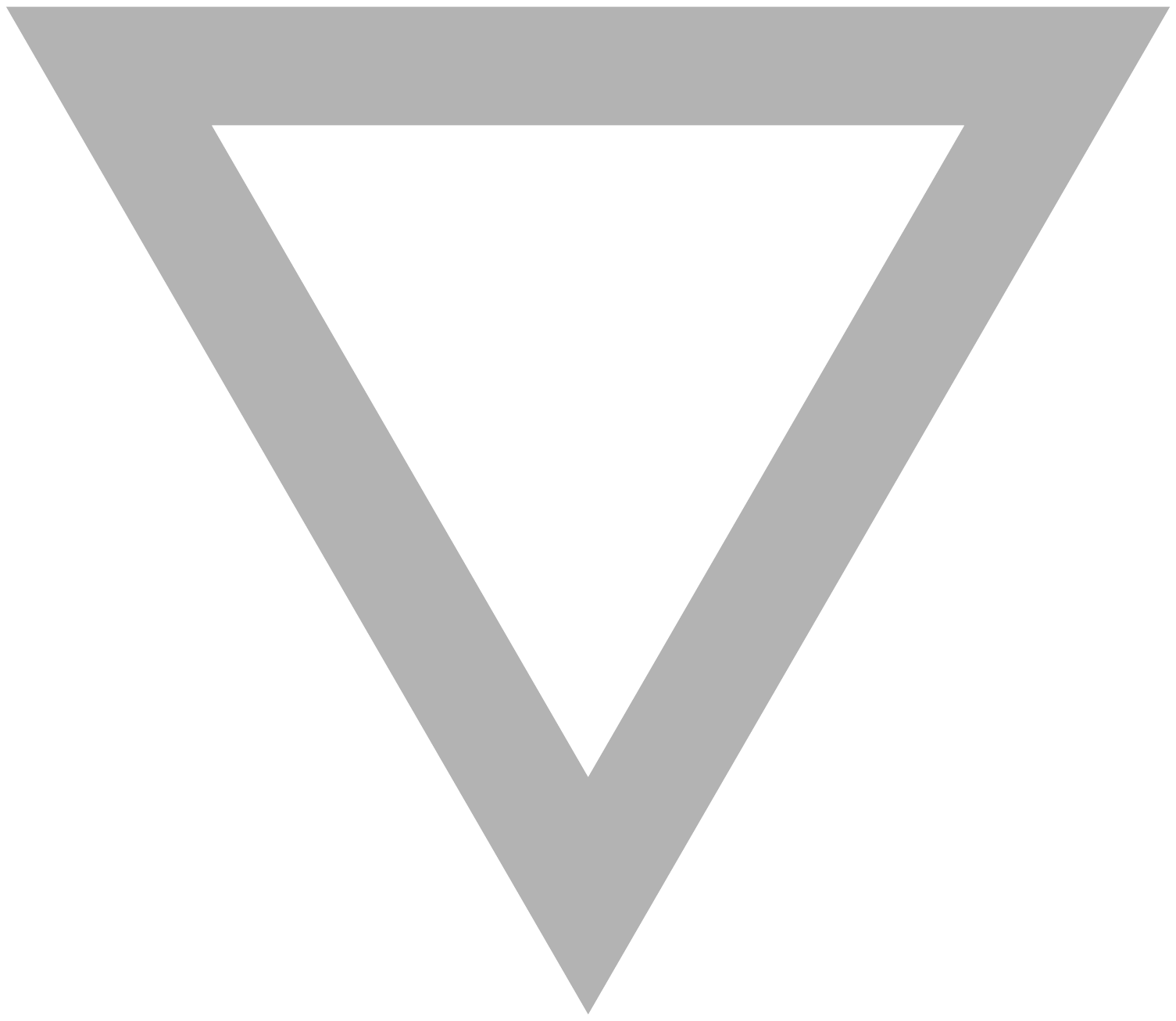}
\end{minipage} & $fc/U_\infty = 0.15 ; h_o = 1.00 ; \theta_o = 55^{\circ}$ & 0.20 & 6 & S\\
\begin{minipage}{0.03\textwidth}
 \includegraphics[width=\linewidth]{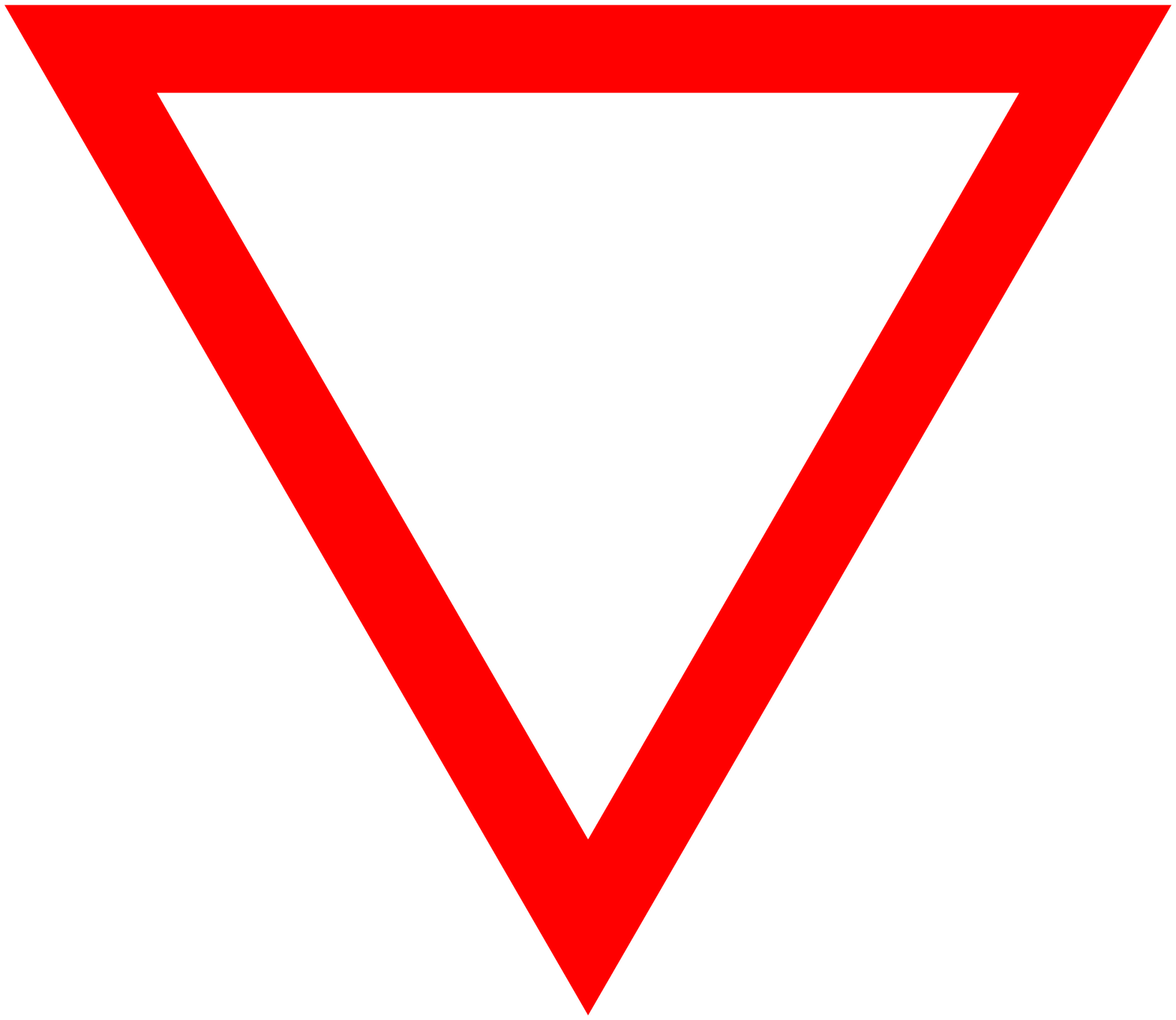}
\end{minipage} & $fc/U_\infty = 0.15 ; h_o = 1.25 ; \theta_o = 65^{\circ}$ & 0.27 & 6 & S\\
\begin{minipage}{0.03\textwidth}
 \includegraphics[width=\linewidth]{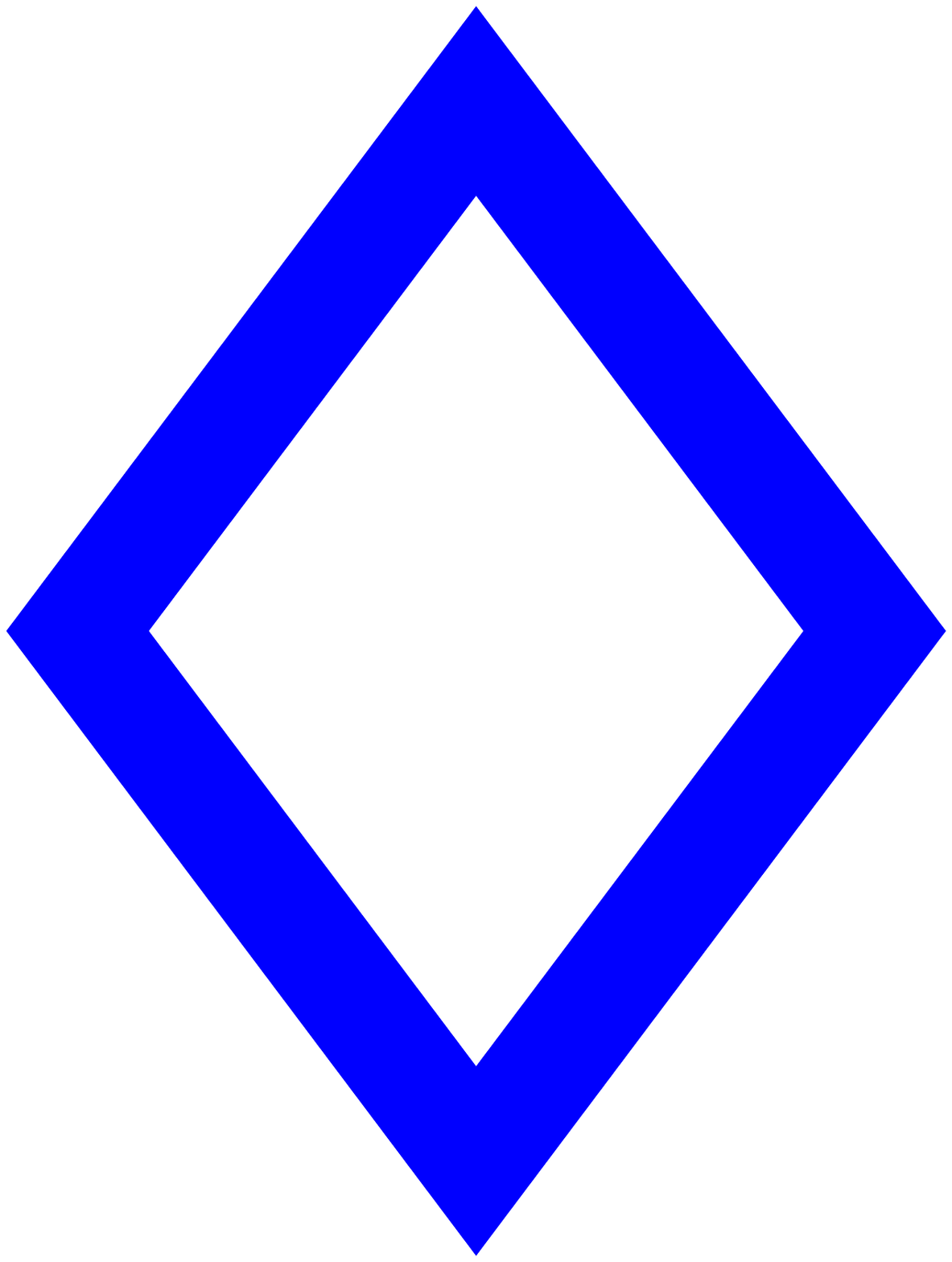}
\end{minipage} & $fc/U_\infty = 0.12 ; h_o = 1.00 ; \theta_o = 55^{\circ}$ & 0.31 & 6 & S\\
\begin{minipage}{0.03\textwidth}
 \includegraphics[width=\linewidth]{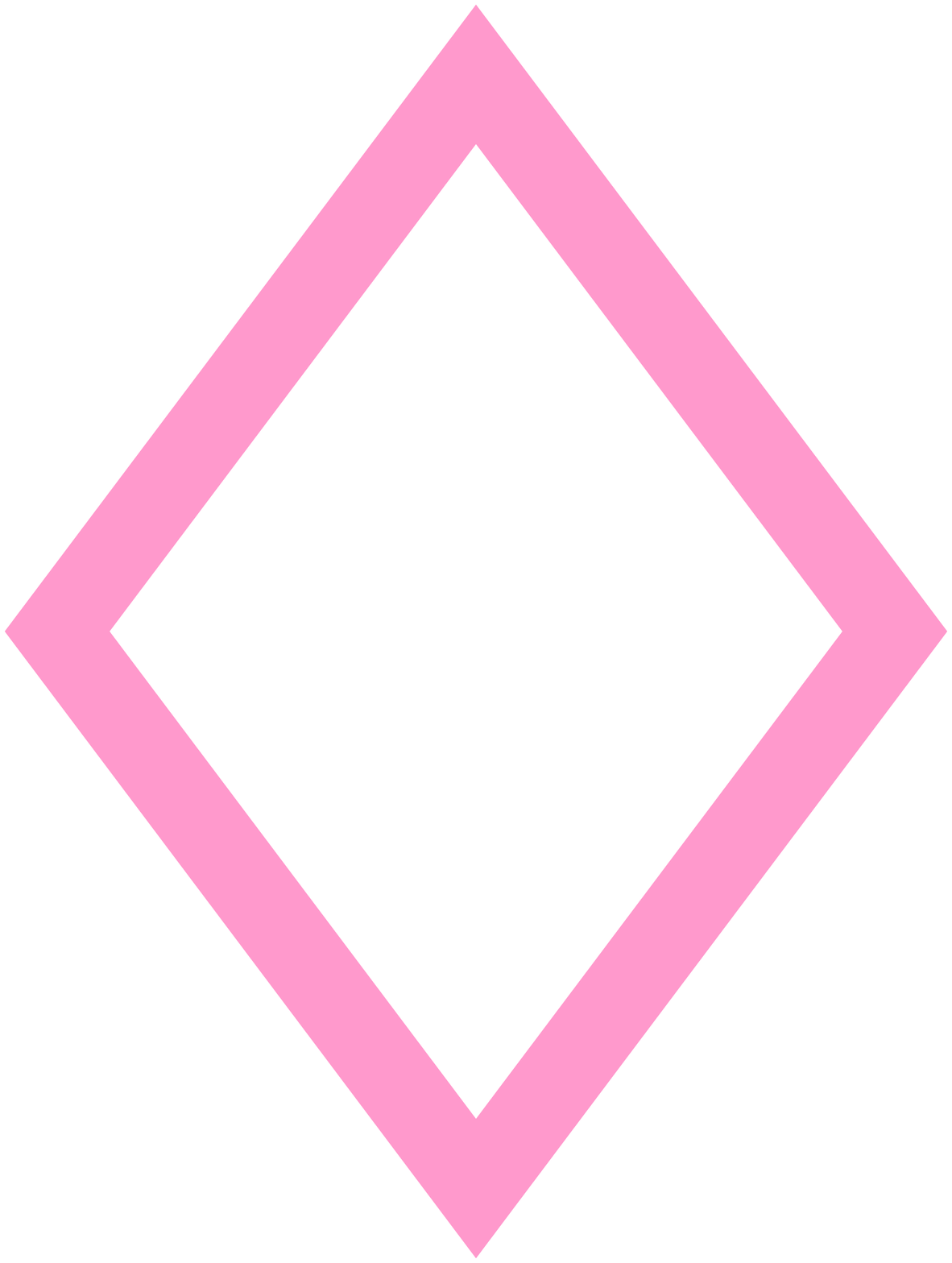}
\end{minipage} & $fc/U_\infty = 0.12 ; h_o = 1.25 ; \theta_o = 65^{\circ}$ & 0.38 & 6 & S\\
\begin{minipage}[t]{0.18\textwidth}
 S \raisebox{-0.2\height}{\includegraphics[width=0.18\linewidth]{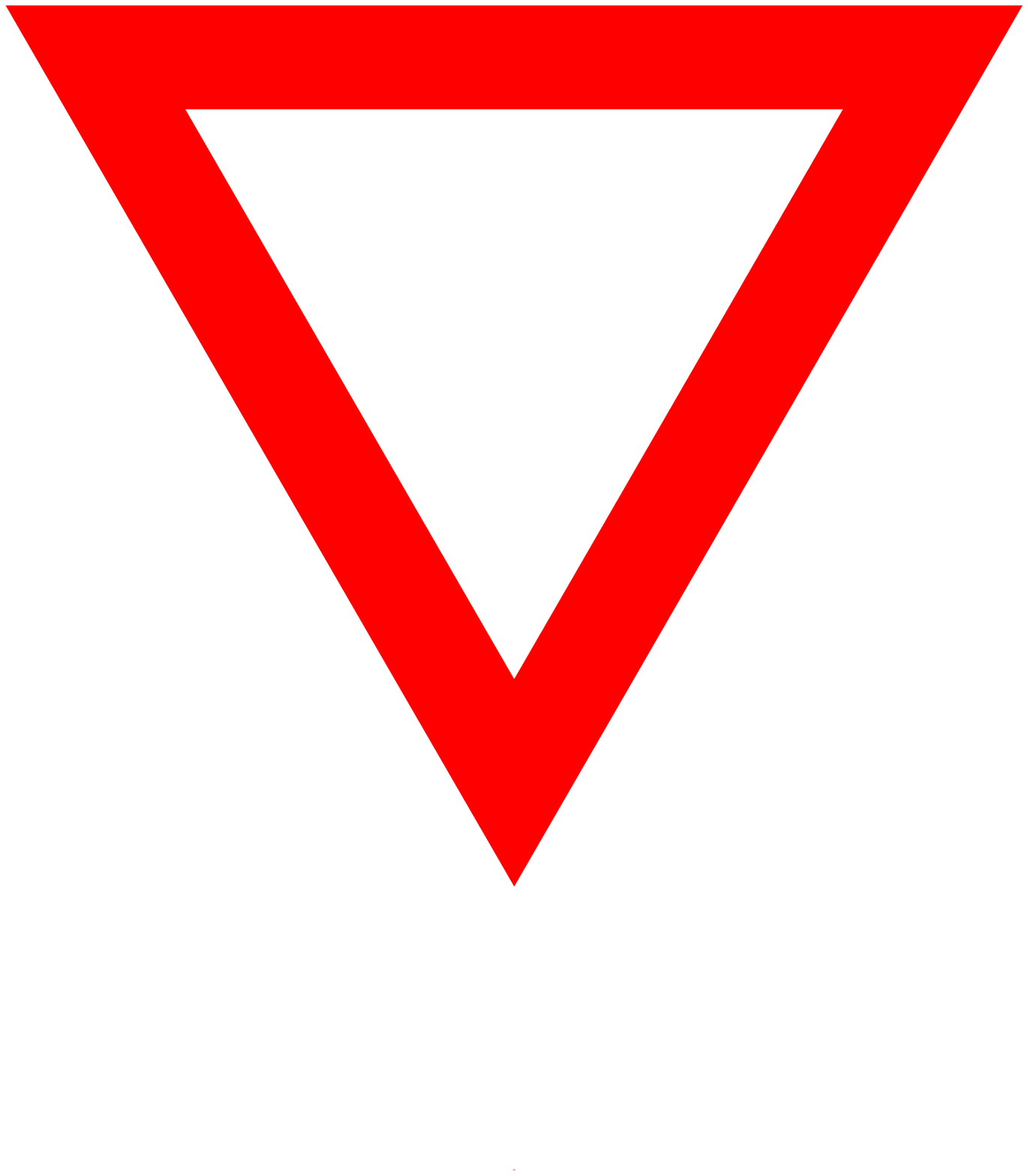}}
 E \raisebox{-0.23\height}{\includegraphics[width=0.18\linewidth]{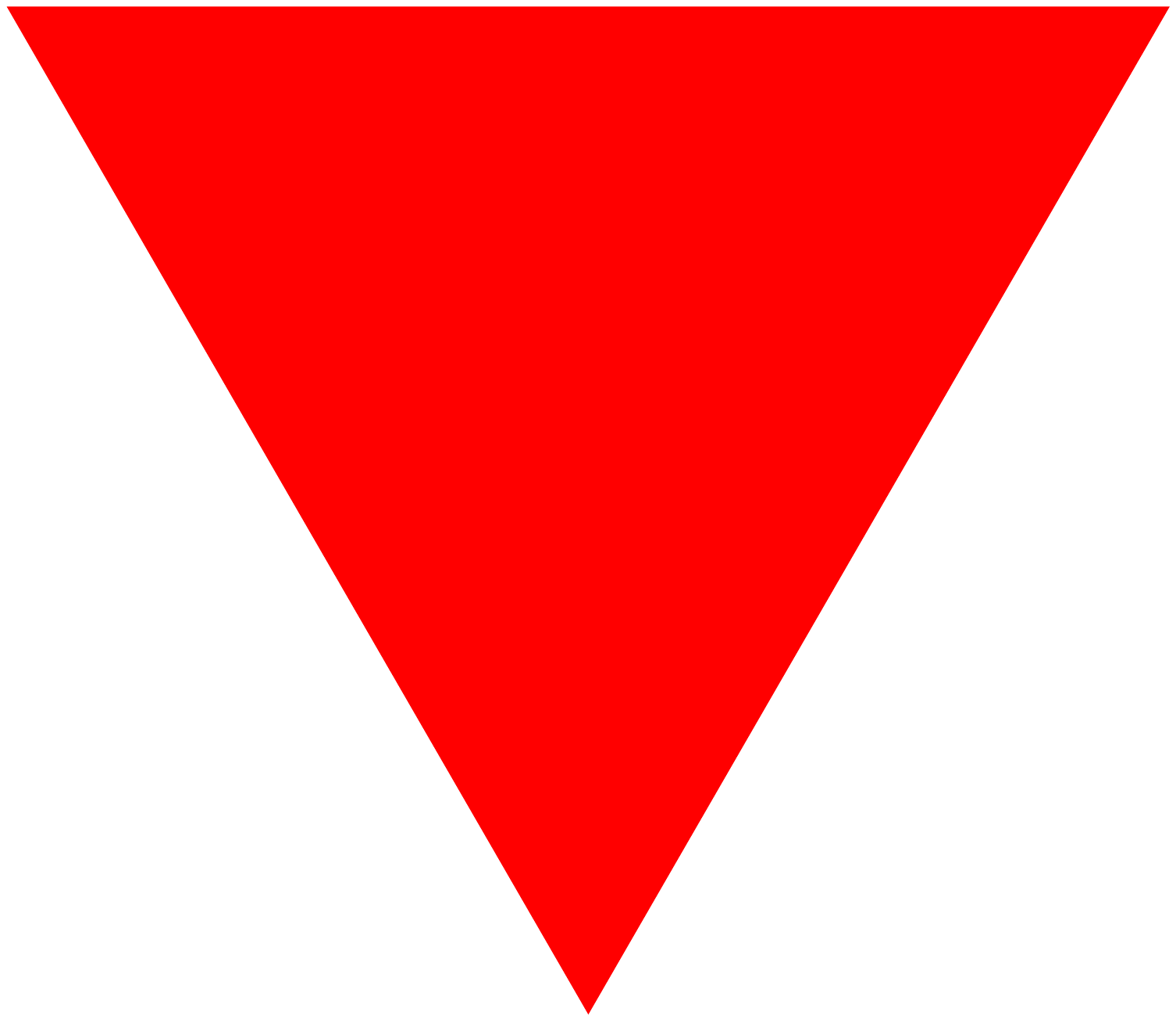}}
\end{minipage} & $fc/U_\infty = 0.15 ; h_o = 1.00 ; \theta_o = 65^{\circ}$ & 0.38 & 6 & S,E,PIV\\
\begin{minipage}{0.03\textwidth}
 \includegraphics[width=\linewidth]{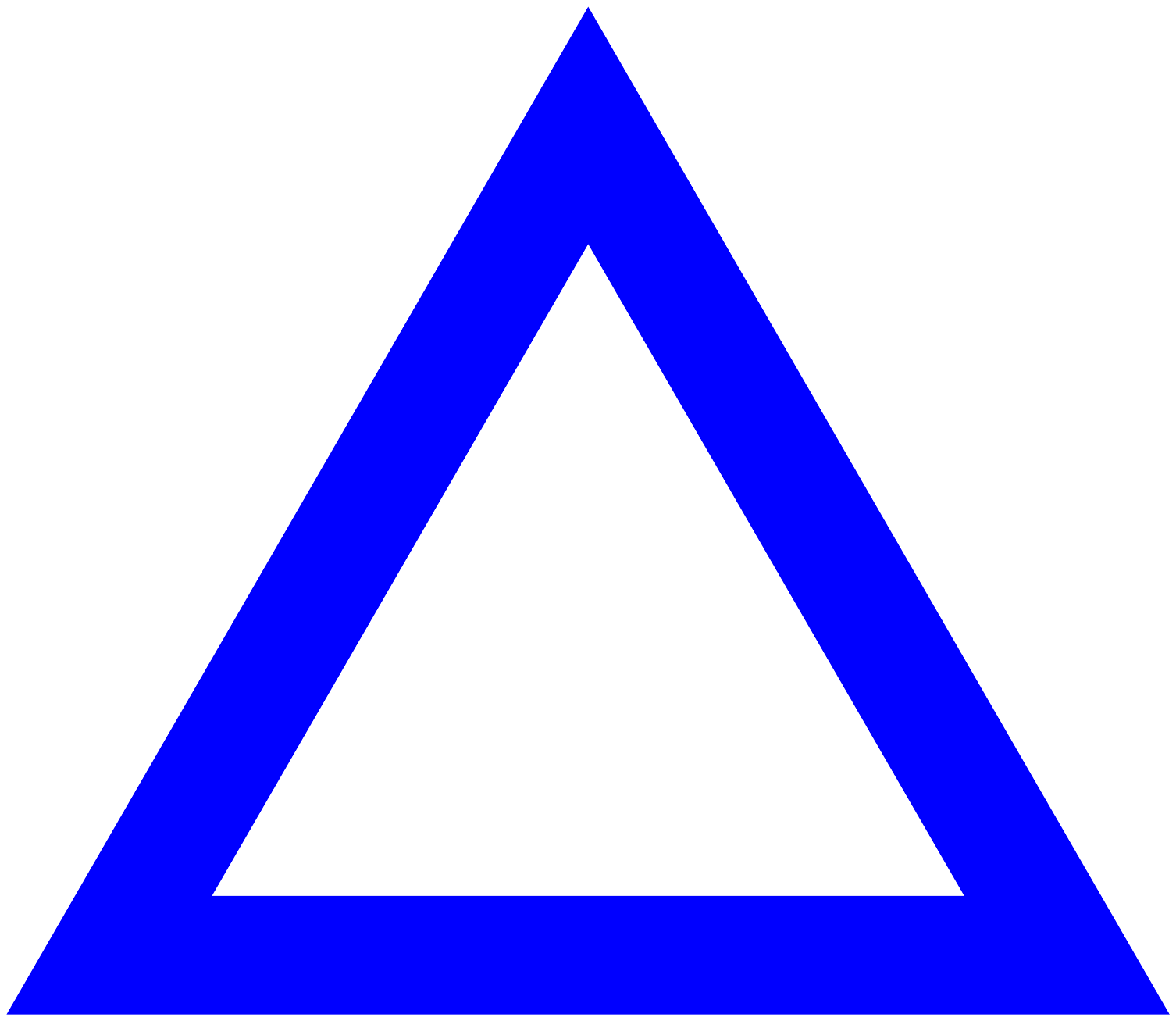}
\end{minipage} & $fc/U_\infty = 0.10 ; h_o = 1.00 ; \theta_o = 55^{\circ}$ & 0.40 & 6 & S\\
\begin{minipage}[t]{0.18\textwidth}
 S \raisebox{-0.2\height}{\includegraphics[width=0.18\linewidth]{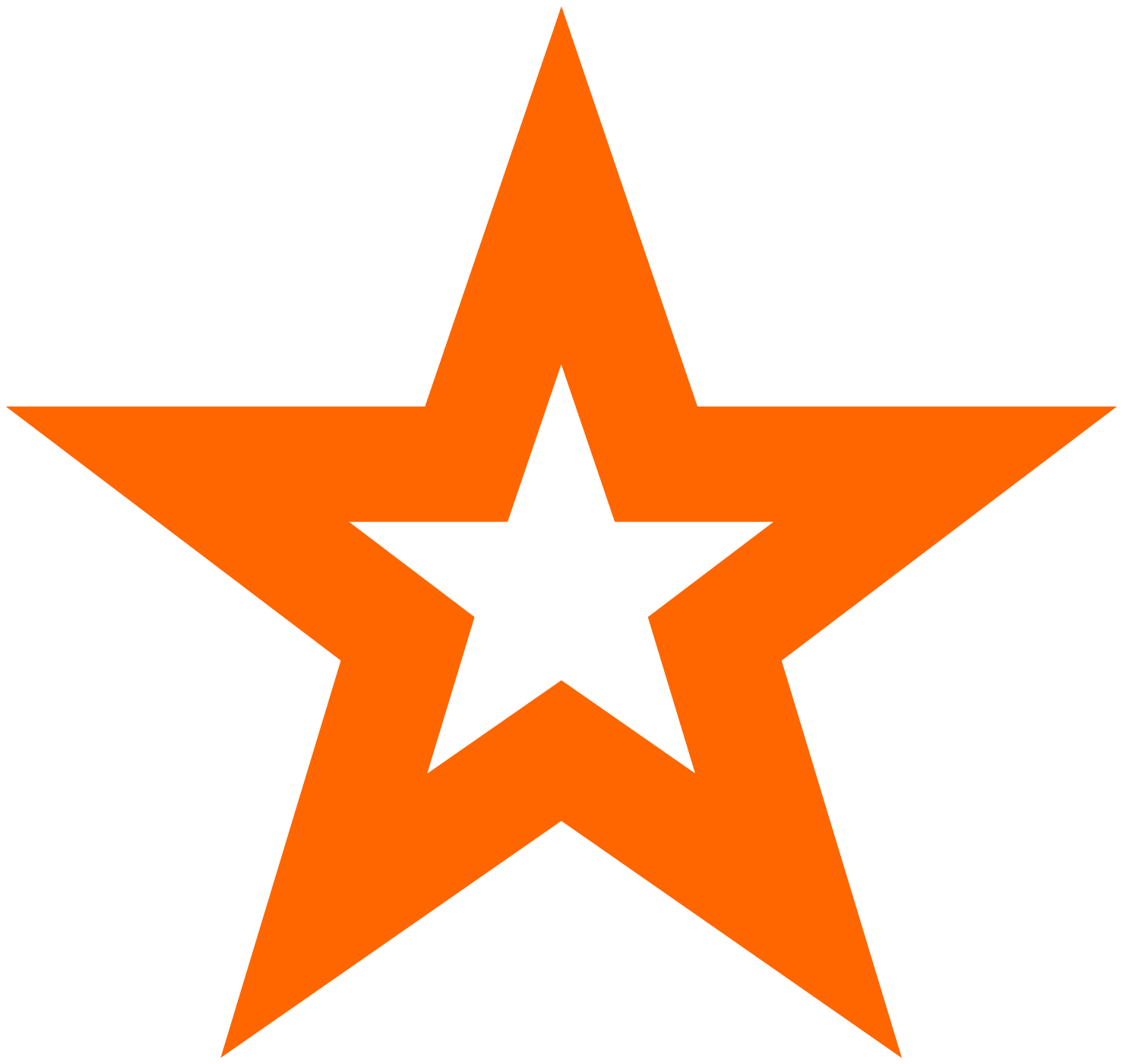}}
 E \raisebox{-0.2\height}{\includegraphics[width=0.18\linewidth]{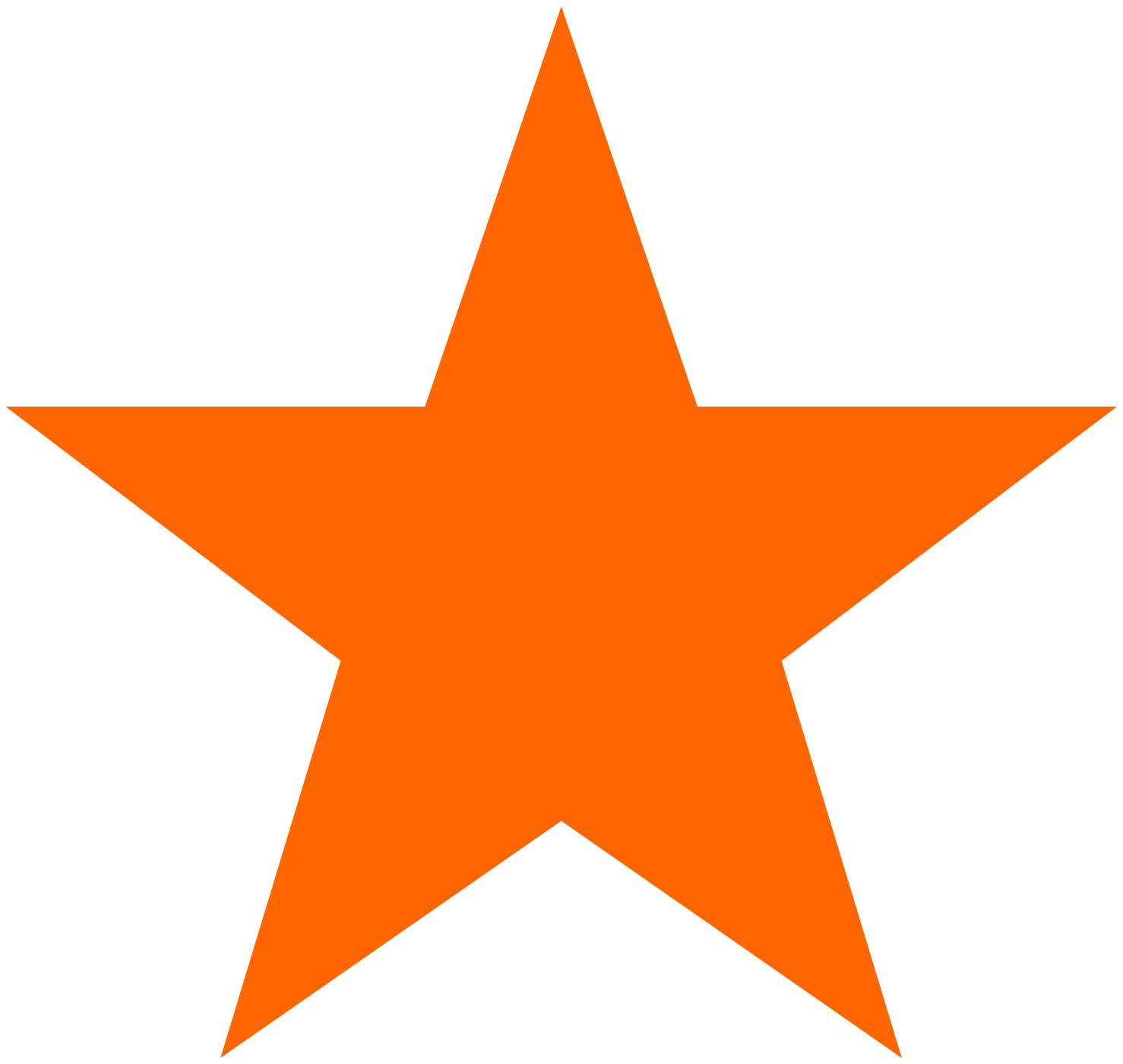}}
\end{minipage}  & $fc/U_\infty = 0.14 ; h_o = 1.00 ; \theta_o = 65^{\circ}$ & 0.41 & 6 & S,E\\
\begin{minipage}[t]{0.18\textwidth}
 S \raisebox{-0.2\height}{\includegraphics[width=0.18\linewidth]{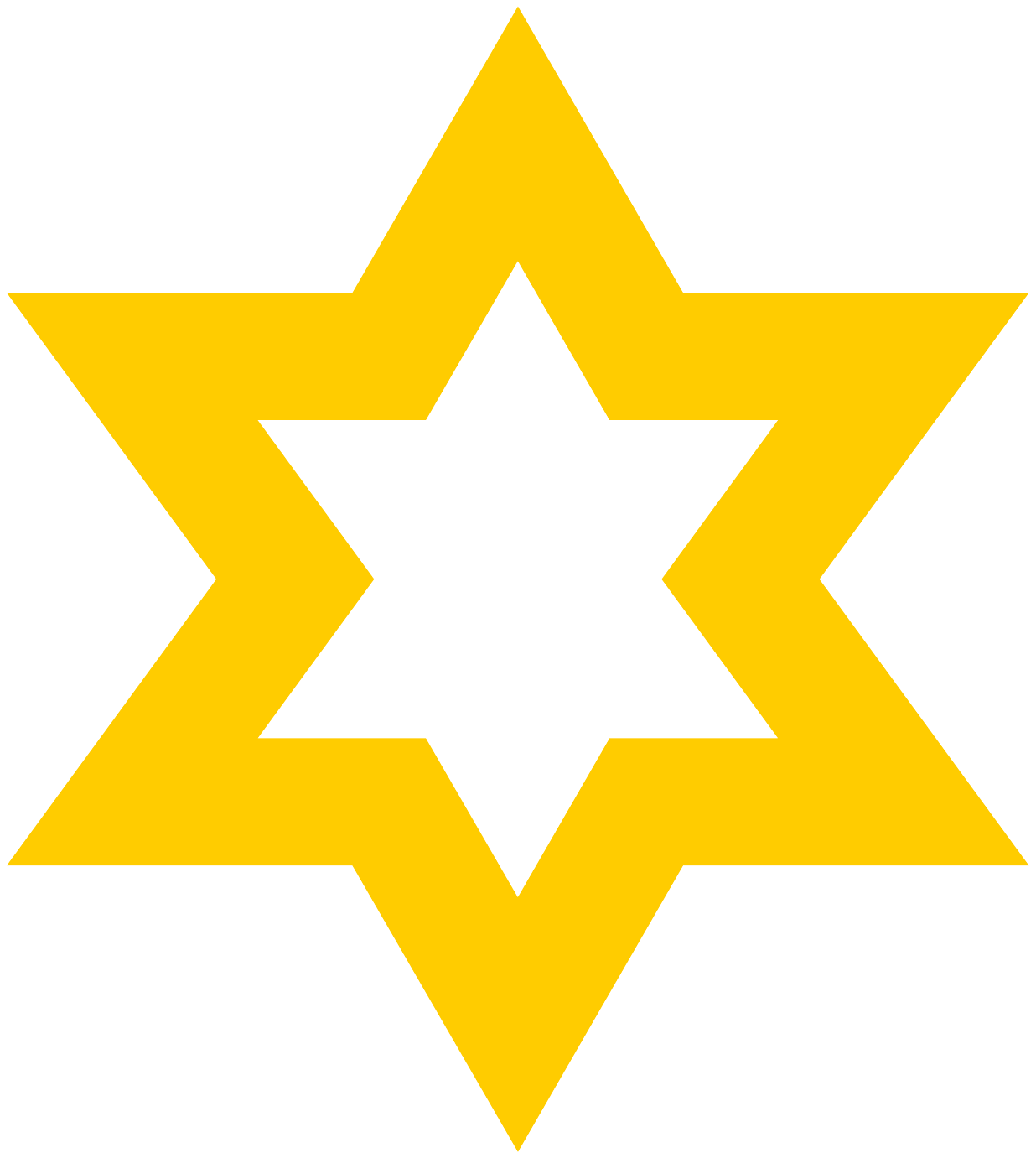}}
 E \raisebox{-0.22\height}{\includegraphics[width=0.18\linewidth]{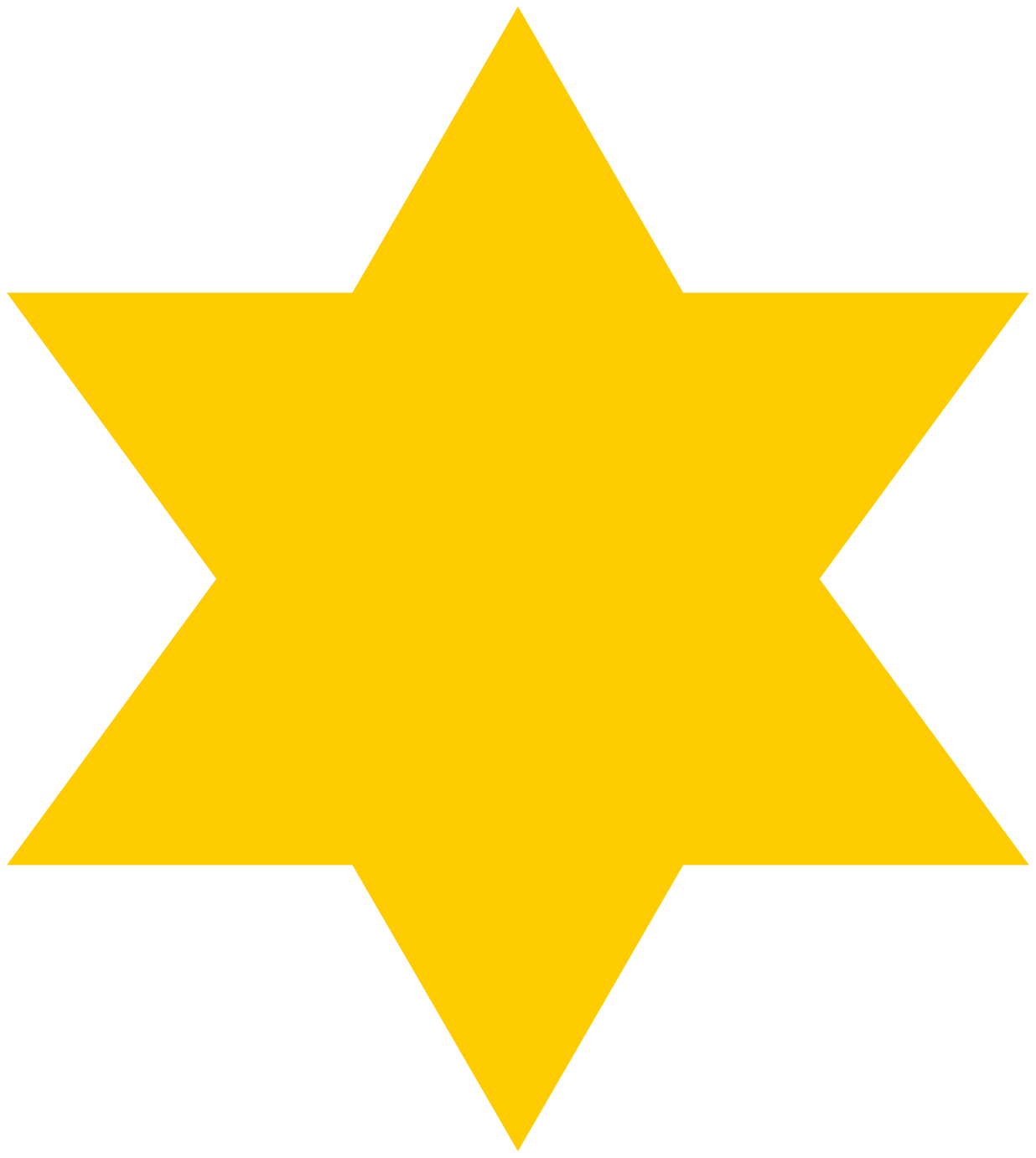}}
\end{minipage} & $fc/U_\infty = 0.13 ; h_o = 1.00 ; \theta_o = 65^{\circ}$ & 0.45 & 6 & S,E \\
\begin{minipage}{0.03\textwidth}
 \includegraphics[width=\linewidth]{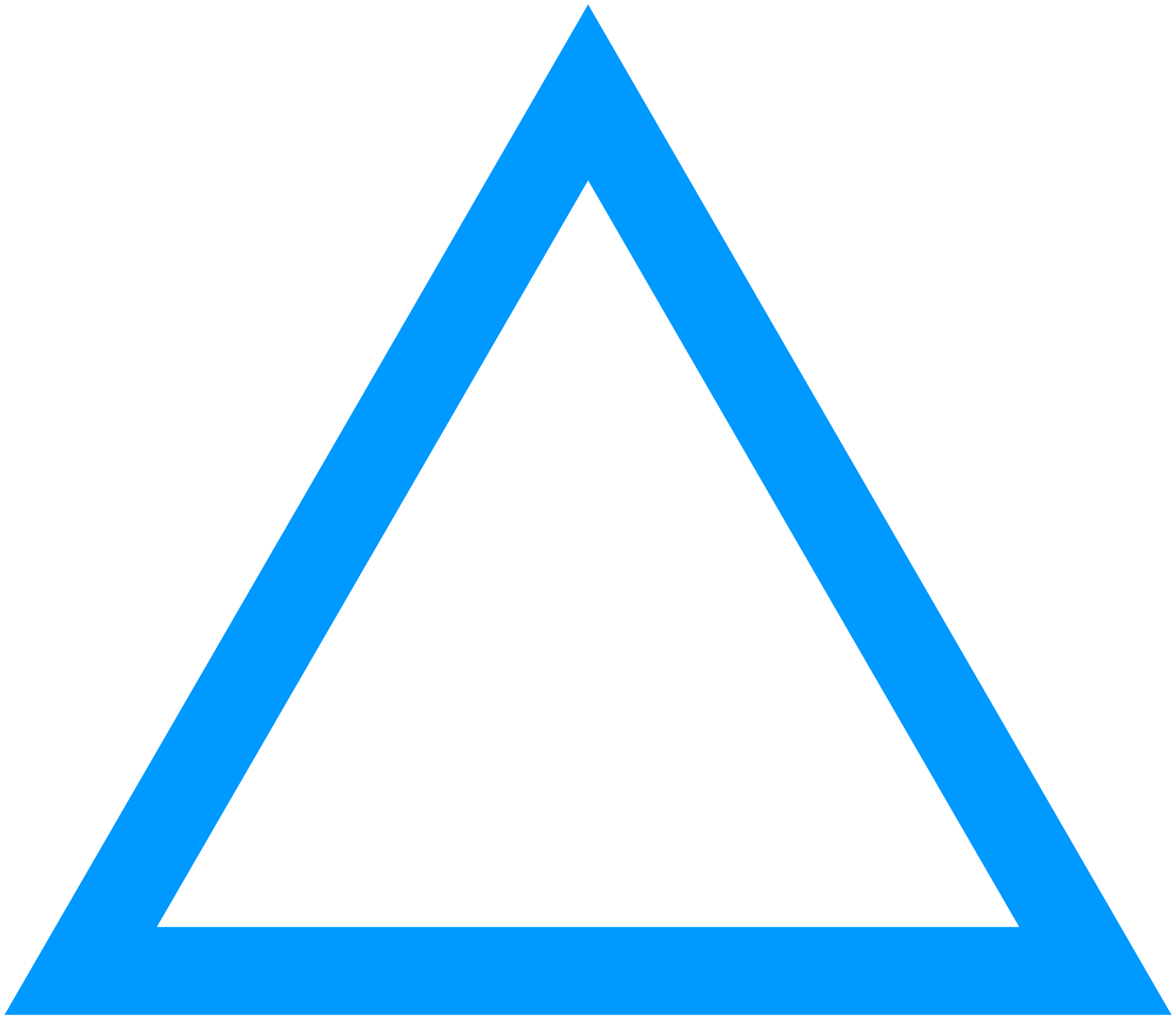}
\end{minipage} & $fc/U_\infty = 0.10 ; h_o = 1.25 ; \theta_o = 65^{\circ}$ & 0.47 & 6 & S\\
\begin{minipage}{0.03\textwidth}
 \includegraphics[width=\linewidth]{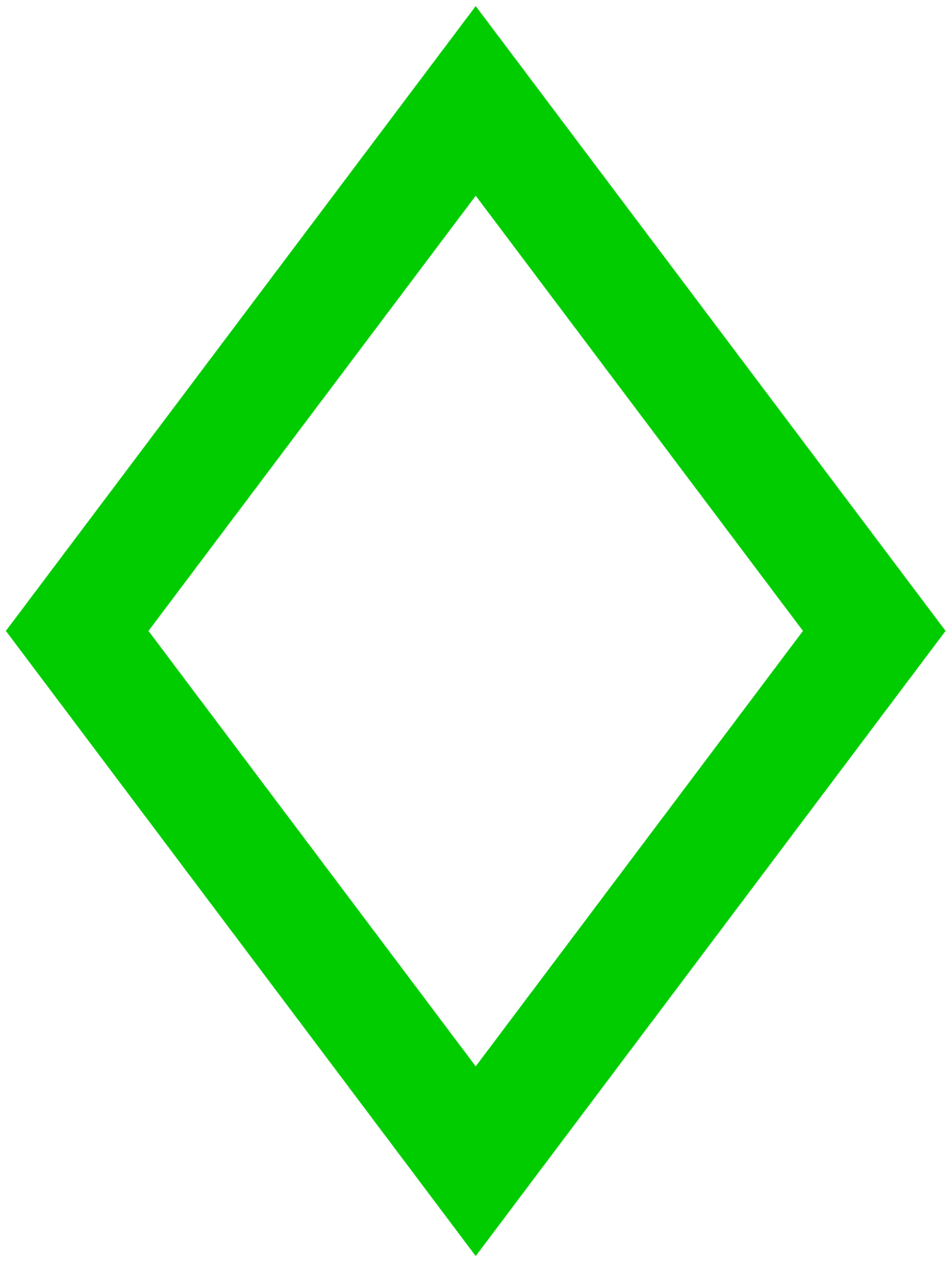}
\end{minipage} & $fc/U_\infty = 0.12 ; h_o = 1.00 ; \theta_o = 65^{\circ}$ & 0.49 & 4 & S\\
\begin{minipage}{0.03\textwidth}
 \includegraphics[width=\linewidth]{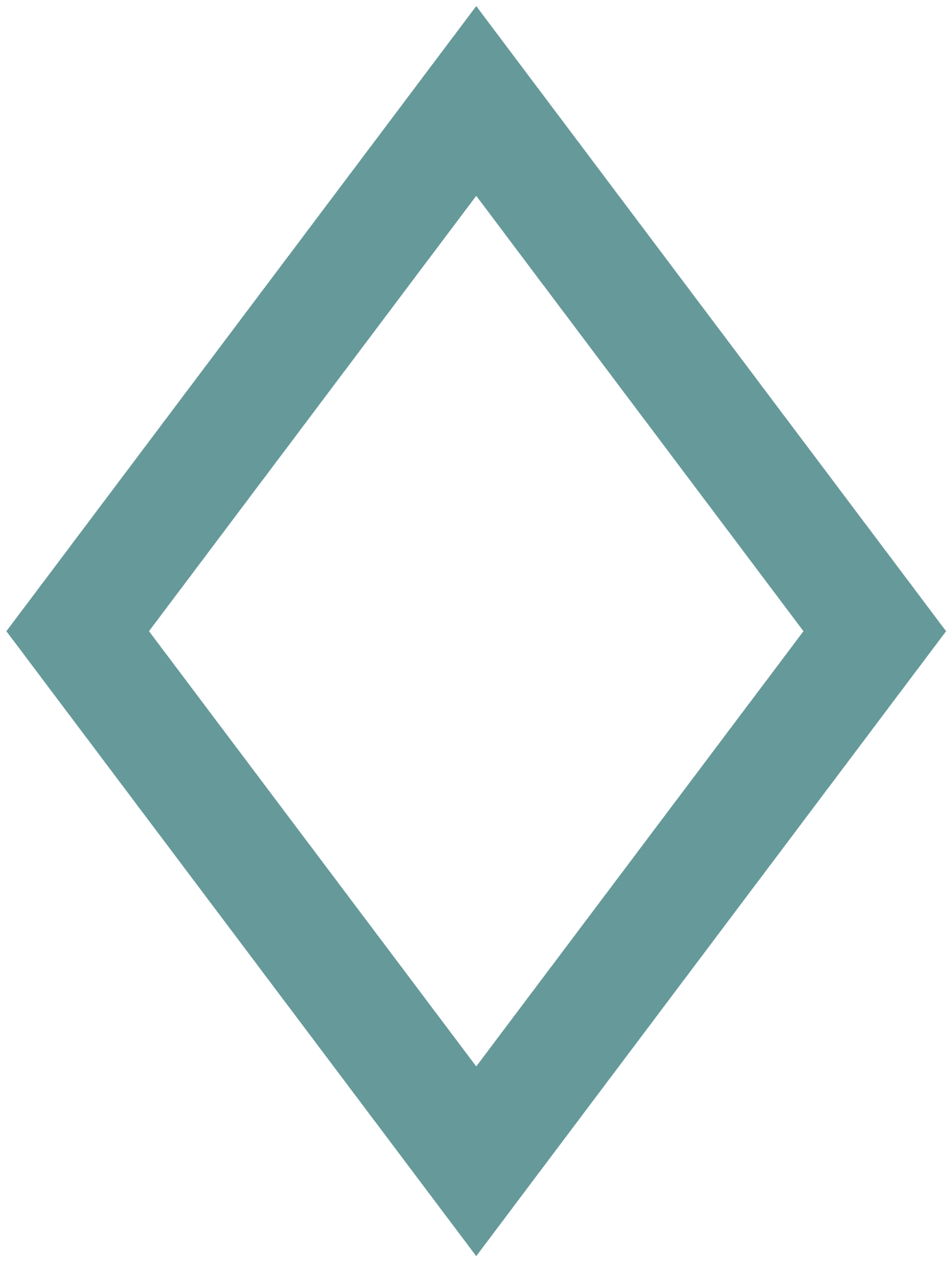}
\end{minipage} & & & 5 & S\\
\begin{minipage}[t]{0.18\textwidth}
 S \raisebox{-0.2\height}{\includegraphics[width=0.18\linewidth]{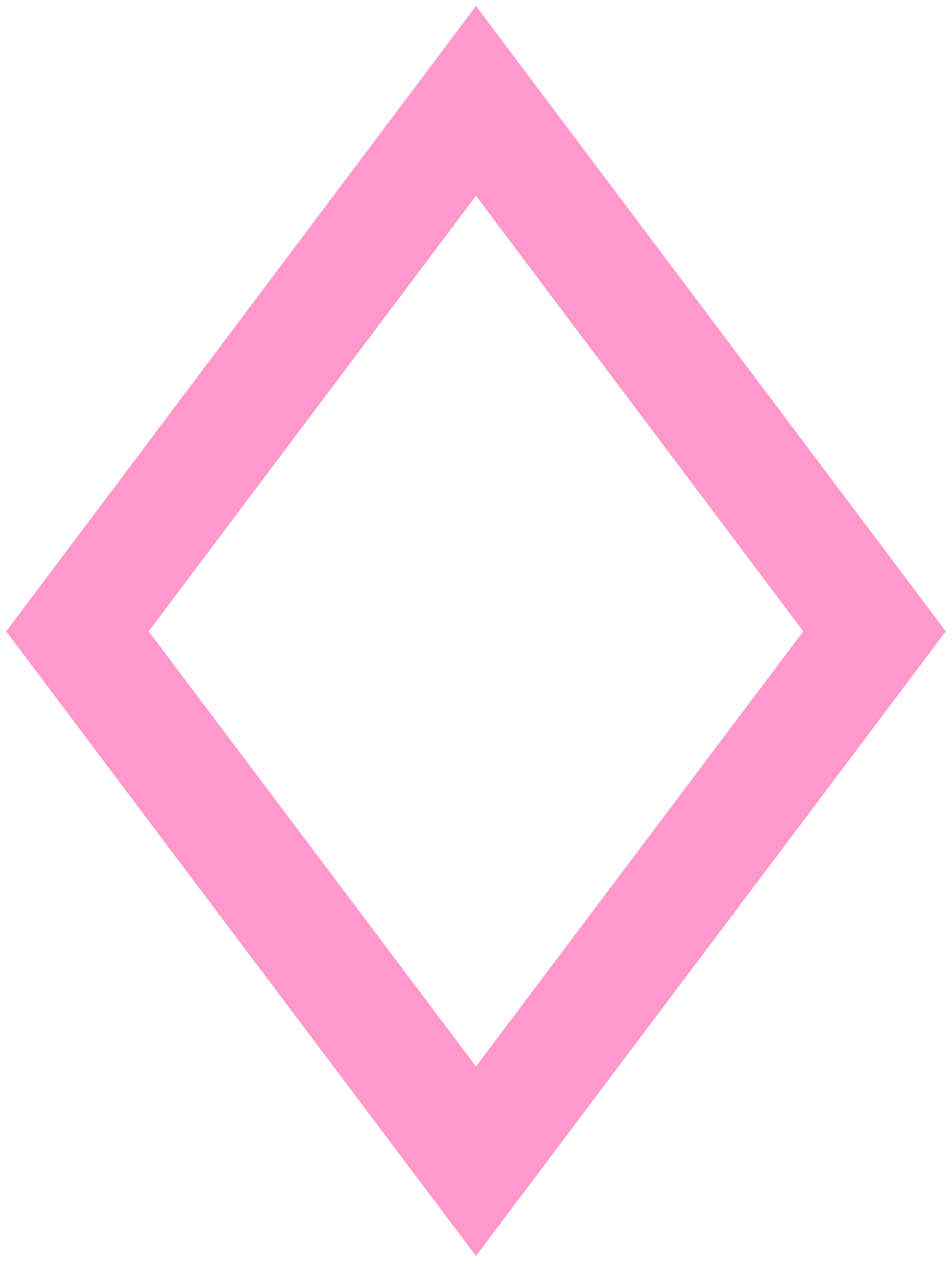}}
 E \raisebox{-0.2\height}{\includegraphics[width=0.18\linewidth]{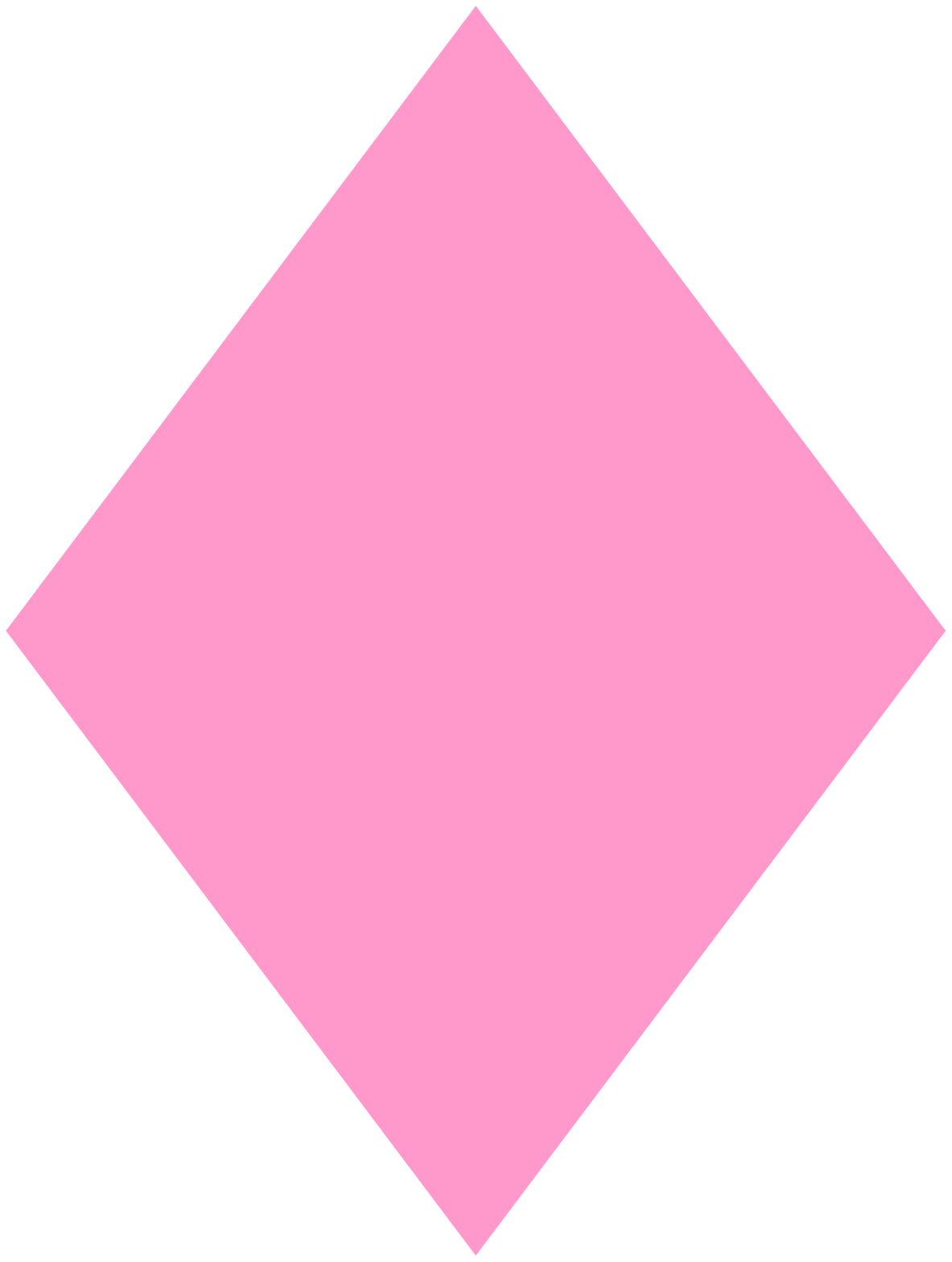}}
\end{minipage} & & & 6 & S,E,PIV\\
\begin{minipage}{0.03\textwidth}
 \includegraphics[width=\linewidth]{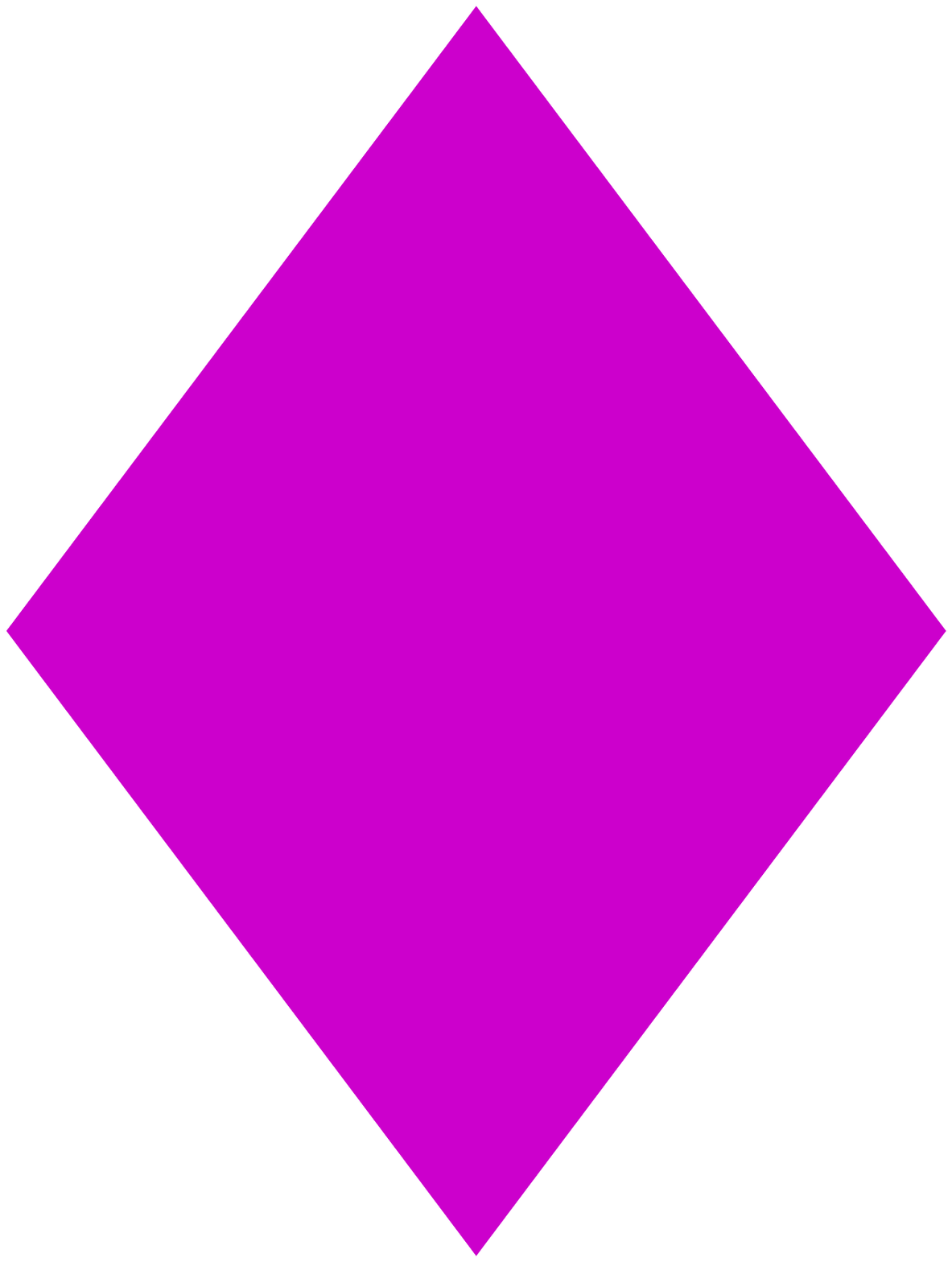}
\end{minipage} & & & 8 & E\\
\begin{minipage}{0.03\textwidth}
 \includegraphics[width=\linewidth]{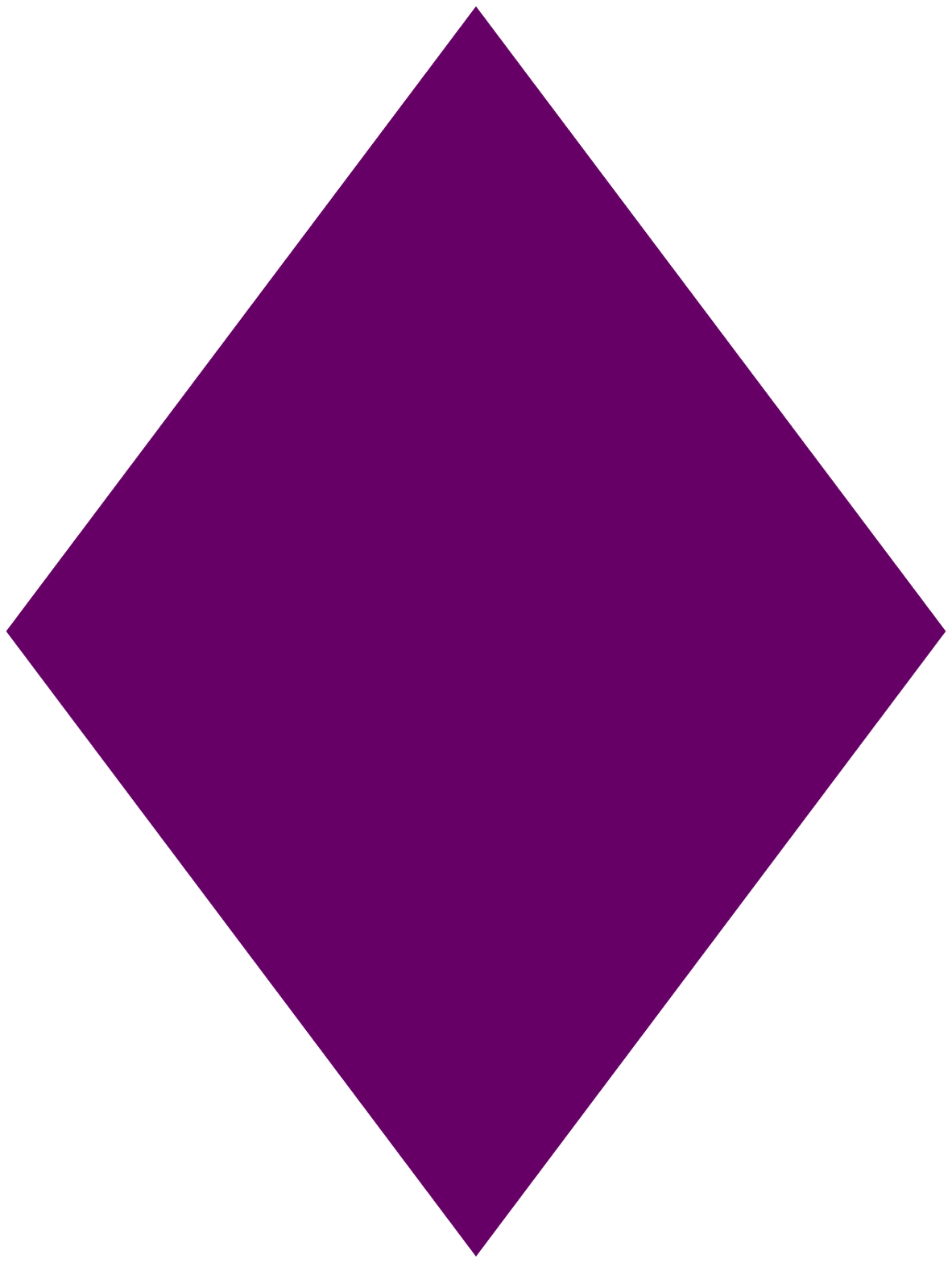}
\end{minipage} & & & 9 & E\\
\begin{minipage}{0.03\textwidth}
 \includegraphics[width=\linewidth]{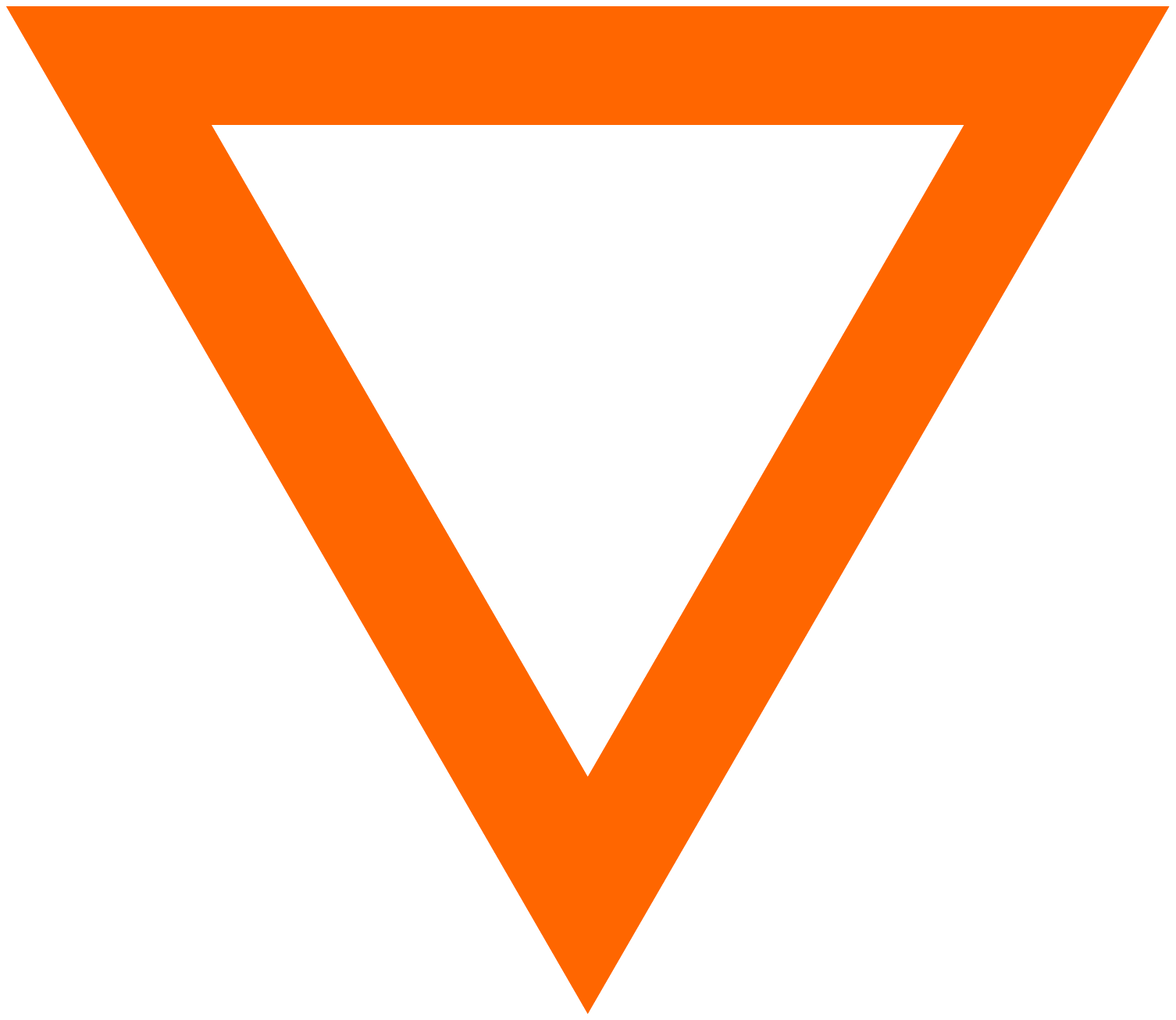}
\end{minipage} & $fc/U_\infty = 0.15 ; h_o = 0.75 ; \theta_o = 65^{\circ}$ & 0.52 & 6 & S\\
\begin{minipage}[t]{0.18\textwidth}
 S \raisebox{-0.2\height}{\includegraphics[width=0.18\linewidth]{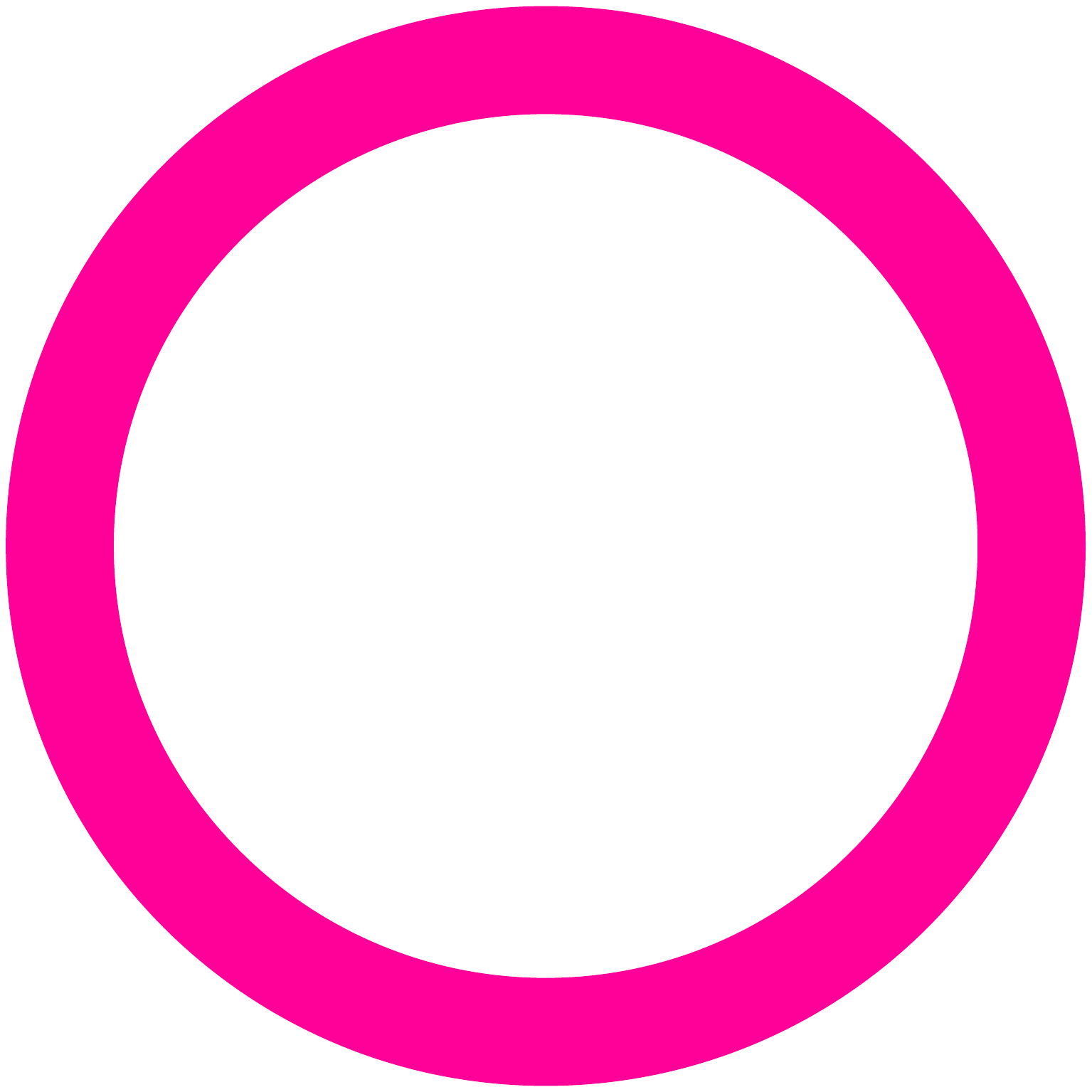}}
 E \raisebox{-0.2\height}{\includegraphics[width=0.18\linewidth]{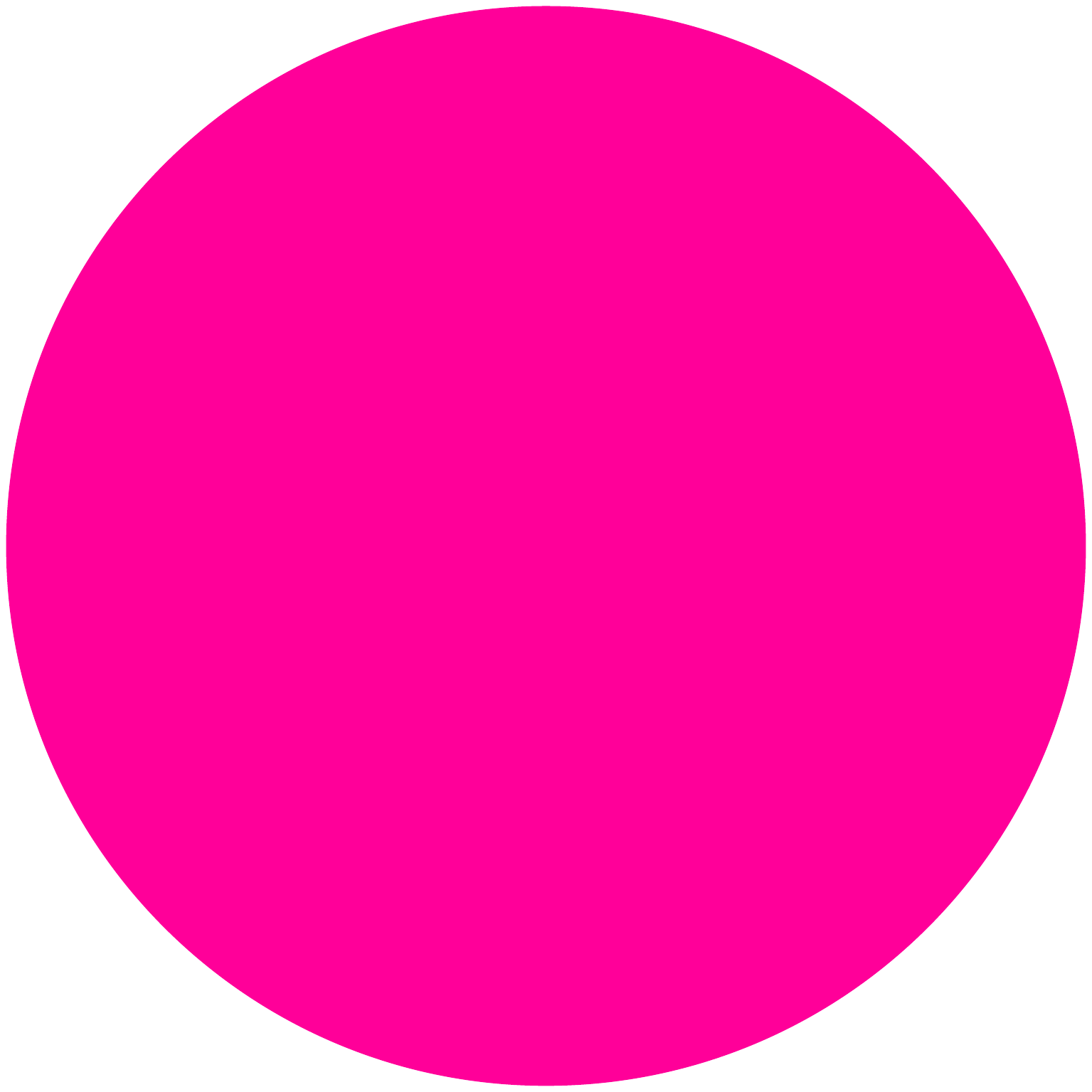}}
\end{minipage} & $fc/U_\infty = 0.11 ; h_o = 1.00 ; \theta_o = 65^{\circ}$ & 0.53 & 6 & S,E \\
\begin{minipage}{0.03\textwidth}
 \includegraphics[width=\linewidth]{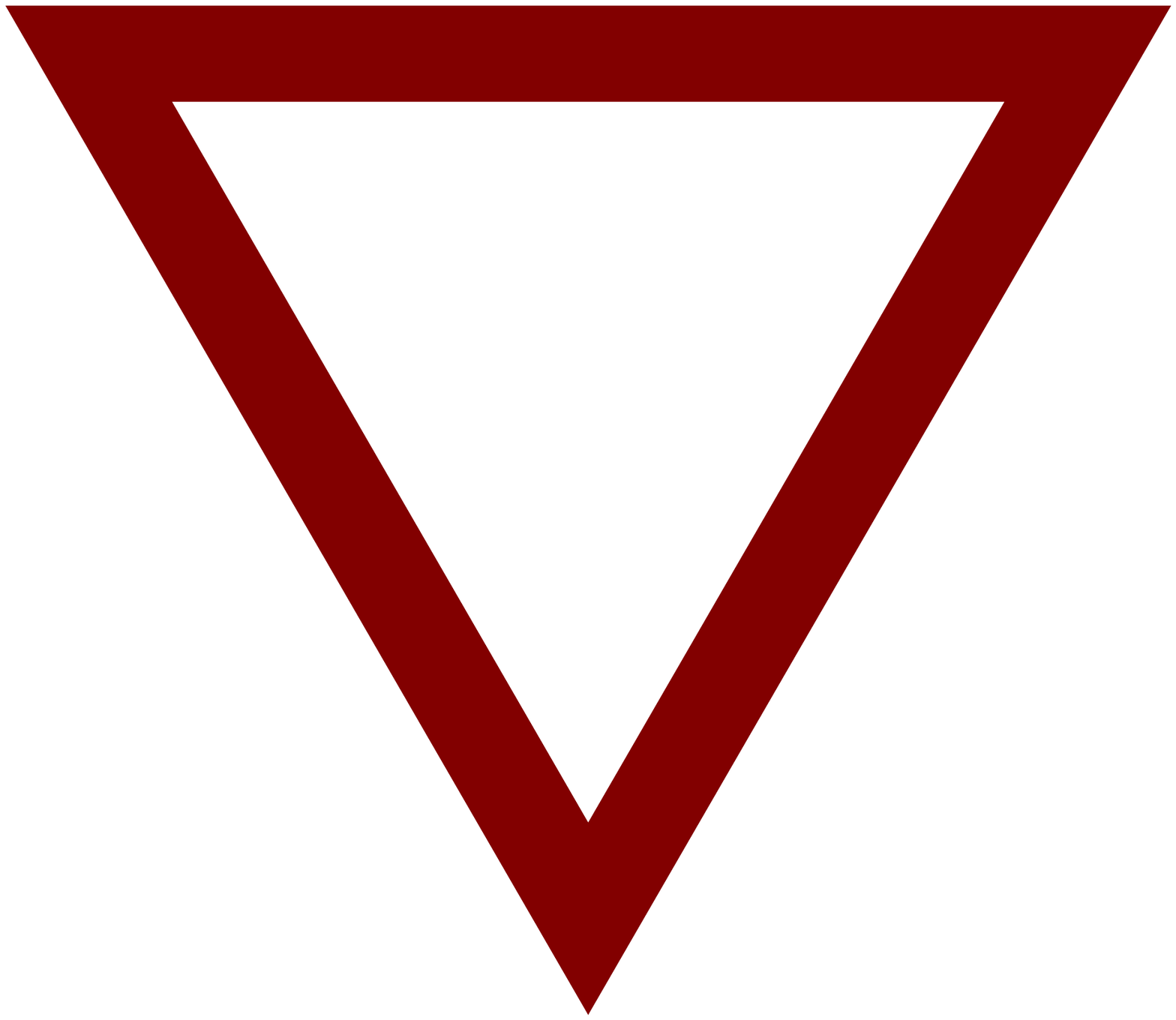}
\end{minipage} & $fc/U_\infty = 0.15 ; h_o = 1.00 ; \theta_o = 75^{\circ}$ & 0.55 & 6 & S\\
\begin{minipage}[t]{0.18\textwidth}
 S \raisebox{-0.2\height}{\includegraphics[width=0.17\linewidth]{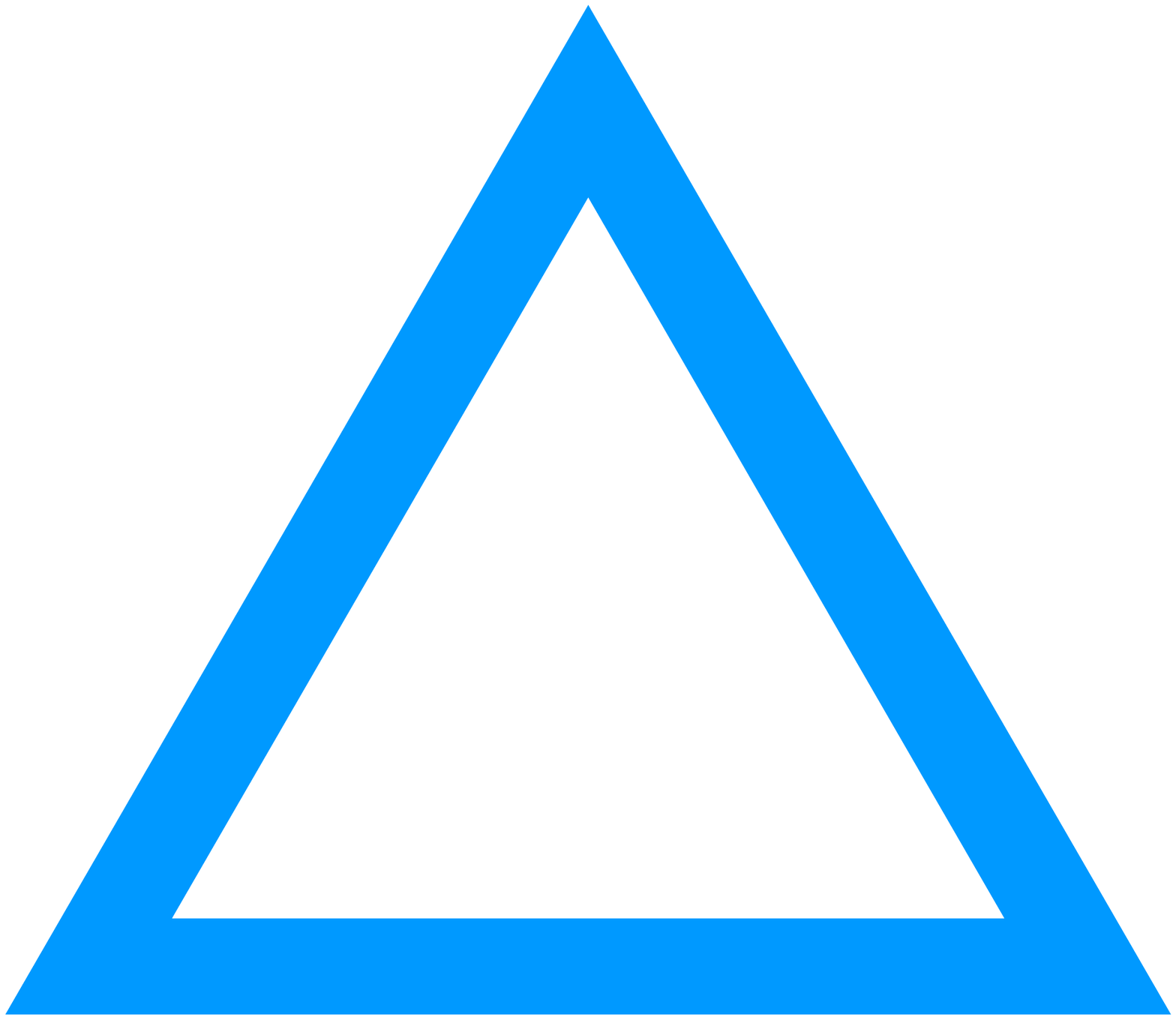}}
 E \raisebox{-0.28\height}{\includegraphics[width=0.17\linewidth]{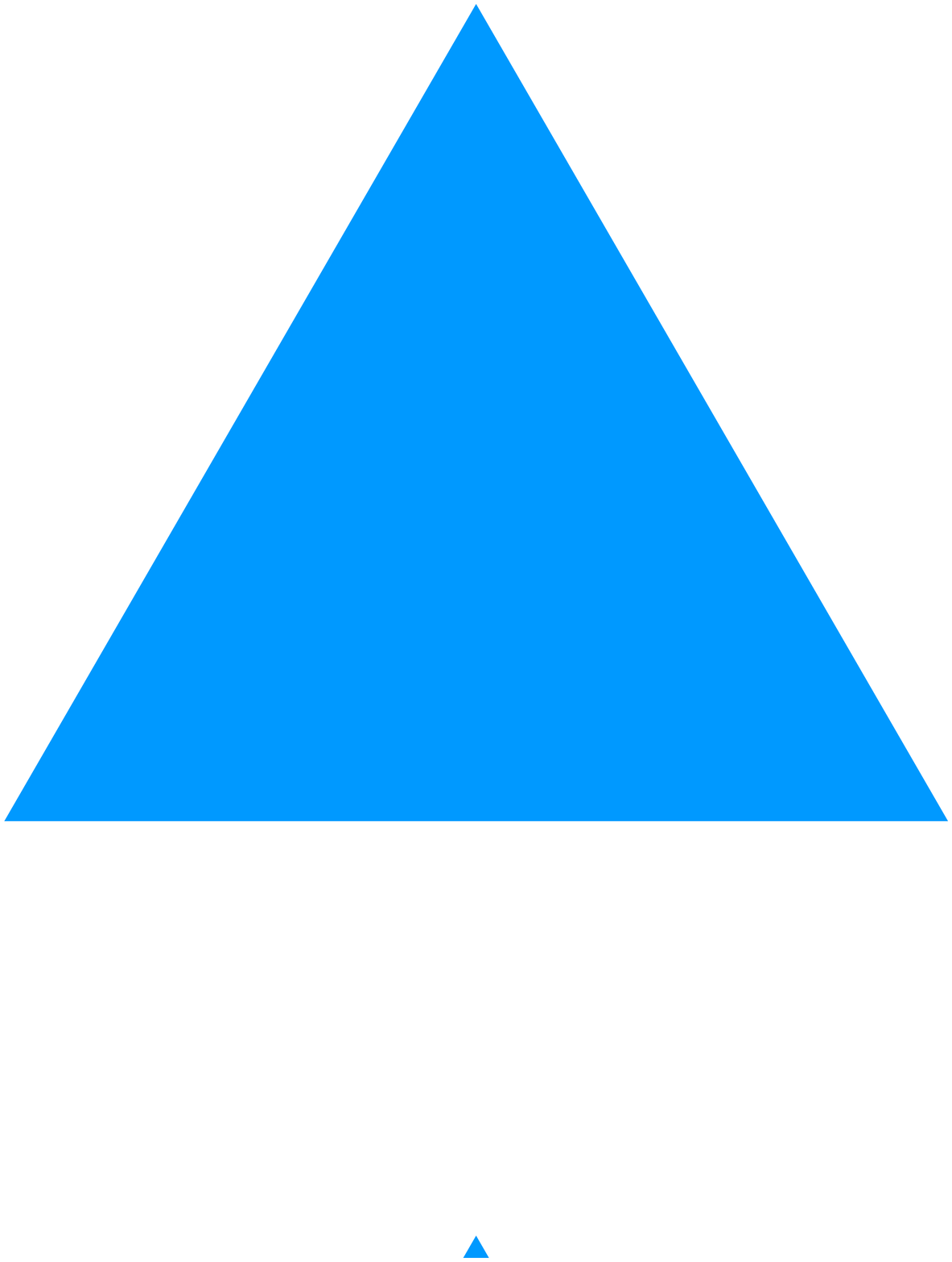}}
\end{minipage}  & $fc/U_\infty = 0.10 ; h_o = 1.00 ; \theta_o = 65^{\circ}$ & 0.57 & 6 & S,E,PIV \\
\begin{minipage}{0.03\textwidth}
 \includegraphics[width=\linewidth]{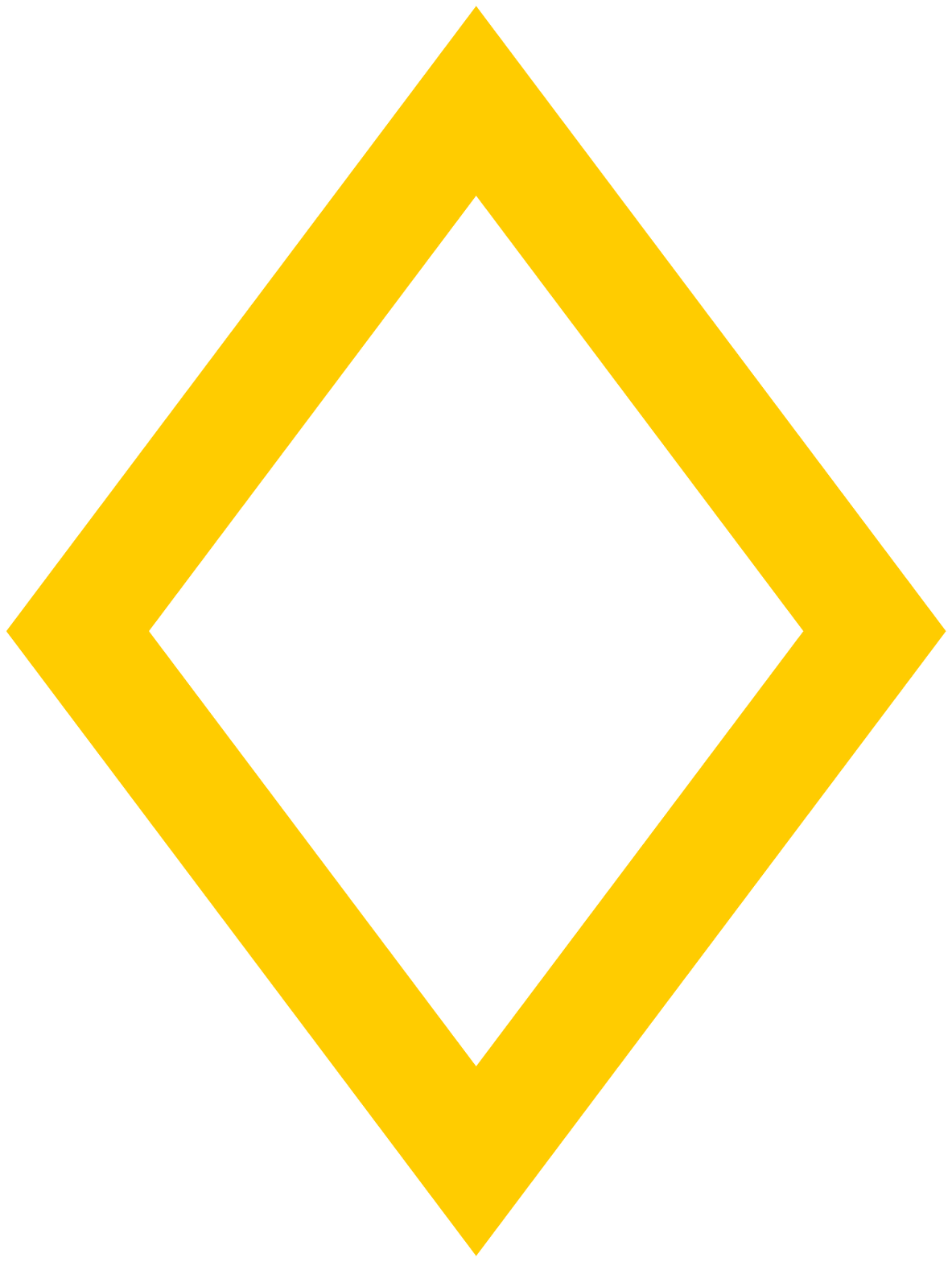}
\end{minipage} & $fc/U_\infty = 0.12 ; h_o = 0.75 ; \theta_o = 65^{\circ}$ & 0.62 & 6 & S\\
\begin{minipage}{0.03\textwidth}
 \includegraphics[width=\linewidth]{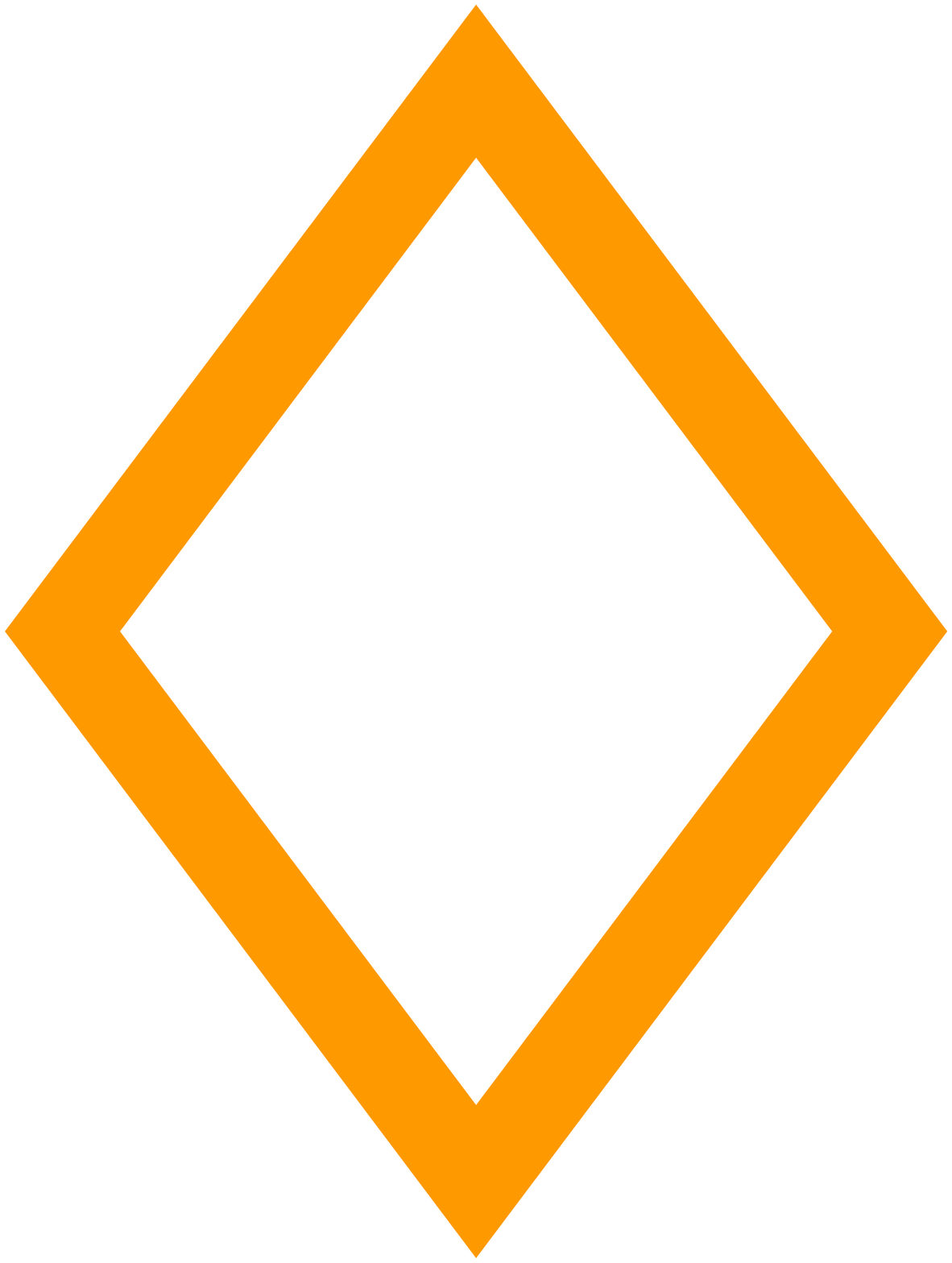}
\end{minipage} & $fc/U_\infty = 0.12 ; h_o = 1.00 ; \theta_o = 75^{\circ}$ & 0.66 & 6 & S\\
\begin{minipage}{0.03\textwidth}
 \includegraphics[width=\linewidth]{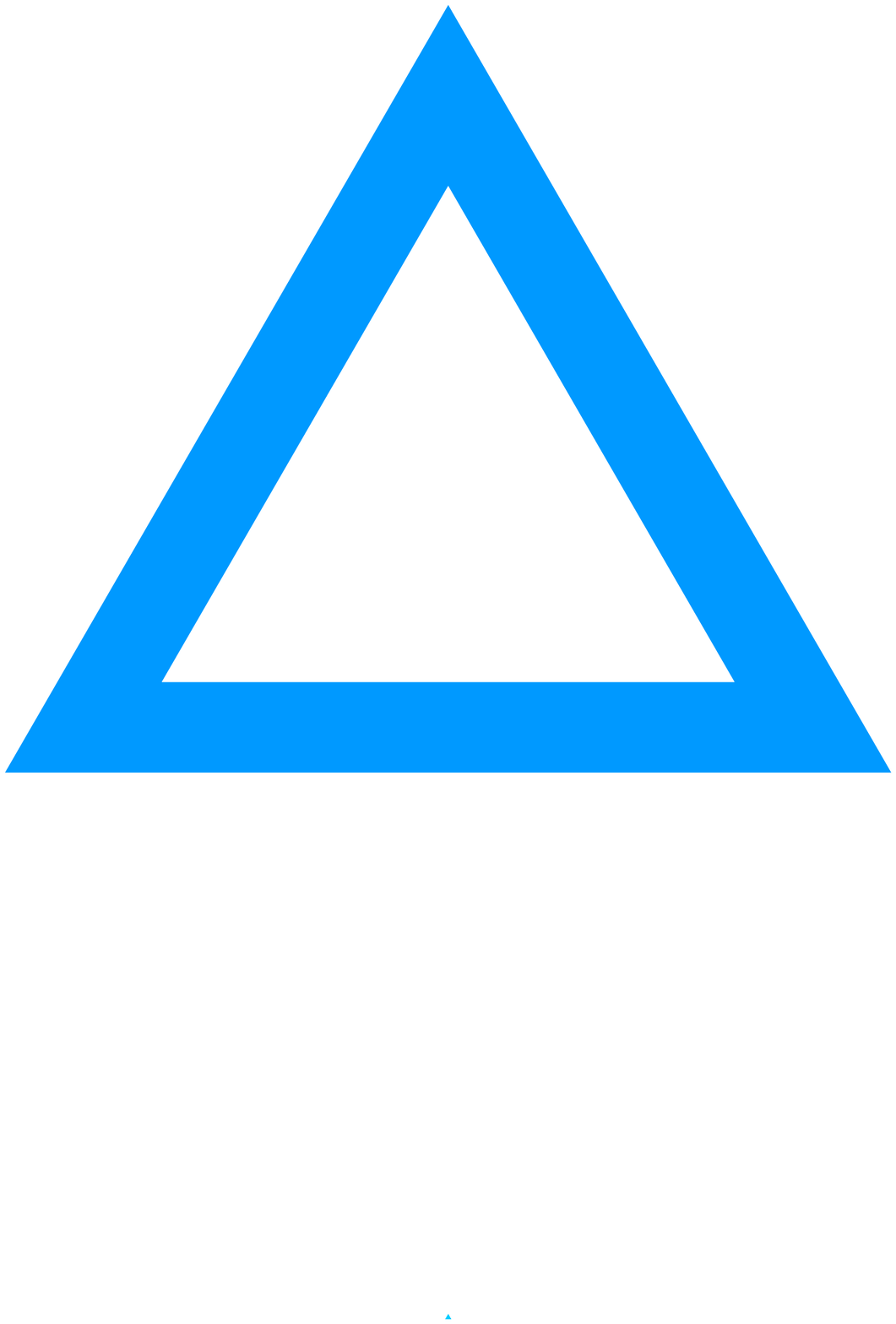}
\end{minipage} & $fc/U_\infty = 0.10 ; h_o = 0.75 ; \theta_o = 65^{\circ}$ & 0.69 & 6 & S\\
\begin{minipage}{0.03\textwidth}
 \includegraphics[width=\linewidth]{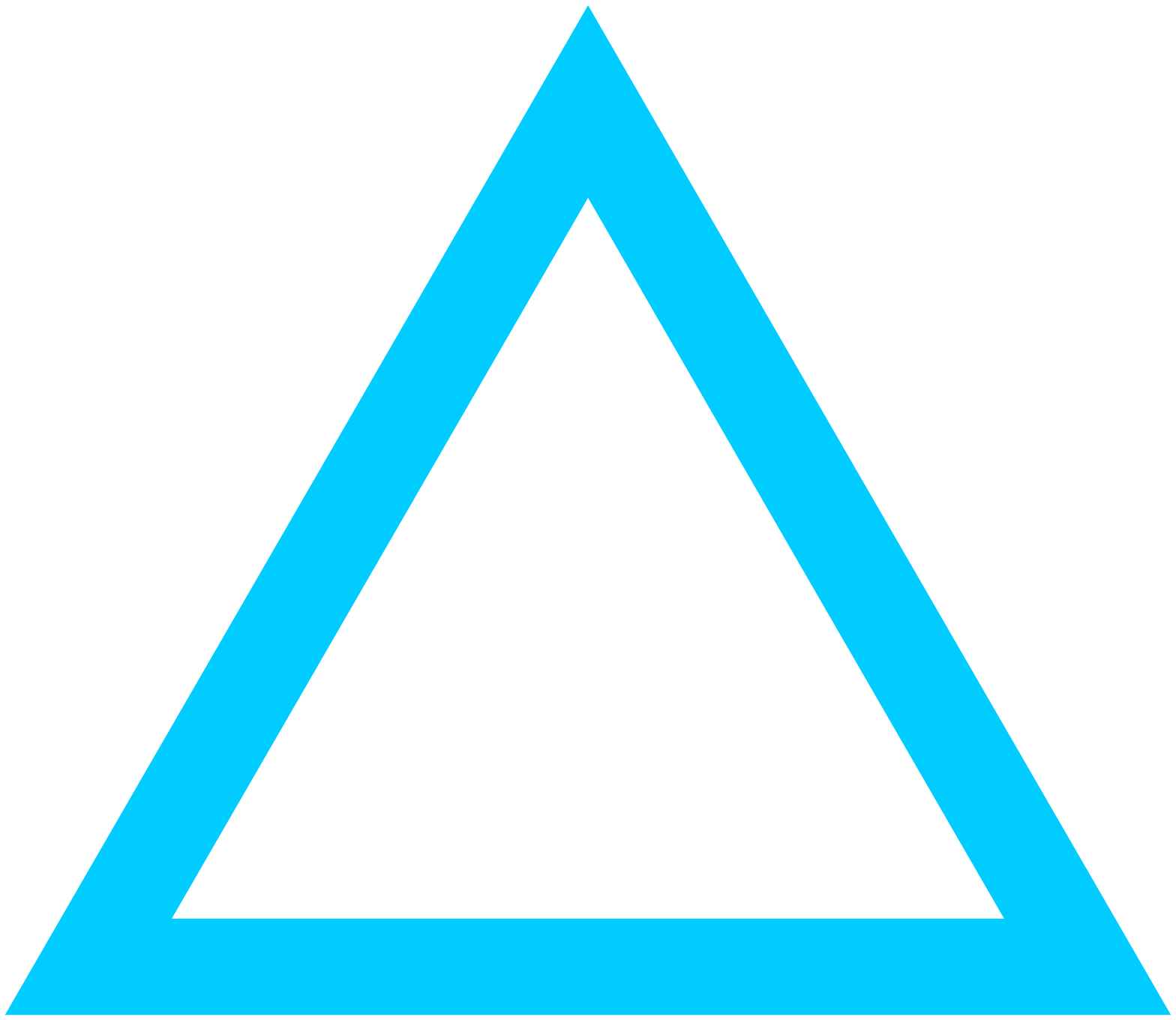}
\end{minipage} & $fc/U_\infty = 0.10 ; h_o = 1.00 ; \theta_o = 75^{\circ}$ & 0.75 & 6 & S\\
\hline
\end{tabular}
\label{t:tablekin}
\end{table}

\subsection{Numerical methods}

\subsubsection{Governing Equations and Numerical Techniques}

The numerical simulations utilize a direct numerical simulation (DNS) to solve the incompressible Navier-Stokes equations,

\begin{equation}
\rho \left (\frac{\partial \mathbf{u}}{\partial t}  + \mathbf{u} \cdot \mathbf{\nabla}\mathbf{u}\right ) + \mathbf{\nabla} p = \mathbf{\nabla} \cdot \mu \left (\mathbf{\nabla}\mathbf{u}+\mathbf{\nabla}\mathbf{u}^T\right )
\label{eq:NS}
\end{equation}

\begin{equation}
\nabla \cdot \mathbf{u} = 0
\label{eq:incomp}
\end{equation}

\noindent where $\mathbf{u}$ is the velocity vector, $p$ is the pressure, $\rho$ is the fluid’s density, and $\mu$ is the fluid’s dynamic viscosity. All numerical simulations are performed using a second-order accurate finite volume, pressure-implicit split-operator (PISO) method \cite{issa1986} implemented in $\textit{OpenFOAM}$ \cite{weller1998}.

In order to impose motion on the foils, a dynamic meshing algorithm is implemented. The mesh motion is initiated by prescribing the position of the cells on the foil's boundary, which align with the desired foil kinematics. The location of all the mesh nodes in the domain are solved at every time step using a solid body rotation (SBR) mesh motion equation \cite{Johnson1994}, 

\begin{equation}
\mathbf{\nabla} \cdot \left( \lambda \mathbf{\nabla}\mathbf{x_m}\right) + \mathbf{\nabla} \left( \lambda \left(  \mathbf{\nabla}\mathbf{x_m} - \mathbf{\nabla}\mathbf{x_m}^T \right) \right) - \nu Tr\left(\mathbf{\nabla}\mathbf{x_m}\right) = 0
\label{eq:SBR}
\end{equation}

\noindent where $\nu$ is the Lamé constant, $Tr$ represents the mathematical trace operation and $\lambda = 1/r$ is a diffusion constant for the motion of mesh nodes relative to the boundary motion where $r$ represents the distance from the foil (i.e. solid body). The mesh motion methodology has been previously validated for propulsive oscillating foils \cite{DaveSpauldingFranck2020}.

\begin{figure}[htbp]
\centering
\includegraphics[width=0.8\textwidth]{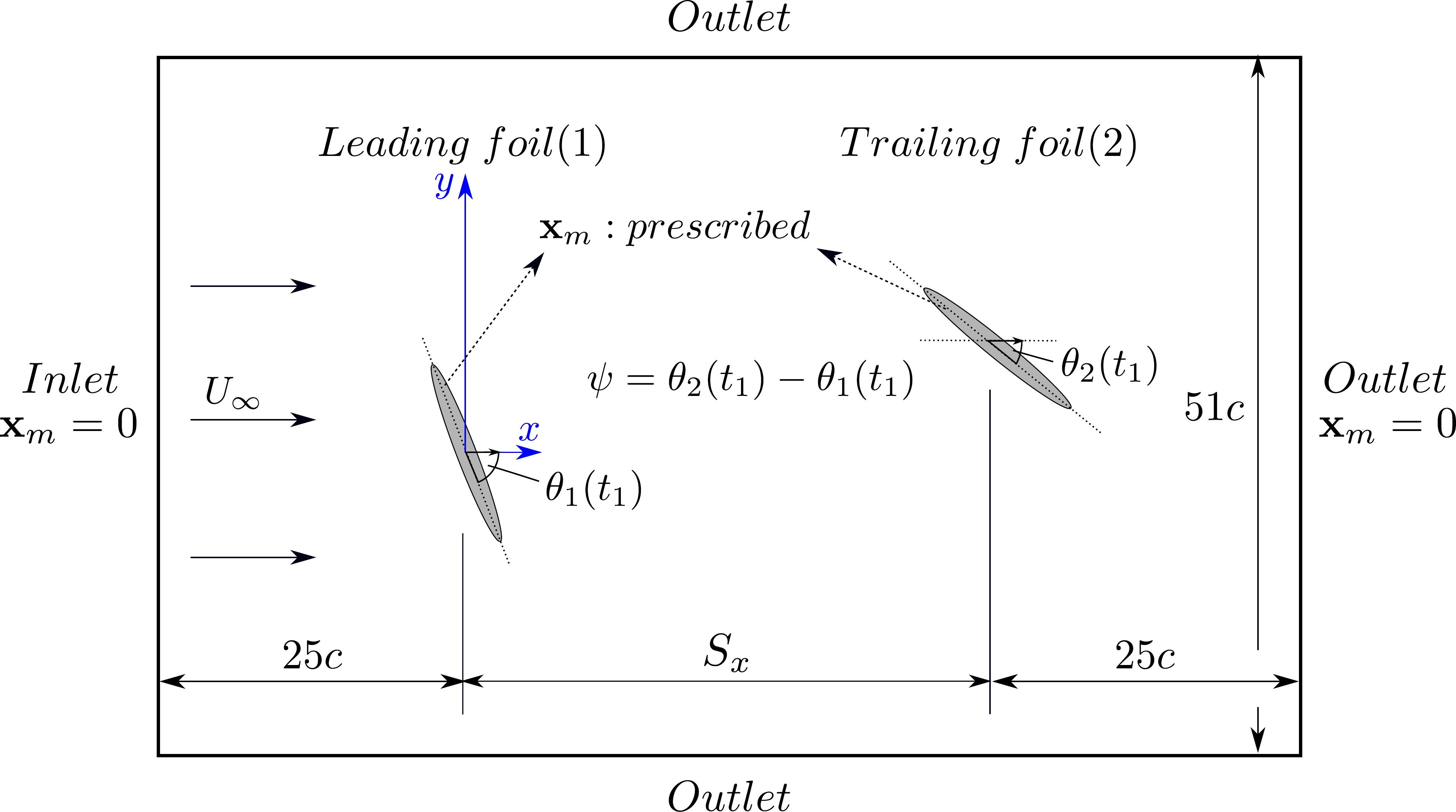}
\caption{Two-foil schematic with boundary conditions and configuration parameters.}
\label{f:twofoil}
\end{figure}

A schematic of the computational domain is shown in Figure \ref{f:twofoil}. The domain is $51c$ in the vertical direction with $25c$ upstream of the first foil and $25c$ downstream of the second foil in the horizontal direction. Inlet boundary conditions are imposed on the left side, and outlet conditions on the top, bottom and right sides. A non-slip wall condition is imposed on the foil surface. The mesh motion is constrained to zero at all outer boundaries. The two foils are at rest position (no mesh deformation) at the bottom of the stroke when $\theta=0^{\circ}$, and positioned so the heave stroke is vertically centered in the domain.

\subsubsection{Mesh Details}

\begin{figure}[htbp]
\centering
\includegraphics[width=0.9\textwidth]{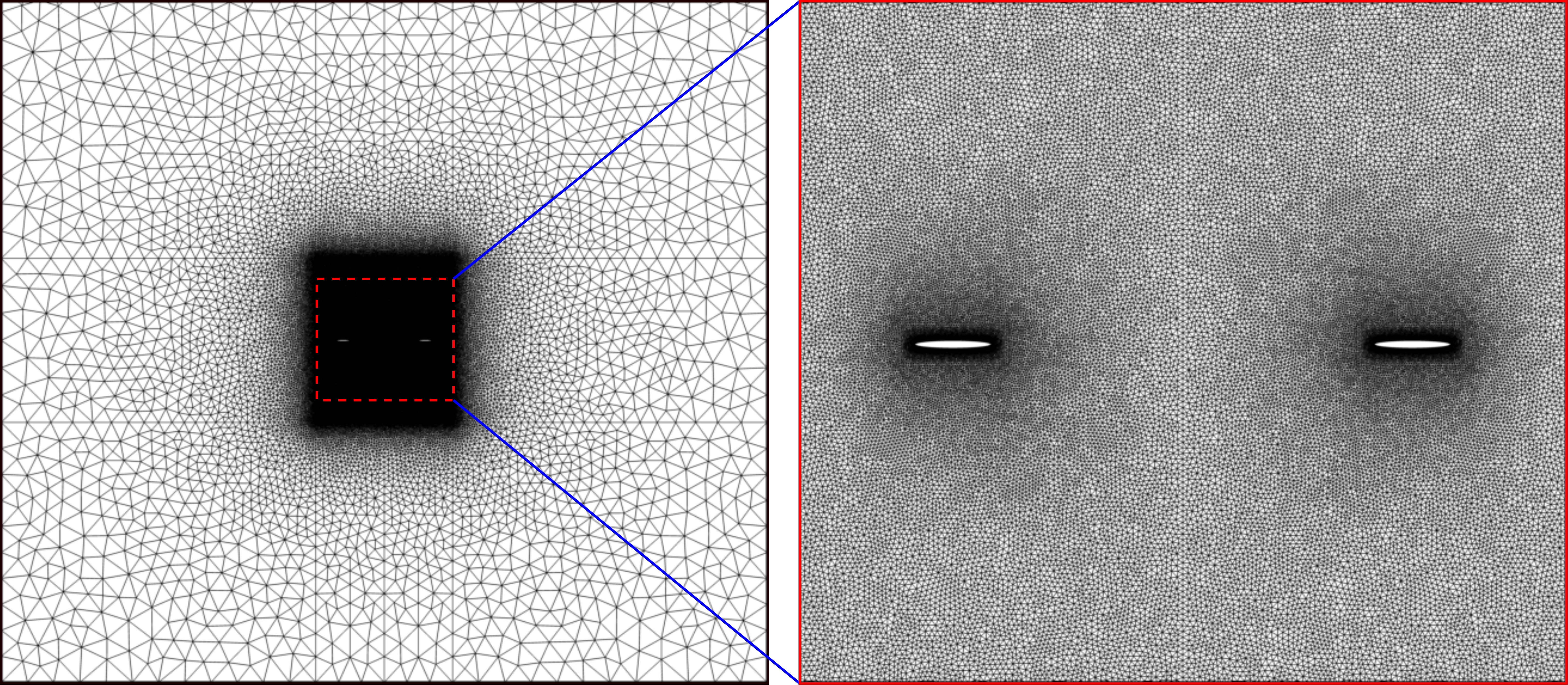}
\caption{Mesh 3B; whole domain (left); immediate vicinity of the two foils (right).}
\label{f:mesh}
\end{figure}

\begin{figure}[htbp]
\centering
		\includegraphics[width=1\linewidth]{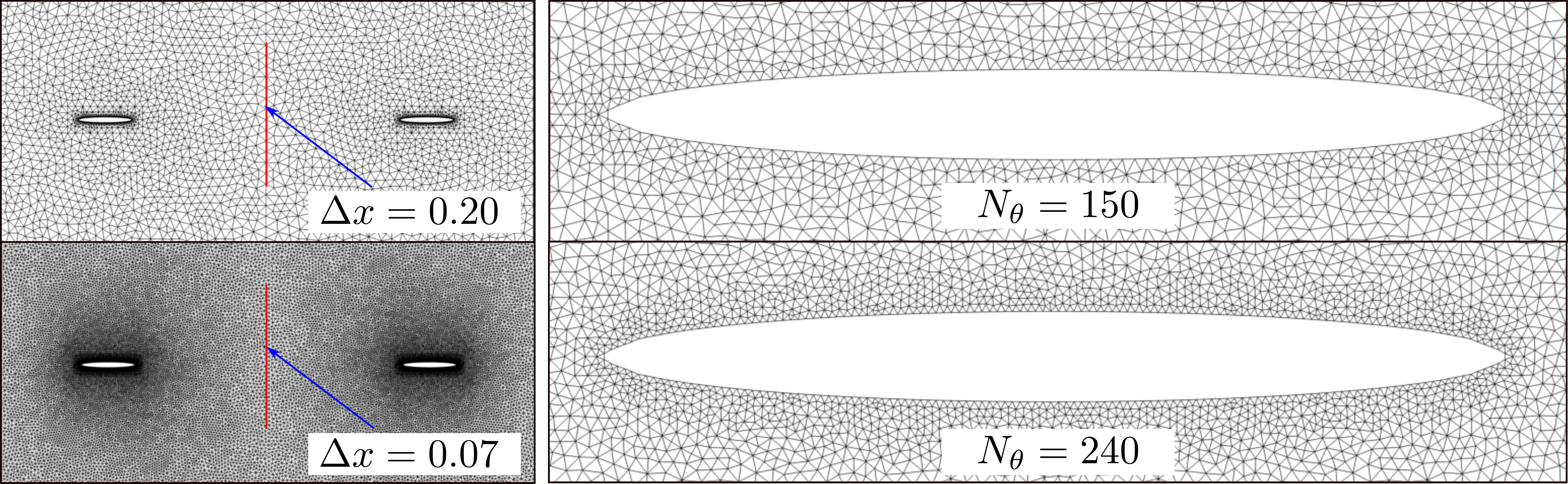}
\caption{Cell sizes at an equidistant position between foils and number of nodes around foil. Top left: Mesh 1B. Top right: Mesh 3A. Bottom left: Mesh 3B. Bottom right: Mesh 3B.}
\label{f:meshchar}
\end{figure}

A 2D unstructured mesh is generated using Gmsh \cite{gmsh} for each simulation as shown in Figure \ref{f:mesh}. To properly capture the boundary layer and separation phenomena there is a higher clustering of points around the foils compared to the outer domain, as well as a higher concentration of points within the region between the two foils. Mesh independence is analyzed on eight different meshes with varying resolution as measured in the boundary layer and in the wake between the two foils. Table \ref{t:table1} summarizes the characteristics for each of the eight meshes. Meshes $1$ to $4$ have decreasing $\Delta x$ values in the wake and close to the foil, and the letters A and B denote the number of nodes along the foil circumference. Figure \ref{f:meshchar} illustrates the cell sizes in the wake and the number of nodes around the foil. For comparison, results from a DNS of a stationary mesh is also displayed on Table \ref{t:table1}. The stationary mesh uses a non-inertial reference frame to prescribe motion and is only computed with a single foil \cite{RibeiroFranck2020}. 

\begin{table}[htbp]
\centering
\caption{Mesh characteristics. $N$ is the number of nodes. The wake $\Delta x$ corresponds to the cell size at a equidistant position between foils. $N_\theta$ represents the number of nodes in the azimuthal direction on foil's surface. The foil $\Delta x$ corresponds to the cell size at around $0.20c$ from each foil.}
\setlength{\tabcolsep}{6pt}
\begin{tabular}{ccccc}
\hline
 & & Wake & & Foil \\
Mesh   & $N$ & $\Delta x$ & $N_\theta$ & $\Delta x$ \\ \hline
Mesh 1A & $0.21 \times 10^5$ & 0.20 & 150 & 0.15 \\
Mesh 2A & $0.61 \times 10^5$ & 0.10 & 150 & 0.05 \\
Mesh 3A & $1.07 \times 10^5$ & 0.07 & 150 & 0.03 \\
Mesh 4A & $2.46 \times 10^5$ & 0.05 & 150 & 0.01 \\ \hline
Mesh 1B & $0.22 \times 10^5$ & 0.20 & 240 & 0.15 \\
Mesh 2B & $0.64 \times 10^5$ & 0.10 & 240 & 0.05 \\
Mesh 3B & $1.10 \times 10^5$ & 0.07 & 240 & 0.03 \\
Mesh 4B & $2.55 \times 10^5$ & 0.05 & 240 & 0.01 \\ \hline
Stationary & $1.08 \times 10^5$ & - & 240 & 0.02 \\ \hline
\end{tabular}
\label{t:table1}
\end{table}

\begin{figure}[htbp]
\centering
	\begin{subfigure}{0.49\textwidth}
	\centering
      	\includegraphics[width=1\linewidth]{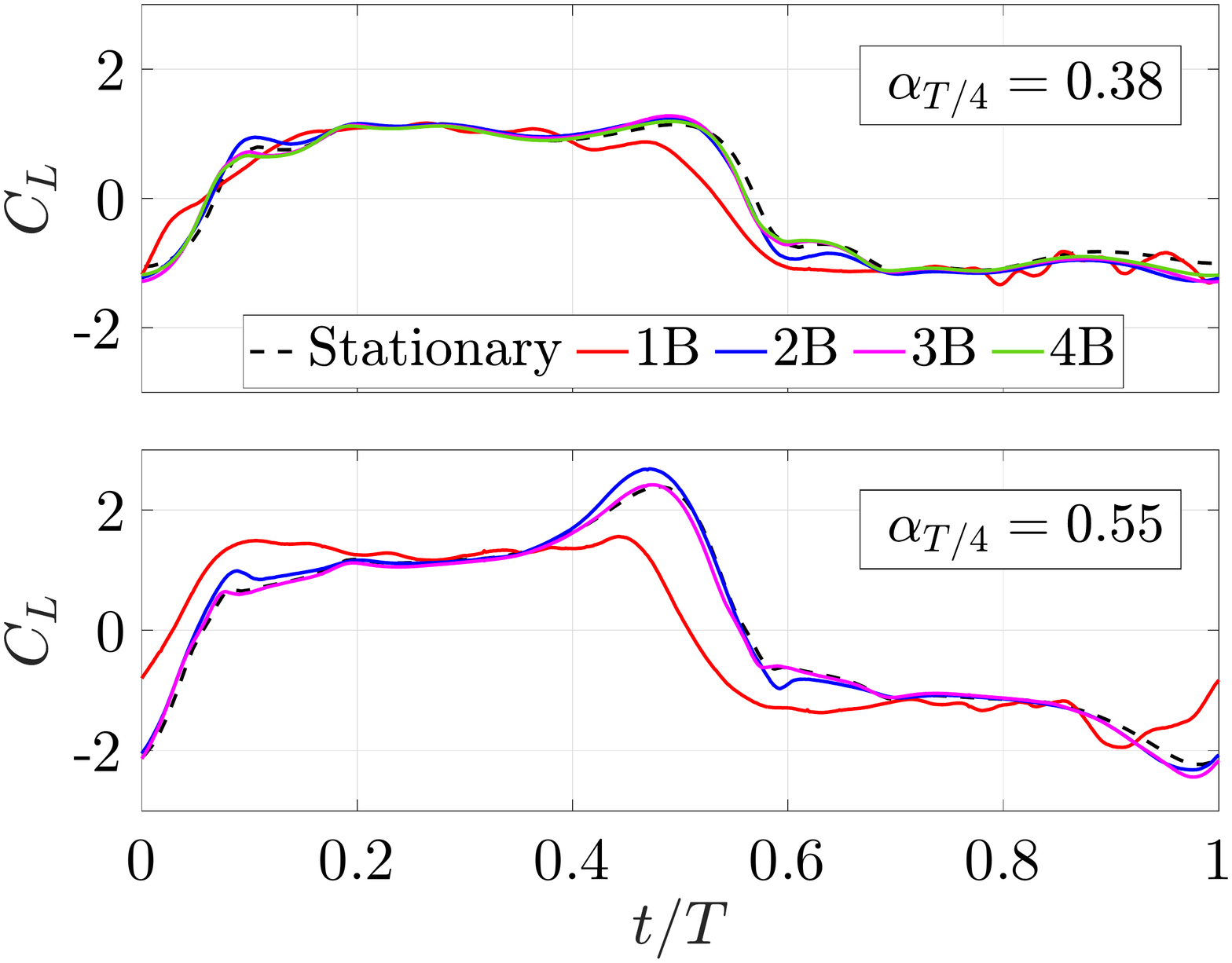}
		\caption{}
	\label{f:5a}
	\end{subfigure}
	\begin{subfigure}{0.49\textwidth}
	\centering
		\includegraphics[width=1\linewidth]{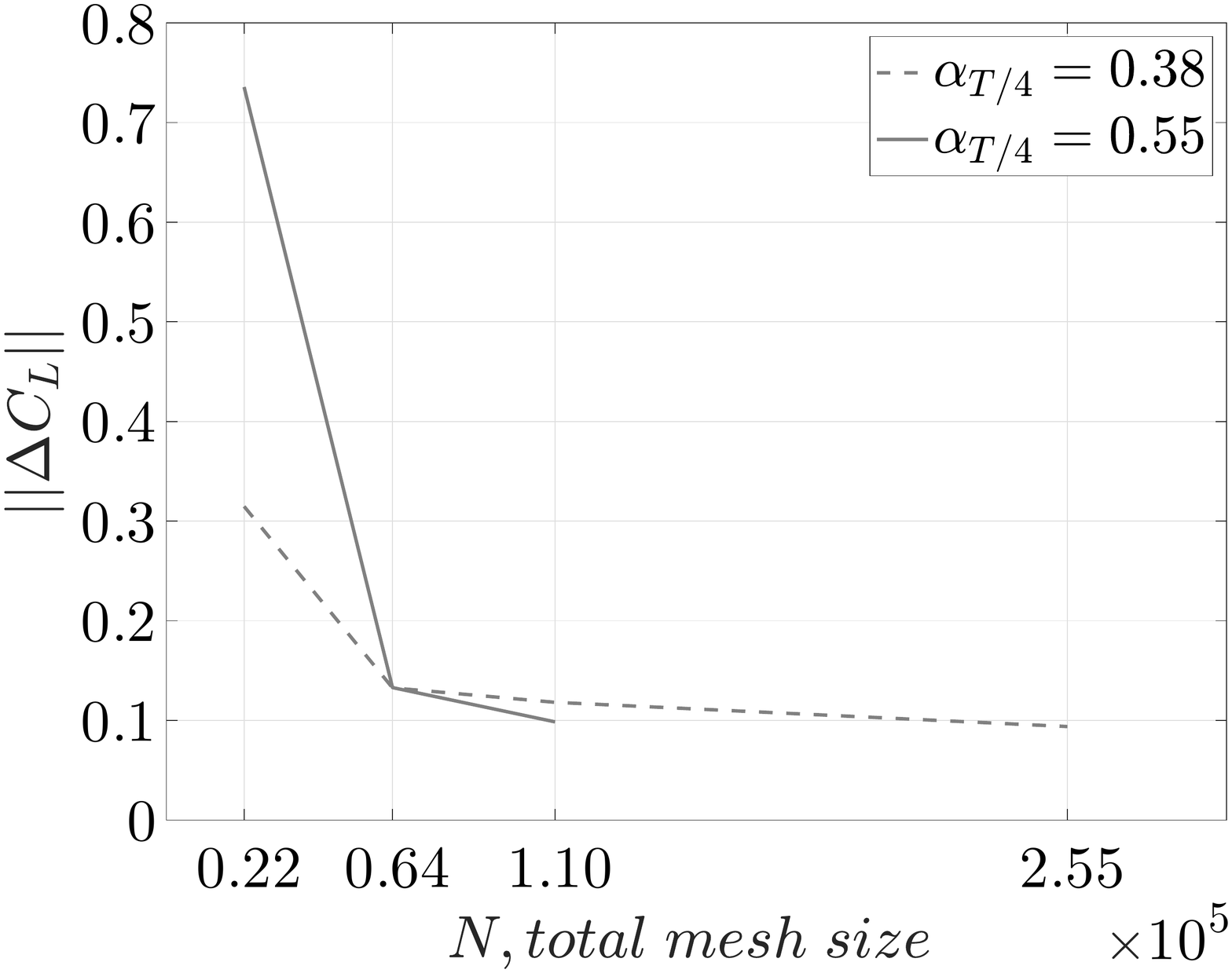}
		\caption{}
	\label{f:5b}
	\end{subfigure}
\caption{(a) Phase-averaged lift coefficient, $C_L$, of the leading foil as a function of non-dimensional time, $t/T$, for each dynamic mesh 'B' compared to stationary mesh; (b) $L^2$-norm of the difference between the dynamic mesh and the stationary mesh for increasing mesh size, $N$ (Please refer to Table \ref{t:table1} for mesh characteristics).}
\label{f:meshvalind}
\end{figure}

Mesh sensitivity is evaluated by comparing the forces directly on the foil and the resolution of flow structures within the wake. Figure \ref{f:5a} shows the phase-averaged lift coefficient of the leading foil computed with meshes 1B, 2B, 3B, and 4B, each compared to the stationary mesh. Since high heave and high pitch amplitudes increase mesh deformation, the forces are evaluated at a representative low ($\alpha_{T/4}=0.38$) and high ($\alpha_{T/4}=0.55$) relative angles of attack. As the number of mesh cells $N$ increases, the solution converges to the stationary mesh solution, with minor differences between mesh 3B ($N \approx 1.1\times10^5$) and mesh 4B ($N \approx 2.55 \times 10^5$). Due to the high resolution and high deformation, mesh 4B does not work with the higher angle of attack. Figure \ref{f:5b} demonstrates the convergence of the solution with increasing resolution by comparing the $L^2$-norm of the difference between the stationary mesh and the dynamic mesh, computed as

\begin{equation}
||\Delta C_L|| = \sqrt{\frac{1}{n} \sum_{k=1}^n (C_{L,stationary}^k - C_{L, dynamic}^k)^2}.
\label{eq:L2}
\end{equation}

\noindent A similar analysis is performed comparing the `A' and `B' meshes, yielding small differences between 3A and 3B. Thus the final mesh is chosen to be mesh 3B.

\begin{figure}[htbp]
\centering
\includegraphics[width=0.6\textwidth]{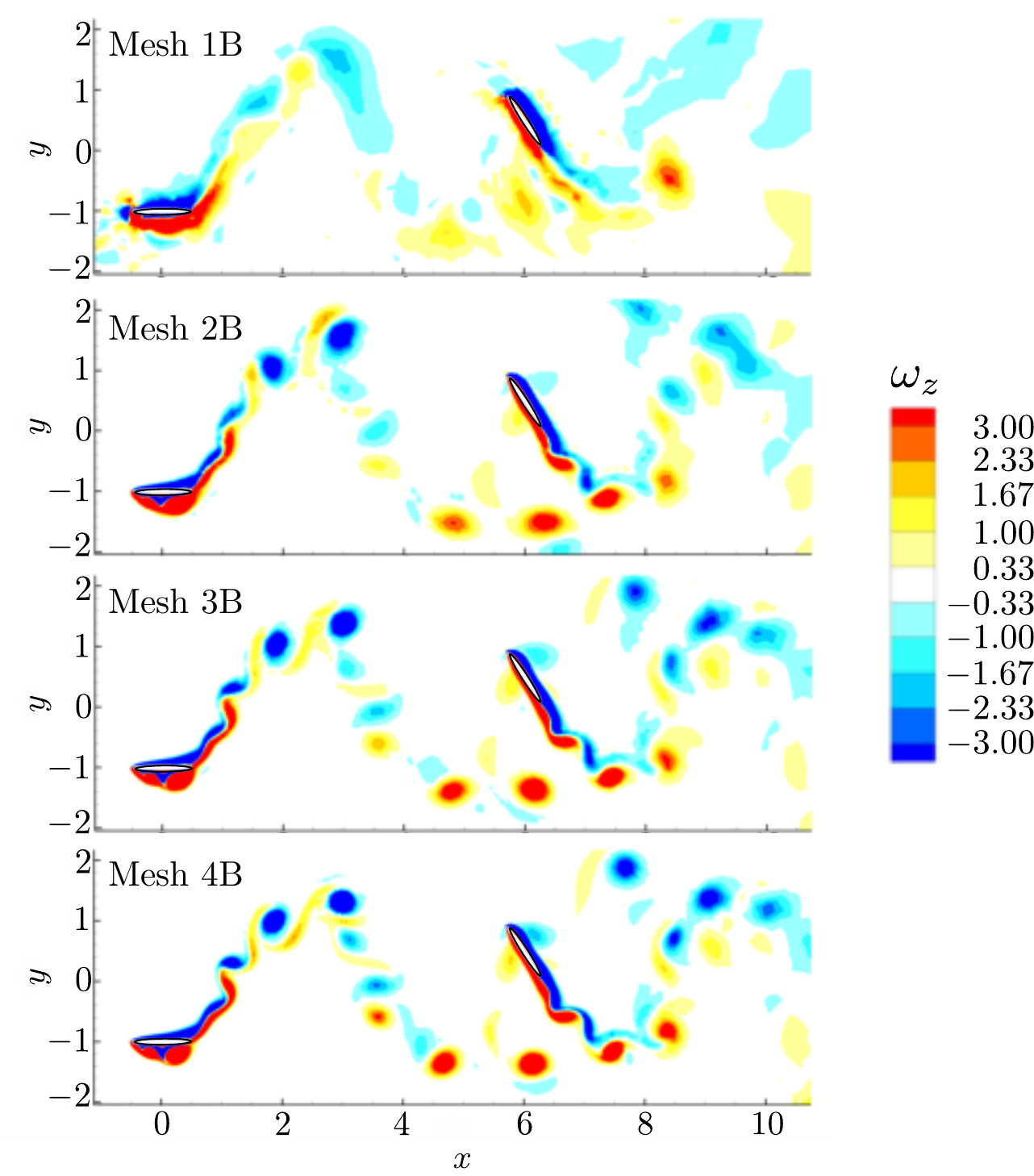}
\caption{Instantaneous vorticity field in the $z$ direction, $\omega_z$. Kinematics: $f = 0.15; h_o = 1.00; \theta_o=65^{\circ}$}
\label{f:vortmesh}
\end{figure}

Figure \ref{f:vortmesh} shows the instantaneous spanwise vorticity field, $\omega_z$, for all B meshes at $\alpha_{T/4}=0.38$. For the chosen kinematics there are multiple vortices present within the wake. To ensure the mesh resolution between the two foils is adequate to capture the vortices interacting with the trailing foil, a qualitative comparison is shown between the four resolutions. There is a significant difference in vorticity strength between mesh 1B and 2B, indicating that 1B is under-resolved. Meshes 2B, 3B, 4B show the same number of vortices in the wake, however some resolution effects are still detected in 2B, whereas 3B and 4B demonstrate very similar vorticity fields. 

In summary, mesh 3 shows minimal force and flow-field differences with respect to mesh 4, and mesh 3B is only $6\%$ more computationally expensive than mesh 3A. Thus, mesh 3B is selected for all simulations. With mesh 3B selected for all simulations, 8 cycles are simulated. As a benchmark for computational time, the simulations took 9 hours to run 10 convective time units on an Intel Scalable 2.6GHz processor for a serial run. Job arrays are used to automate the process of running the simulations.

\subsection{Experimental methods}

\begin{figure}[htbp]
\centering
\includegraphics[width=0.5\textwidth]{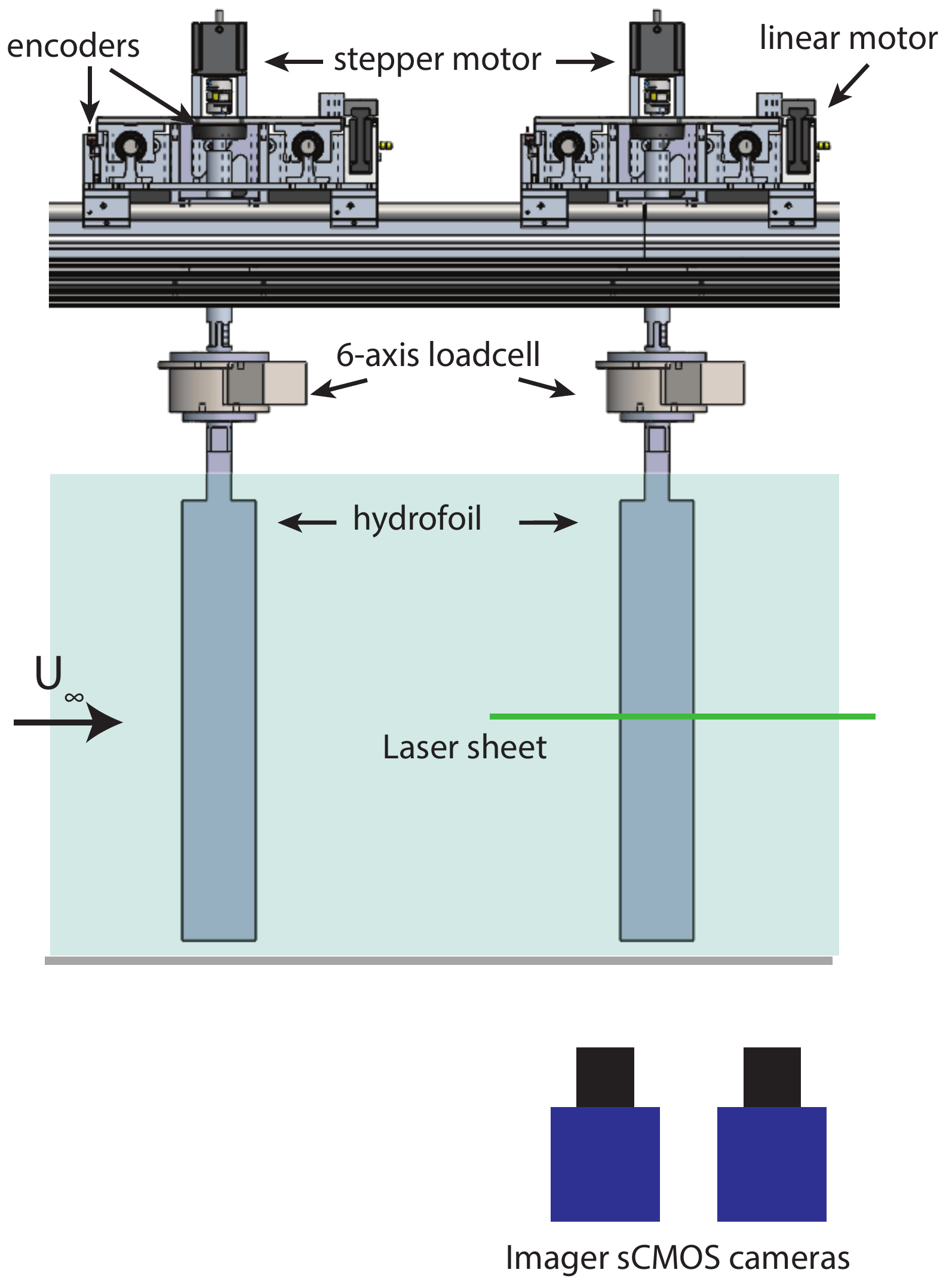}
\caption{Side view of tandem-foil system in the flume, with identical leading (left) and trailing (right) foils, force transducers, encoders, actuators, laser sheet and cameras for PIV measurements.}
\label{f: expsetup}
\end{figure}

The experiments are conducted in a closed-circuit water flume (cross section: 0.8 m (wide) x 0.6 m (deep)) at Brown University. As shown in Figure \ref{f: expsetup} (side view), two rectangular hydrofoils (chord, $c$ = 0.076 m, span $b$ = 0.457 m, thickness $\sigma$ = 0.007 m) are vertically mounted in a tandem configuration in the flume with a uniform freestream velocity, $U_{\infty}$, measured by an Acoustic Doppler Velocimeter (Vectrino) ($Re=30,000$ based on the chord length). The motions of each foil (Equations \ref{eq:heave} and \ref{eq:pitch}) are controlled by two servo motors (AeroTech) for heaving motion, and two stepper motors (Applied Motion Products) for pitching motion. The heaving and pitching positions ($h$ and $\theta$) are recorded by Optical encoders (US Digital). The lift ($F_y$) and torque ($M_z$) on the foils are measured by two six-axis force transducers (ATI IP65) mounted directly on the foils. With the measured forces and the corresponding foil positions, the energy harvesting efficiency is evaluated using Equations \ref{eq:eta} and \ref{eq:power}. To account for the finite size of the flume a blockage correction based on the method described by Houlsby \cite{houlsby} is performed on the data. For more details on blockage correction procedure, please see Appendix A.

Particle Image Velocimetry (PIV) measurements are performed to visualize the flow field around the tandem foils. As shown in Figure \ref{f: expsetup}, the laser sheet, generated using a double-pulsed Nd:YAG laser, (Quantel Evergreen, $\lambda$ = 532 nm, upto 200 mJ/pulse) is positioned the mid-span of the foil, with four cameras (LaVision Imager sCMOS, 2560 pixels by 2160 pixels) and 35-mm lenses (Nikon) placed beneath the test section to capture the flow field. The flow is seeded with silver-coated hollow ceramic spheres of 100 $\mu m$ (Potter Industries). The image pairs are acquired at a frame rate of 25 Hz and then processed for velocity vectors by DaVis 8 (LaVision) using sequential cross-correlations with decreasing interrogation window sizes (initial: 64 pixels by 64 pixels, final: 32 pixels by 32 pixels), producing a field of view measuring around 5 chords x 4 chords after stitching together the images from the four cameras.

\subsection{Comparison between experiments and simulations for a single foil}

\begin{figure}[htbp]
\centering
\includegraphics[width=0.6\textwidth]{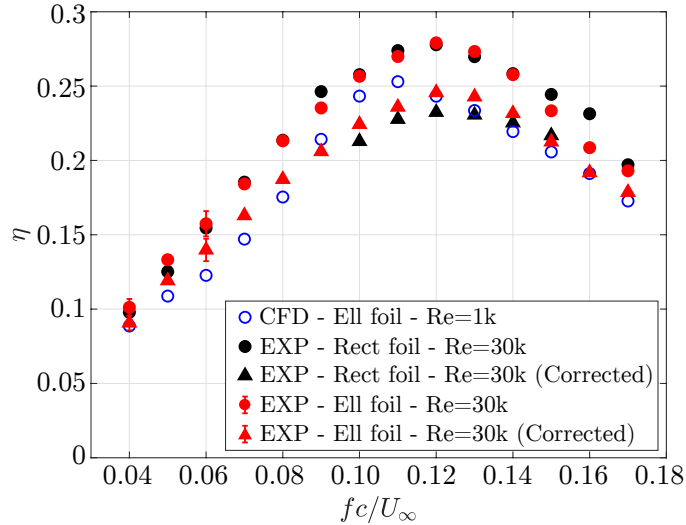}
\caption{Foil efficiency, $\eta$, as a function of reduced frequency, $fc/U_\infty$. Corrected data correspond to the experimental blockage correction (Appendix A). Kinematics: $h_o = 1.00; \theta_o=65^{\circ}$.}
\label{f:validation}
\end{figure}

Using a single foil, the energy harvesting performance is compared over a range of reduced frequencies to validate the computational and experimental set-ups. Figure \ref{f:validation} shows the foil's efficiency at heave amplitude $h_o=1.00$ and pitch amplitude $\theta_o=65^{\circ}$. 
Experiments are performed with a rectangular cross-sectional foil shape. To demonstrate the insensitivity of foil shape and make a direct comparison with the CFD, experiments are also performed with an elliptical cross-sectional foil, which yield very similar results to the rectangular foil. 
Once a blockage correction is applied, the experiments and simulations have a strong agreement, with the maximum efficiency peaking at a slightly earlier reduced frequency for the simulations. The small discrepancies between the simulations and the experiments are likely caused by differences in Reynolds number ($1000$ for simulations and $30,000$ for experiments) \cite{RibeiroFranck2020}, and the three-dimensional effects inherent to experiments but not captured by the two-dimensional simulation.

\section{Results and Discussion}

\subsection{Relative angle of attack as a predictive quantity}

\begin{figure}[htbp]
\centering
\includegraphics[width=0.6\linewidth]{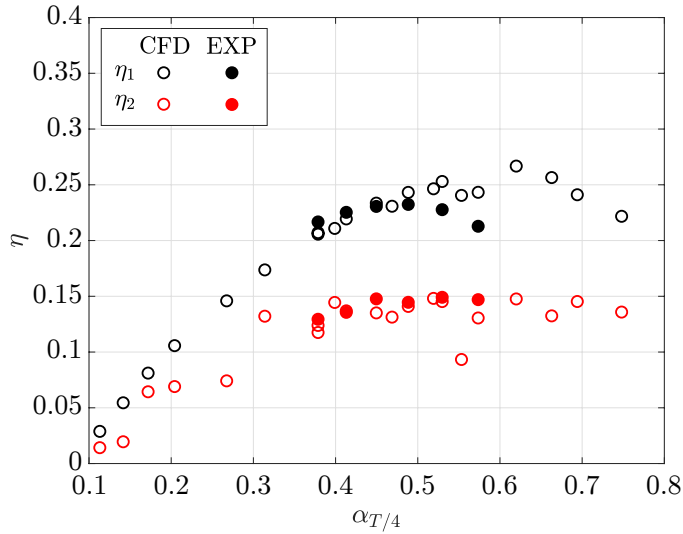}
\caption{Leading and trailing foil efficiencies, $\eta_1$ and $\eta_2$ respectively, with respect to various $\alpha_{T/4}$ for both simulations and experimental data.}
\label{f:etaUinf}
\end{figure}

Given the large kinematic parameter space, it is convenient to reduce frequency, pitch and heave amplitude into a single parameter and utilize the relative angle of attack at maximum pitch ($\alpha_{T/4}$), as defined by Equation \ref{eq:alpharel}, as a representative of the sinusoidal kinematics \cite{kindum2008, Kim2017, RibeiroFranck2020}. Figure \ref{f:etaUinf} shows the leading and trailing foil efficiency of all kinematics listed in Table \ref{t:tablekin}. All points correspond to cases with the same inter-foil distance ($S_x=6$) and each point for the trailing foil corresponds to the inter-foil phase that provides the highest efficiency.

As shown by Figure \ref{f:etaUinf}, $\alpha_{T/4}$ is strongly correlated with the foil's energy harvesting efficiency \cite{Kim2017,Su2019,RibeiroFranck2020, kinseytandem2012}, and shows strong agreement between the experiments and simulations. The leading foil efficiency increases as $\alpha_{T/4}$ increases, until around $\alpha_{T/4}=0.50$ where it reaches a maximum efficiency of approximately $25\%$. Surpassing $\alpha_{T/4}=0.60$, leading foil efficiency decreases slightly, indicating the foil kinematics are receding from the optimal energy harvesting range. The trailing foil follows a similar trend at lower angles of attack, but due to less available kinetic energy in the wake between foils it plateaus at an efficiency of $15\%$ at $\alpha_{T/4}=0.40$. However, after this angle of attack the efficiency remains constant with increasing $\alpha_{T/4}$ as opposed to the slight decay seen for the leading foil. The data for both experiments and simulations agree that $\alpha_{T/4}$ is a predictive kinematic quantity in terms of evaluating the energy harvesting efficiency over a range of diverse kinematics, but also that the trailing foil suffers from lower efficiency due to the reduced velocity in the immediate wake region.

\subsection{Wake velocity, wake width and the relationship with foil kinematics}

In Figure \ref{f:etaUinf}, both the leading and trailing foil efficiency are computed as percent of power extracted from that available in the freestream within the swept area $Y_p$. However, an alternative definition for the trailing foil involves computing the available power immediately upstream, in the wake between the two foils; this can be performed with a more detailed analysis of the wake properties and characteristics. 

\begin{figure}[htbp]
\centering
\includegraphics[width=0.6\textwidth]{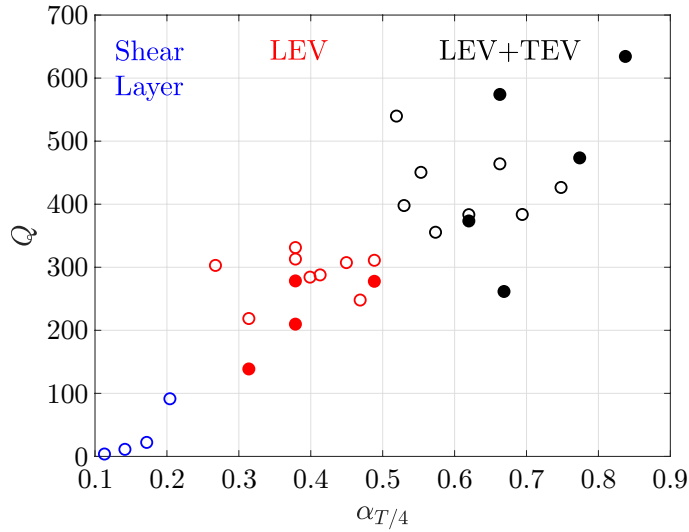}
\caption{Maximum Q value of the primary vortex for each $\alpha_{T/4}$. Open symbols correspond to numerical data. Filled circles are PIV data.}
\label{f:Q}
\end{figure}

First, the formation and strength of the leading edge vortices are analyzed. Depending on the specific kinematics, there are one or multiple vortices shed per half-stroke. To understand the wake's dependence on foil kinematics, the strength of the primary vortex is computed as a function of $\alpha_{T/4}$ in Figure \ref{f:Q}. The primary vortex is defined as the strongest vortex developed on the suction side of the foil during a half-stroke. The Q criterion, defined as $Q=\frac{1}{2} \left(\left \| \mathbf{\Omega} \right \|^2 - \left \|\mathbf{S} \right \|^2 \right )$, is computed as a vortex identification method \cite{Hunt1988}, with $\mathbf{\Omega}$ representing the vorticity tensor and $\mathbf{S}$ as the rate-of-strain tensor. Since the vortex strength generally decreases as it convects downstream, only the maximum strength is reported for each set of kinematics in both PIV and numerical data. Thus, Figure \ref{f:Q} displays the maximum non-dimensionalized Q ($Q/U_\infty^2$) with respect to $\alpha_{T/4}$. 

In general, with increasing relative angle of attack the strength of the primary leading edge vortex increases, which is consistent throughout the PIV and simulation data with few exceptions. As $\alpha_{T/4}$ increases, there is more scatter in the maximum Q values between various kinematics, which is likely explained by the flow physics of the higher relative angles of attack described below. 

\begin{figure}[htbp]
\centering
\includegraphics[width=1\textwidth]{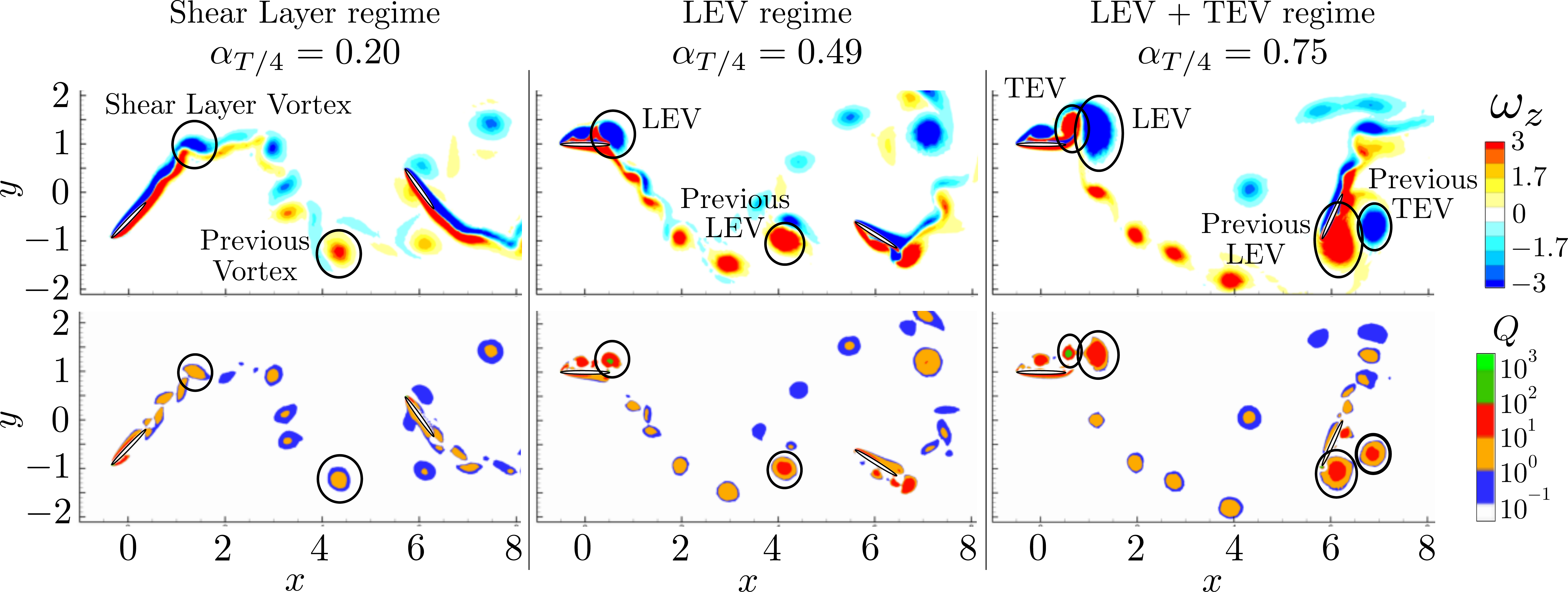}
\caption{Instantaneous vorticity, $\omega_z$, and Q flow fields for a representative $\alpha_{T/4}$ of each regime. The inter-foil phases selected are the cases of maximum trailing foil efficiency. The primary vortex is highlighted (whose path is shown in Figure \ref{f:kvortu}), as well as the alternate signed primary vortices from the previous half-stroke.}
\label{f:Qvortinst}
\end{figure}

Analyzing the maximum Q values and how they are correlated with $\alpha_{T/4}$, three regimes are identified and defined by low (shear layer regime), medium (LEV regime), and high (LEV+TEV regime) $\alpha_{T/4}$ values, as shown in Figure \ref{f:Q}. To better understand how these regimes are defined, Figure \ref{f:Qvortinst} shows the instantaneous vorticity and Q flow fields for three representative $\alpha_{T/4}$, one for each regime.

The first regime is characterized by low relative angles of attack, within the approximate range of $0<\alpha_{T/4} \leq 0.20$, and is labeled the `shear layer regime'. For $\alpha_{T/4}$ lower than $0.20$, the separation is dominated by a shear layer with very little vortex formation, indicated by Q values close to zero in Figures \ref{f:Q} and \ref{f:Qvortinst}. At $\alpha_{T/4}=0.20$, the shear layer shows a small degree of separation resulting in a relatively small vortex with respect to other kinematics in Figure \ref{f:Qvortinst}, with $Q$ less than or equal to 100.  
As $\alpha_{T/4}$ increases, $Q$ also increases as the primary LEV grows in size and strength. The second regime, labeled as the `LEV regime' is from approximately $0.20<\alpha_{T/4}<0.50$ and has an easily identifiable primary LEV, as highlighted on the second column in Figure \ref{f:Qvortinst}, whose strength varies between $100<Q<350$. At $\alpha_{T/4}>0.50$, the values of $Q$ generally grow, but have a large range from $250$ to $650$. A distinguishing feature for this last and final regime is that the primary LEV is paired with a strong trailing edge vortex (TEV), thus labeled `LEV+TEV regime'. Due to the addition of more vortices in the wake, the wake-foil interactions are more prevalent for this final regime as demonstrated by the stronger vortices forming around trailing foil on the third column of Figure \ref{f:Qvortinst}.

\begin{figure}[htbp]
\centering
\includegraphics[width=0.95\linewidth]{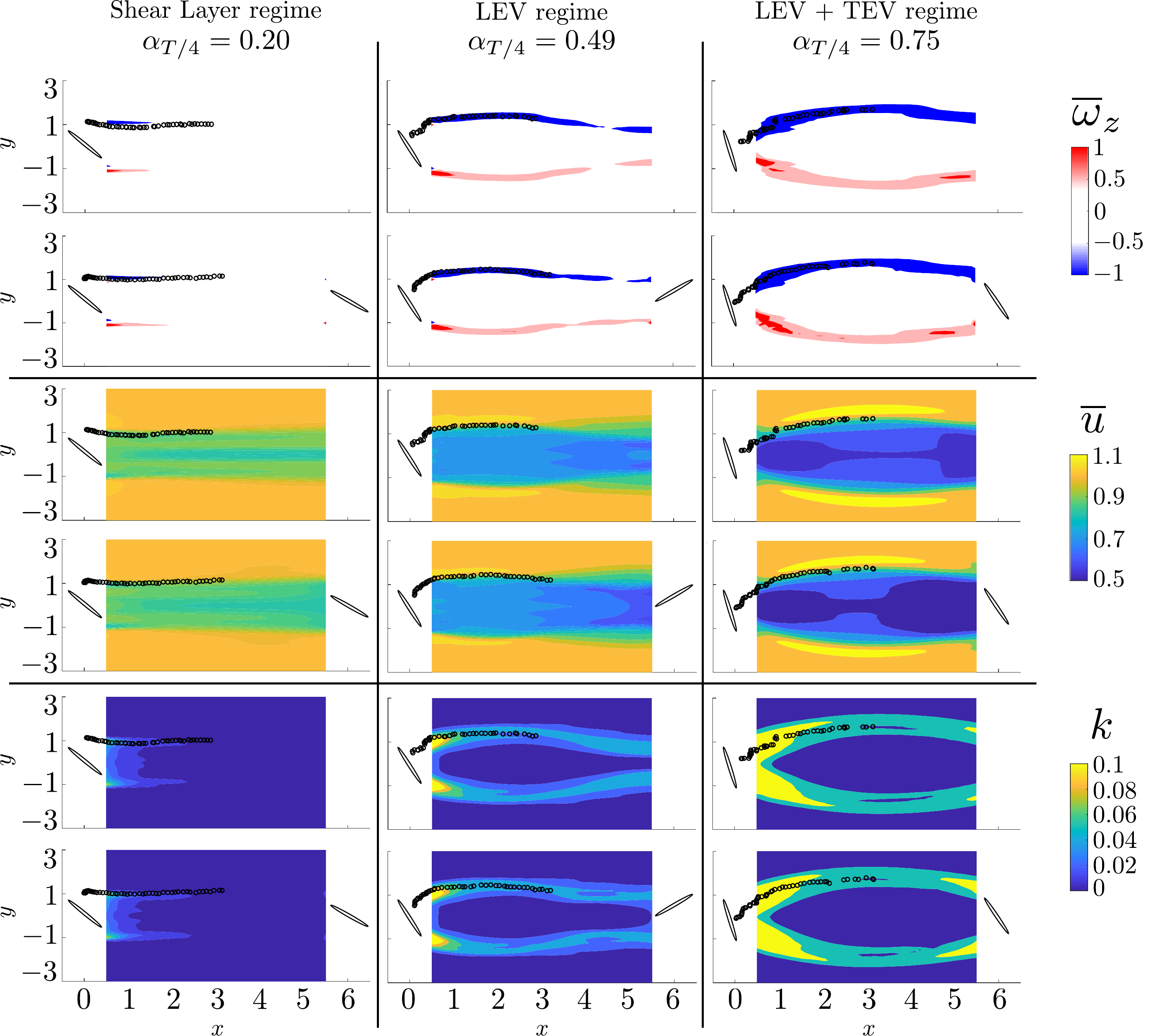}
\caption{Time-averaged vorticity, $\overline{\omega}_z$, streamwise velocity, $\overline{u}$ and turbulent kinetic energy, $k$, for single and two-foil simulations with different $\alpha_{T/4}$ values for the three regimes. Black circles represent the path of the primary vortex shed during foil`s upstroke.}
\label{f:kvortu}
\end{figure}

Next, for each of the three regimes, the velocity in the wake is averaged over time and decomposed into its steady and unsteady components, 
\begin{equation}
 \mathbf{u}(x,y,t) = \mathbf{\overline{u}}(x,y) + \mathbf{u}'(x,y,t),
\label{eq:decomp}
\end{equation}
where the bar ($\bar{\; }$) represents a time-averaged quantity. The energy associated with the unsteady components is computed as a turbulent kinetic energy,
\begin{equation}
 k(x,y) = \frac{1}{2} ( \overline{u'u'} + \overline{v'v'} )
\label{eq:k}
\end{equation}
where $u'u'$ and $v'v'$, correspond to the diagonal terms of the 2D Reynolds stress tensor. It is important to emphasize that the Reynolds decomposition into unsteady components is used to characterize the unsteadiness induced by the large scale vortices in the wake, and not fluctuations due to turbulence.
In analyzing the wake characteristics between tandem foils, the first question is whether the presence of the trailing foil modifies the wake dynamics. Figure \ref{f:kvortu} shows the time-averaged vorticity, $\bar\omega_z$, streamwise velocity, $\bar u$, and turbulent kinetic energy, $k$, for both single and two-foil simulations on three representative values of $\alpha_{T/4}$, one for each regime. The trajectory of the primary clockwise-rotating vortex (circled on Figure \ref{f:Qvortinst}) is also displayed in each plot to check if the presence of a trailing foil changes the vortex trajectory from the leading foil. Similar from Figure \ref{f:Qvortinst}, the inter-foil phases selected represent the cases of maximum trailing foil efficiency. The leading foil is at maximum pitch angle, or $t/T=0.25$ for each case.

There is a strong similarity between the vortex path of single and two-foil simulations, indicating that the presence of the trailing foil does not significantly influence the vortices shed from the leading foil, except in the immediate upstream vicinity of the trailing foil. The flow fields are also very similar between configurations, with the major differences observed in regions after $x=4$, which is within two chord lengths of the trailing foil. This impact that the trailing foil has on the wake is explained by the flow blockage and subsequent increase of pressure caused by the presence of the trailing foil.

Although the wake is similar between the single and two-foil simulations, the wake is significantly different among the three regimes. For the shear layer regime, the flow is more uniform in the x-direction compared to higher $\alpha_{T/4}$, which is a consequence of the low energy extraction of the leading foil, characteristic of foil kinematics of low $\alpha_{T/4}$. Consequently, the energy left in the wake is considerably higher compared to higher $\alpha_{T/4}$, which has a lower streamwise time-averaged velocity in Figure \ref{f:kvortu}. The weak vortex formation at low $\alpha_{T/4}$ is consistent with the mean $k$ field, which is negligible compared to that at higher $\alpha_{T/4}$. At higher $\alpha_{T/4}$, $k$ concentrates in regions close to the leading foil, which correspond to the maximum strength of the vortices as they shed from the foil's surface. The vortices carve out a path of higher $k$ compared to the bulk flow directly behind the foil, as also observed by Young et al. \cite{young2020} at similar kinematics ($fc/U_\infty=0.14;h_o=1.00;\theta_o=76.3^{\circ}$). These streamwise time-averaged velocity and turbulent kinetic energy fields in the wake are used to quantify the energy available in the oncoming flow to the trailing foil.

\begin{figure}[htbp]
\centering
\includegraphics[width=0.6\linewidth]{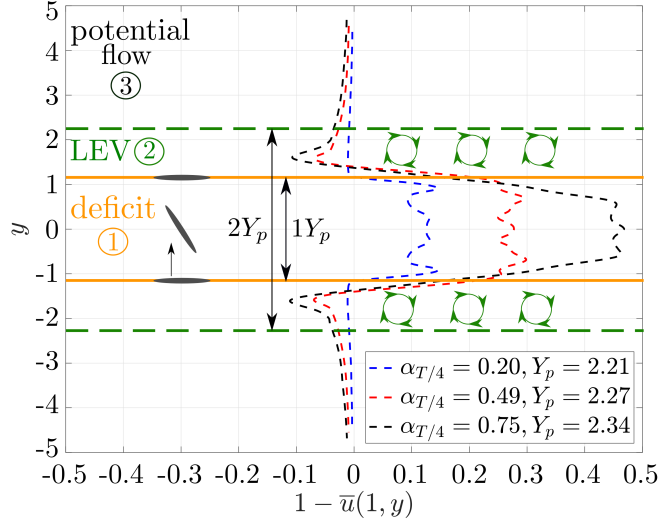}
\caption{Wake deficit, $1 - \overline{u}(x,y)$, as measured at $x=1$ for three different $\alpha_{T/4}$ values, representative of the shear layer, LEV and LEV+TEV regimes. The wake profile is divided into a pure deficit region, a LEV region, and a potential flow region. The orange lines roughly define the deficit region and represents $Y_p \approx 2.3$, close in value to the cases shown. A schematic of the foil motion in the $x-y$ plane is shown for reference.}
\label{f:wakewidth}
\end{figure}

In order to quantify the total kinetic energy in the wake, a wake width is defined. Figure \ref{f:wakewidth} shows the non-dimensionalized wake deficit, defined as the difference between the freestream and the mean streamwise velocity downstream of the first foil, $1 - \overline{u}(x,y)$. The wake deficit is measured at $x=1$ for the same three characteristic $\alpha_{T/4}$ values from Figure \ref{f:kvortu}. The $y-z$ plane at $x=1$ is chosen since it is close to the leading foil and assumed to be sufficiently far from any trailing foil interference. Data is interpolated from the simulations and discretized into $100$ equally spaced points between $y=-2Y_p$ and $y=+2Y_p$.

Examining the profiles in Figure \ref{f:wakewidth}, three regions can be identified. Region `$1$' is defined as the wake deficit region as only positive wake deficit values ($U_\infty > \overline{u}$) are found. A stronger wake deficit is observed with increasing $\alpha_{T/4}$, in accordance with Figure \ref{f:kvortu}. Region `$2$' shows negative wake deficit values ($\overline{u} > U_\infty$) at the same vertical location as the vortices shown by Figure \ref{f:kvortu}. The stronger negative wake deficit corresponds with higher levels of $k$, likely indicating  stronger vortices and increasing $\alpha_{T/4}$. Region `$3$' is characterized by the wake velocity approaching freestream velocity ($U_\infty \approx \overline{u}$). Since the freestream region is uniform, constant, and located far away from the foil, it can be approximated as irrotational and inviscid, and labelled as a potential flow region.

\begin{figure}[htbp]
\centering
	\begin{subfigure}{0.49\textwidth}
	\centering
      	\includegraphics[width=1\linewidth]{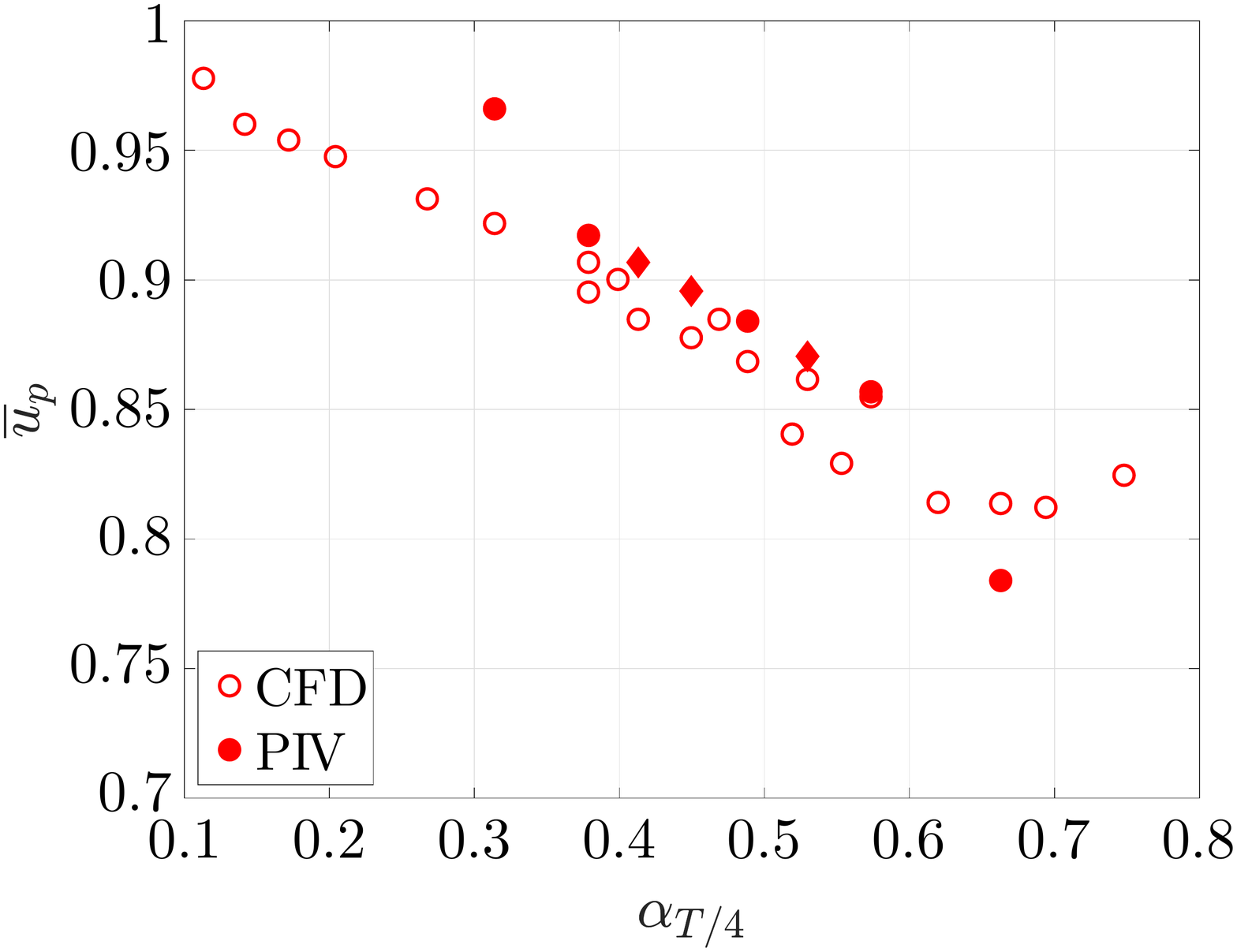}
		\caption{$\overline{u}_p$ vs $\alpha_{T/4}$}
	\label{f:upvsalpha}
	\end{subfigure}
	\begin{subfigure}{0.49\textwidth}
	\centering
		\includegraphics[width=1\linewidth]{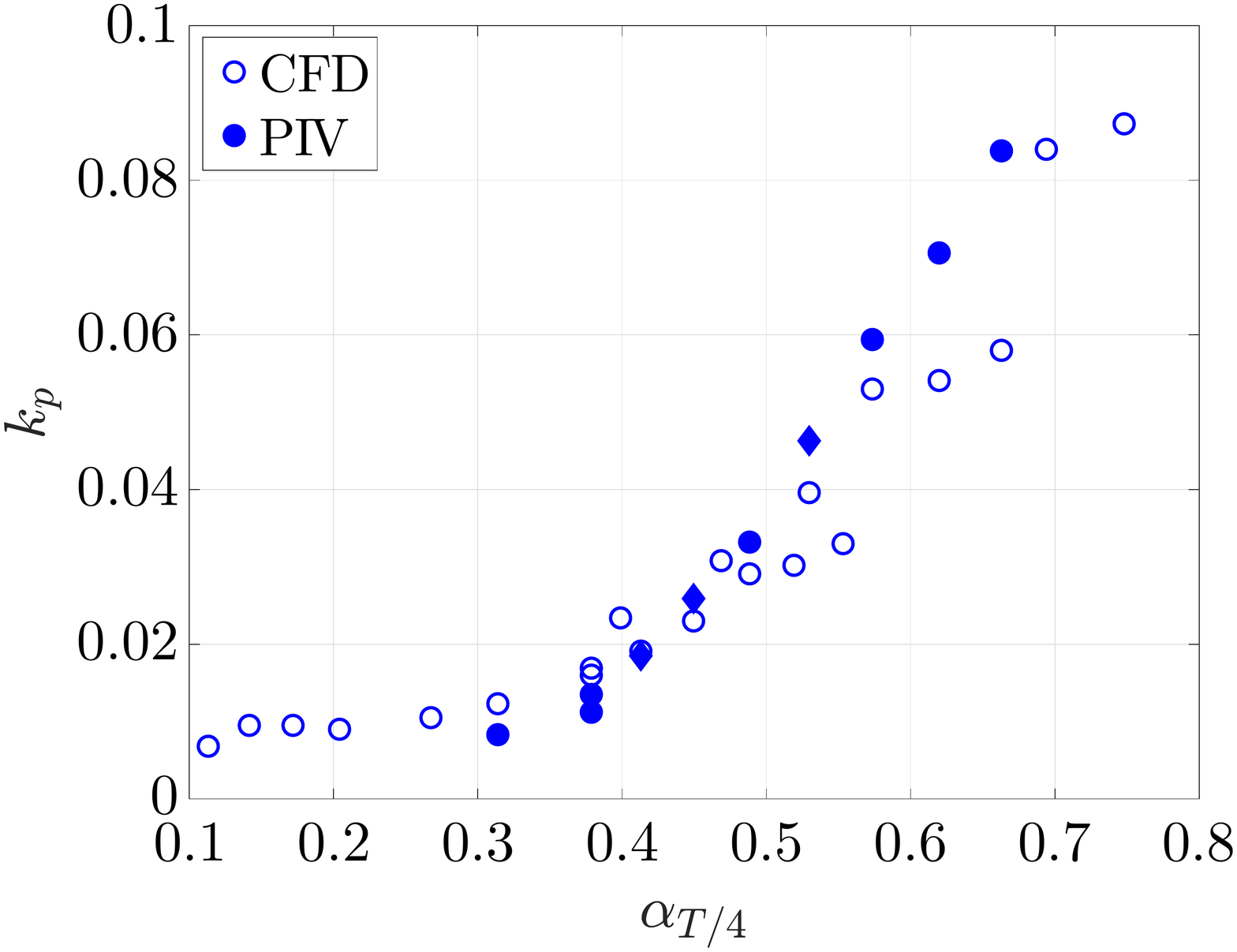}
		\caption{$k_p$ vs $\alpha_{T/4}$}
	\label{f:kvsalpha}
	\end{subfigure}
\caption{Mean wake velocity ($\overline{u}_p$) and turbulent kinetic energy ($k_p$) from numerical and PIV data. The diamond shape marker corresponds to interpolated PIV data (See Appendix A).}
\label{f:upk}
\end{figure}

As highlighted in Figure \ref{f:wakewidth}, the width for regions `$1$' and `$1+2$' are approximately $1Y_p$ and $2Y_p$, respectively, where $Y_p$ is the height of the swept area. The value of $Y_p$ is highly correlated with the heave amplitude, but also slightly dependent on the pitch amplitude (see Figure \ref{f:single}). Based on this analysis, most of the energy in the wake is within an area defined by $y=-Y_p$ to $+Y_p$. Thus, an averaged wake velocity, $\overline{u}_p$, can be calculated at $x=1$ by integrating the velocity deficit over this region,

\begin{equation}
\overline{u}_p = 1 - \frac{1}{2Y_p} \int_{-Y_p}^{+Y_p} 1-\overline{u}(1,y) \; dy.
\label{eq:up}
\end{equation}
The mean turbulent kinetic energy over a specific $y-z$ plane, $k_p$, can be computed in the same manner,
\begin{equation}
k_p \; = \frac{1}{2Y_p} \int_{-Y_p}^{+Y_p} k(1,y) \;\; dy.
\label{eq:kavg}
\end{equation}
Applying Equations \ref{eq:up} and \ref{eq:kavg} to all kinematics for both experimental and numerical data, Figure \ref{f:upk} displays $\overline{u}_p$ and $k_p$ with respect to $\alpha_{T/4}$. Similar to Figure \ref{f:etaUinf}, each point corresponds to the inter-foil phase with highest trailing foil performance.

The wake velocity in Figure \ref{f:upk} has a roughly linear decay with respect to $\alpha_{T/4}$ until approximately $\alpha_{T/4}=0.60$. Beyond this value, the wake velocity levels off around $0.82$ for the available numerical data. The data from PIV is generally in agreement with CFD although its magnitude at the lowest and highest available $\alpha_{T/4}$ deviate from the CFD values by $+0.04$ and $-0.03$, respectively. The drop at $\alpha_{T/4}=0.66$ can be partially explained by a limitation presented in the blockage correction method used on the experiments, which uses the measured $C_d$ values from the experiments, which are very high at this angle of attack (see Appendix A for further details on the blockage correction method).

The turbulent kinetic energy also shows a strong agreement between numerical and PIV data. As opposed to the wake velocity, $k_p$ remains constant at $k_p=0.01$ and then increases with $\alpha_{T/4}$ until it appears to level off around $\alpha_{T/4}=0.70$ at approximately $k_p=0.09$. The increase of the turbulent kinetic energy as $\alpha_{T/4}$ increases is a consequence of the stronger wake vortices especially in the LEV+TEV regime ($\alpha_{T/4}>0.50$). The higher efficiency of the leading foil in this regime compared to poor efficiency in the shear layer regime explains the decrease in wake velocity as more energy is extracted by the leading foil and hence the stronger wake deficit. This energy is mostly carried by the wake velocity since the energy per unit mass using the wake velocity ($\overline{u}^2_p$) is considerably higher than the turbulent kinetic energy. For instance, at $\alpha_{T/4}=0.69$, $\overline{u}^2_p$ is approximately $7.5$ times higher than the turbulent kinetic energy, $k_p$.

\subsection{Combining foil kinematics and configuration parameters}

In the previous section, Figure \ref{f:upk} demonstrated the steady and unsteady wake characteristics relative to the leading foil kinematics. Next, the relationship between the wake characteristics and the trailing foil performance is analyzed. 

\begin{figure}[htbp]
\centering
	\begin{subfigure}{0.49\textwidth} %0.40
	\centering
      	\includegraphics[width=1\linewidth]{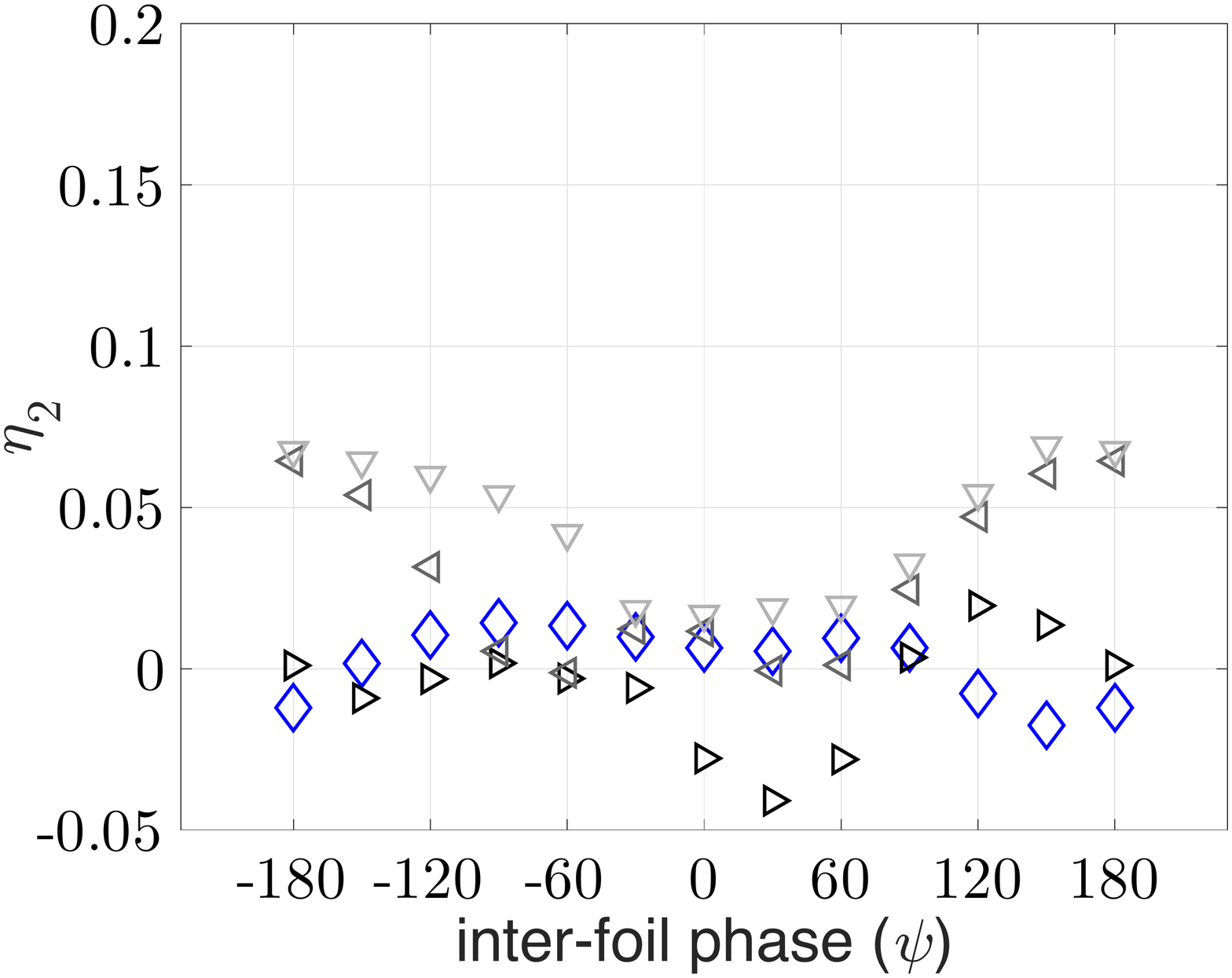}
		\caption{Shear Layer regime}
		\label{f:regimeA-psi}
	\end{subfigure}
	\begin{subfigure}{0.49\textwidth}
	\centering
		\includegraphics[width=1\linewidth]{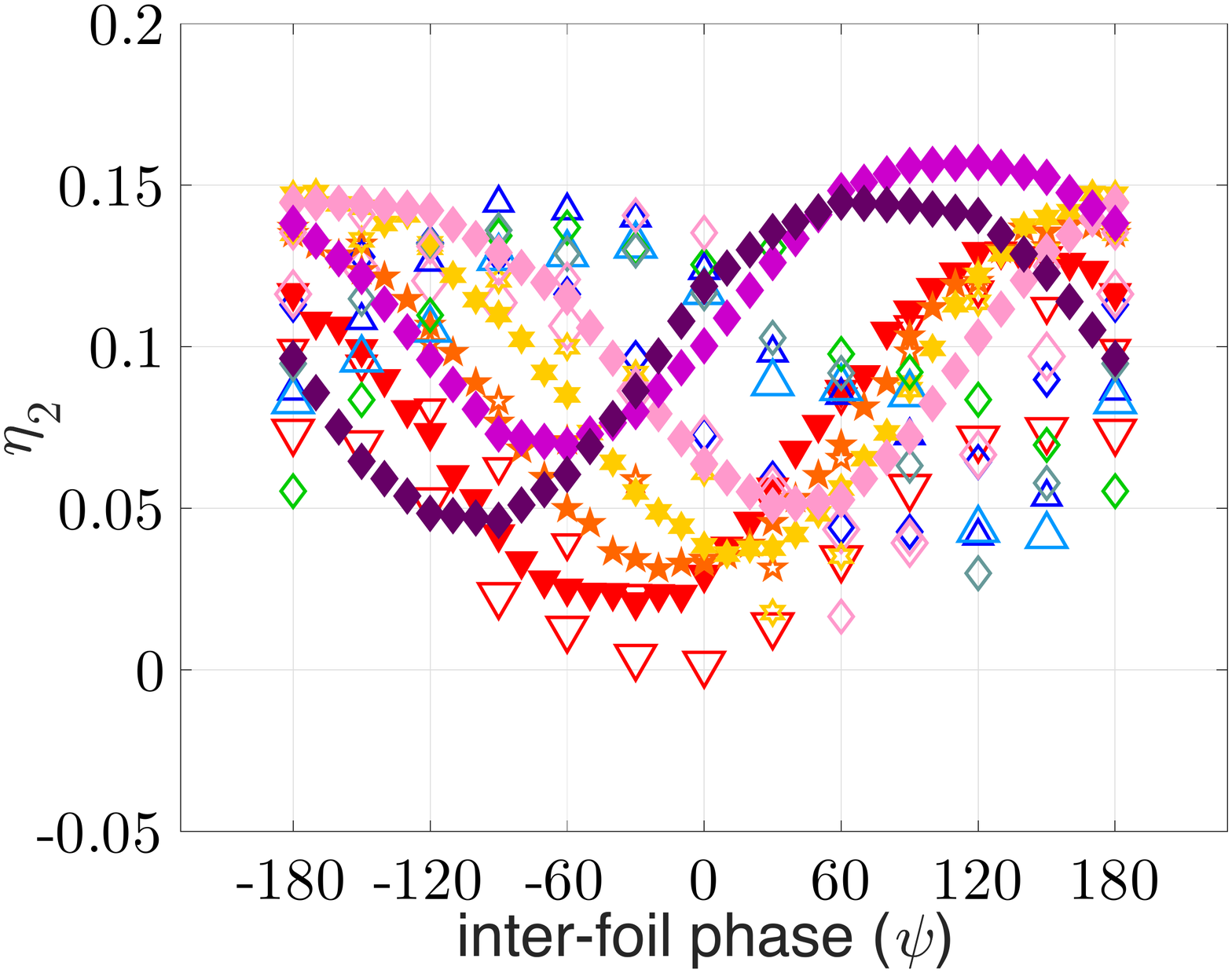}
		\caption{LEV regime}
		\label{f:regimeB-psi}
	\end{subfigure}
	
	\begin{subfigure}{0.49\textwidth}
	\centering
		\includegraphics[width=1\linewidth]{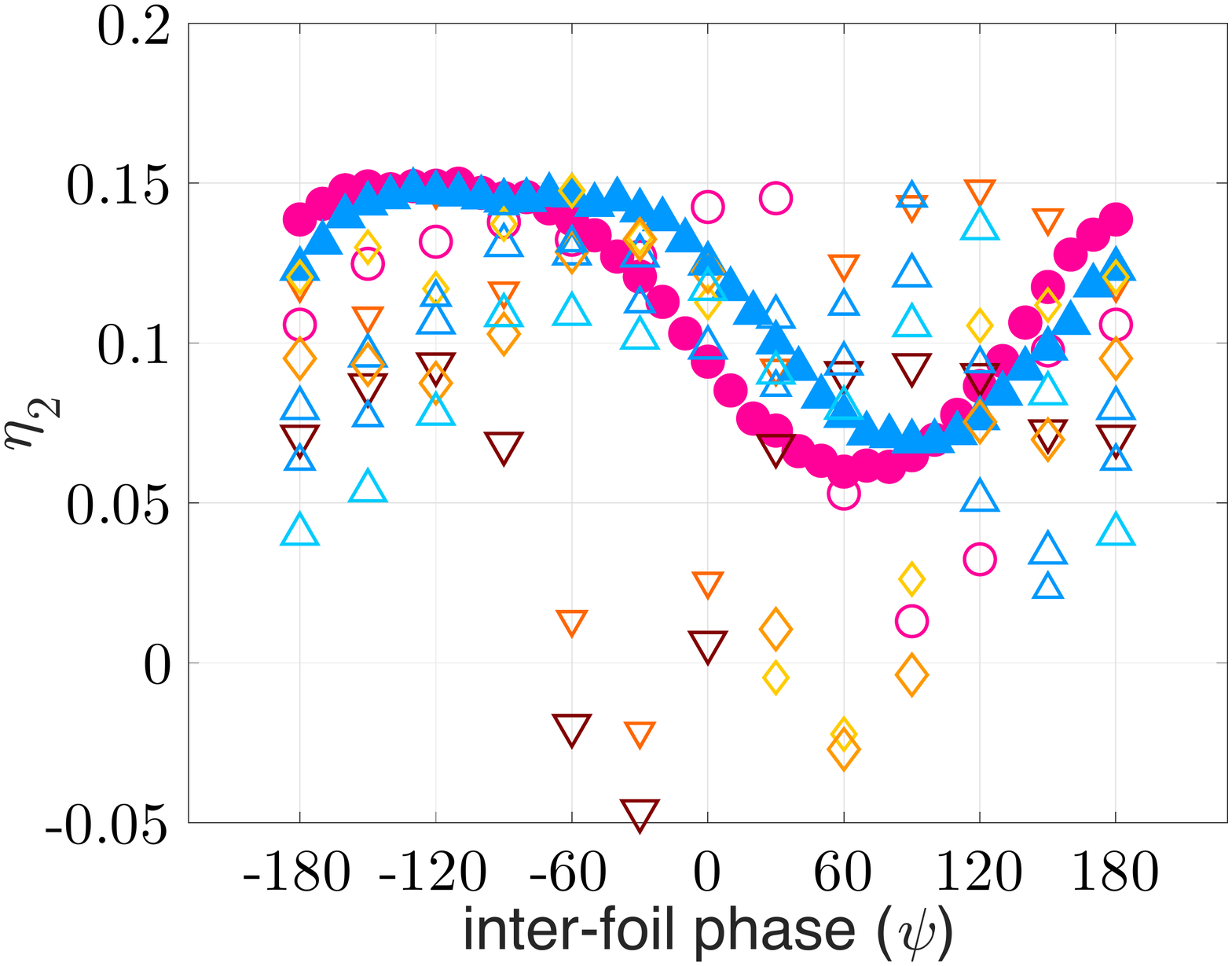}
		\caption{LEV+TEV regime}
		\label{f:regimeC-psi}
	\end{subfigure}
\caption{Trailing foil efficiency, $\eta_2$, with respect to inter-foil phase, $\psi$. Open symbols represent the numerical data, whereas filled ones represent experimental data. Each curve represents a set of kinematics as defined in Table \ref{t:tablekin}.}
\label{f:regimes-psi}
\end{figure}

First, the trailing foil efficiency profiles are plotted with respect to the inter-foil phase for all cases, including all inter-foil distances. This information is split up into the three regimes, and shown in Figure \ref{f:regimes-psi}.

In Figure \ref{f:regimeA-psi}, the maximum trailing foil efficiency achieved in the shear layer regime is approximately $7\%$, whereas for the other two regimes, the trailing foil reaches $15\%$ at optimal inter-foil phase. For the two lowest $\alpha_{T/4}$ values in the shear layer regime, efficiency is close or lower than $0$ at almost any inter-foil phase, which is explained by the kinematics being far from optimal within the energy harvesting range. With the combination of high reduced frequency and low pitch there is very little flow separation and the vortices roll up along the trailing edge at stroke reversal. 

At $\alpha_{T/4}=0.20$, efficiency reaches $7\%$ at the optimal inter-foil phase. Examination of the flow fields indicates that this is a consequence of the trailing foil avoiding the path of shear layer vortices emitted from each half-stroke of the leading foil, as highlighted in Figure \ref{f:Qvortinst}. At this $\alpha_{T/4}$ value, three pairs of clockwise and counter-clockwise vortices shed per half-stroke, each with low strength compared with kinematics at higher $\alpha_{T/4}$. The small and weak vortices follow a horizontal path as shown in Figure \ref{f:kvortu}, and are convected in the same direction as the freestream velocity. Nonetheless, even for the small vortices in this regime, interactions between the primary vortices and trailing foil affects trailing foil's efficiency. The impact on trailing foil efficiency depends on the strength of the vortices and the timing, as a slight shift in inter-foil phase may cause a vortex to directly impinge upon the trailing foil or miss it completely. The timing between the interaction of primary vortices and trailing foil is also described in Kinsey and Dumas \cite{kinseytandem2012} and it is a key feature for the slightly higher efficiency at the optimal inter-foil phase for both $\alpha_{T/4}=0.17$ and $0.20$. For lower $\alpha_{T/4}$ within the shear layer regime, the flow separation is less dramatic and hence the trailing foil is less susceptible to increase or decrease in the energy extraction from the oncoming flow. Consequently, the efficiency is roughly constant throughout the inter-foil phases.

The LEV and LEV+TEV regimes both display a sinusoidal-like profile for the trailing foil efficiency as a function of phase angle. However, there is a significant difference between these two regimes. The LEV regime contains kinematics that generate stronger vortices compared to the shear layer regime (see Figure \ref{f:Qvortinst}). As opposed to the lowest $\alpha_{T/4}$ cases, the path of the recently shed vortices start to separate from the previous half-stroke vortices of opposite sign. Another feature of the LEV regime is the vortex pattern of one pair and two single vortices that shed per half-stroke. The stronger vortices have a larger convective speed, which impact the interaction timing between the primary vortex and trailing foil. For instance, for the kinematics highlighted on the shear layer regime on Figure \ref{f:Qvortinst}, the primary vortex reaches $x=4$ at approximately $t/T=1.00$. The primary vortex for the case shown in the LEV regime reaches same position at $t/T=0.95$ and this values drops to $t/T=0.75$ for the LEV+TEV regime. The result is that the efficiency variation is more dramatic, and appears very close to sinusoidal with respect to inter-foil phase. 

In contrast, in Figure \ref{f:Qvortinst} the LEV+TEV regime shows a clear separation between the current half-stroke and the previous half-stroke of opposite signed vortices. The strength of the primary vortex pair and the four single secondary vortices that shed per half-stroke are significantly higher for this final regime compared to the LEV regime, and hence creates a more chaotic relationship between efficiency and inter-foil phase as observed on regions close to $\psi=0^{\circ}$. In particular, there is a sharp efficiency drop around $\psi=0^{\circ}$, even briefly dropping below the energy harvesting threshold for some kinematics. This break in the sinusoidal trend is explained by the unpredictable interaction of trailing foil with the oncoming flow that includes stronger secondary vortices from the leading foil. The higher convective speed presented in this final regime also has a major role in the sharp efficiency drop. Thus, the LEV+TEV regime has a larger efficiency range from $-5\%$ to $15\%$ compared to the $0\%-15\%$ from the LEV regime.

Although information can be extracted from Figure \ref{f:regimes-psi}, there is no clear efficiency trend with respect to the inter-foil phase especially in the range from $\psi=-120^{\circ}$ to $\psi=120^{\circ}$, as it is compounded by the effect of kinematics and inter-foil distance. Instead of using inter-foil phase, a key variable to establish the relationship between the wake and trailing foil performance is the `wake phase', $\Phi$, defined as

\begin{equation}
\Phi = 2\pi \frac{S_x}{\overline{u}_p}f^{*} + \psi
\label{eq:wakemodel}
\end{equation}

\noindent with $f^{*}=\frac{fc}{U_\infty}$. The `wake phase' is a modification from the originally implemented global phase model proposed by Kinsey and Dumas \cite{kinseytandem2012} and this model describes the phase shift between the wake trajectories of both foils considering both wakes being convected at the freestream velocity. The difference between the `wake phase' and the `global phase' parameters is the use of $\overline{u}_p$ instead of freestream velocity, which more accurately describes the mean flow speed between the two foils, especially for medium to high $\alpha_{T/4}$ cases, where the wake deficit is more apparent.

\begin{figure}[htbp]
\centering
	\begin{subfigure}{0.49\textwidth}
	\centering
      	\includegraphics[width=1\linewidth]{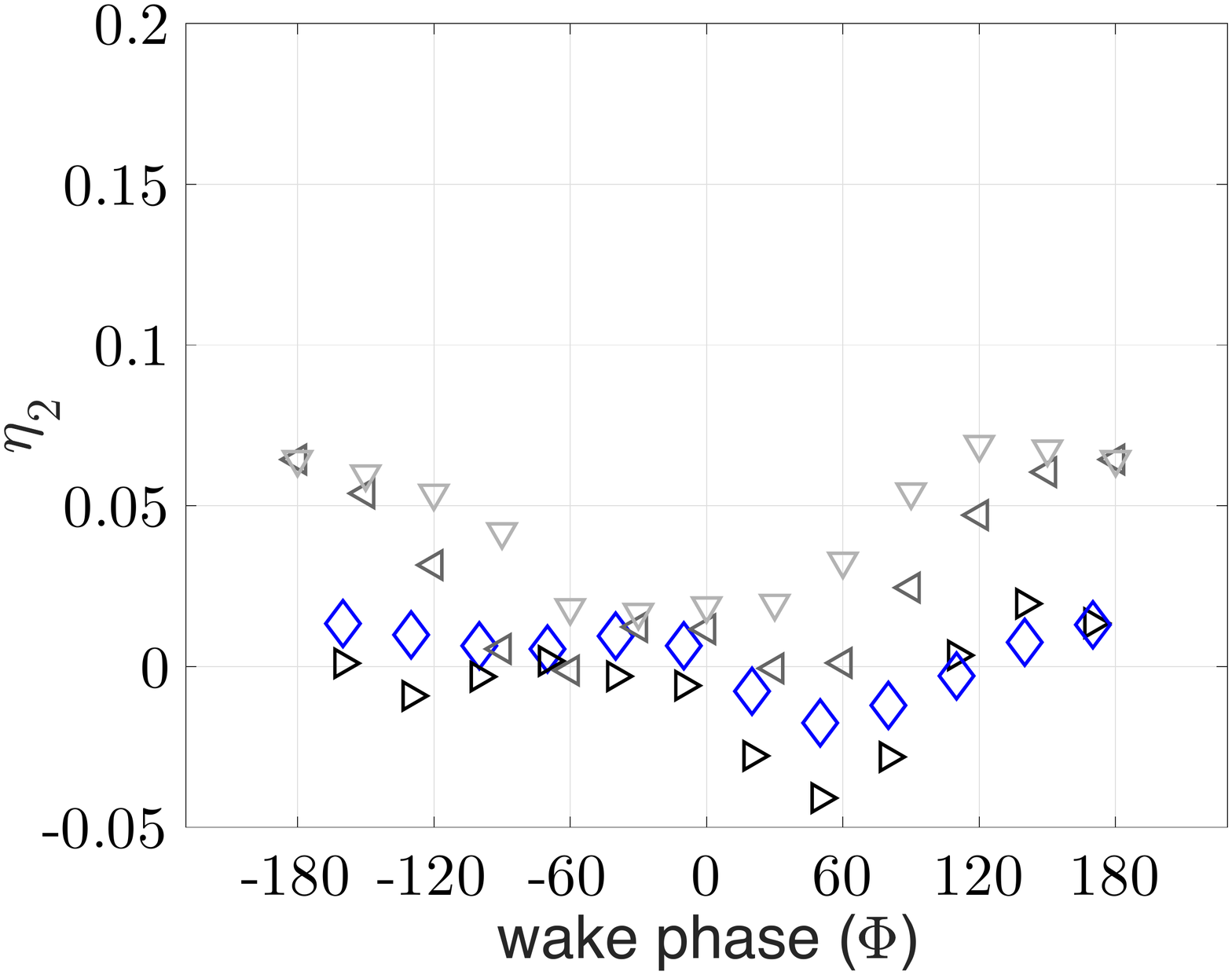}
		\caption{Shear Layer regime}
		\label{f:regimeA}
	\end{subfigure}
	\begin{subfigure}{0.49\textwidth}
	\centering
		\includegraphics[width=1\linewidth]{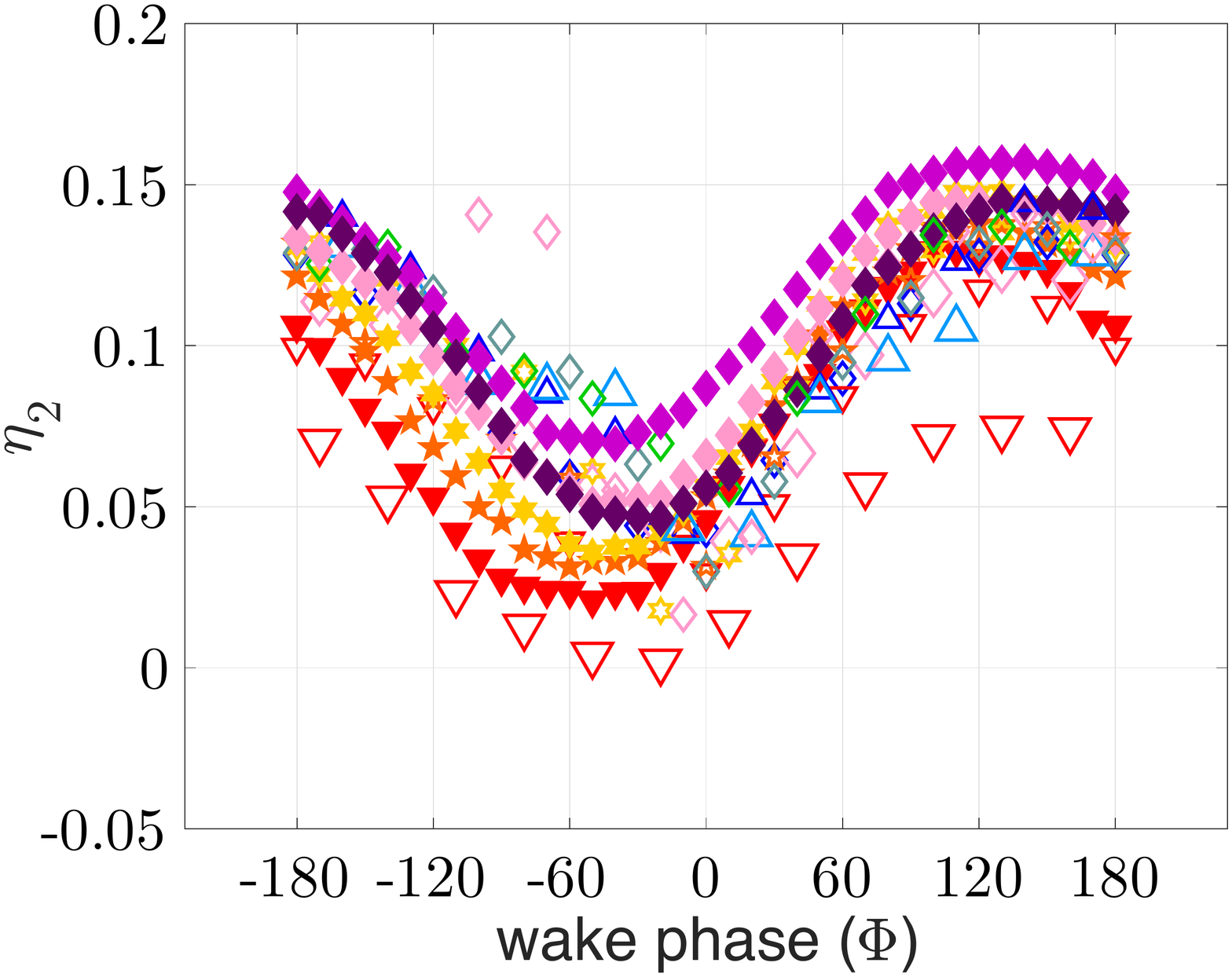}
		\caption{LEV regime}
		\label{f:regimeB}
	\end{subfigure}
	
	\begin{subfigure}{0.49\textwidth}
	\centering
		\includegraphics[width=1\linewidth]{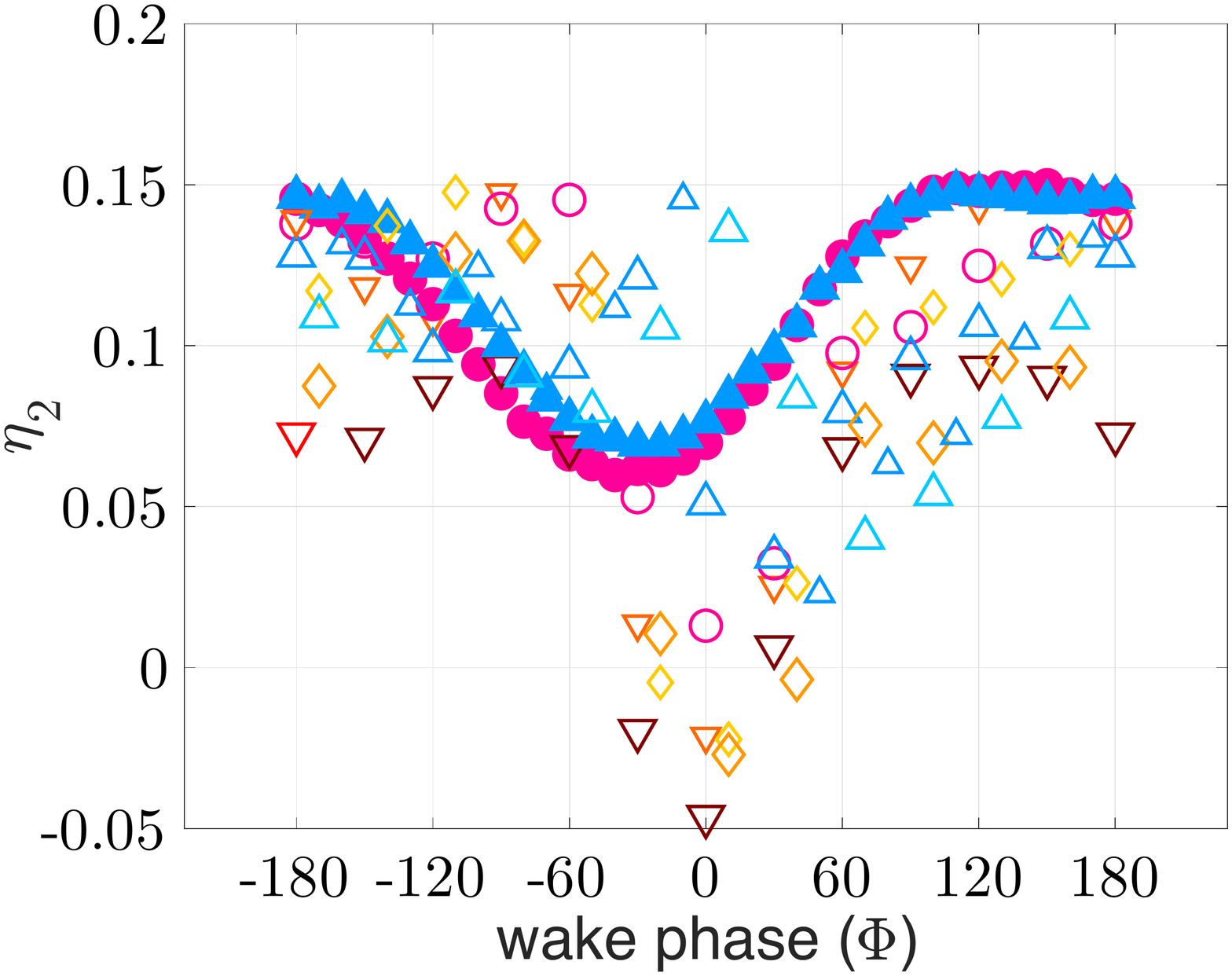}
		\caption{LEV+TEV regime}
		\label{f:regimeC}
	\end{subfigure}
\caption{Trailing foil efficiency ($\eta_2$) regimes using the wake phase model. $\Phi$ is in degrees. Open symbols stands for numerical data, whereas filled ones are experimental data. Each curve represents a set of kinematics as defined in Table \ref{t:tablekin}.}
\label{f:regimes}
\end{figure}

Kinsey and Dumas observed that their original global phase model does not provide any information on the relative angle of attack and the related occurrence of flow separation. However, the proposed wake phase model incorporates the streamwise velocity of the wake which subsequently correlates with the leading foil kinematics, and $\alpha_{T/4}$, as shown in Figure \ref{f:upk}. However, the highly variable vortex dynamics observed throughout the range of kinematics means that the interactions with the trailing foil can again be divided into three regimes. Figure \ref{f:regimes} organizes the trailing foil efficiency with respect to the wake phase for all kinematics within each regime. 

For the shear layer regime in Figure \ref{f:regimeA}, the wake phase model attempts to align the variation in efficiency across the various inter-foil phase angles for the four sets of kinematics tested numerically. By incorporating the reduced frequency and measured wake velocity in the model, the lowest efficiency is between $\Phi=0^{\circ}$ and $\Phi=60^{\circ}$. The two highest $\alpha_{T/4}$ cases of this regime have their minimum efficiency around $\Phi=0^{\circ}$ and the maximum efficiency at approximately $7\%$.

The LEV regime in Figure \ref{f:regimeB} demonstrates a strong collapse of trailing foil efficiency with respect to the wake phase. A roughly sinusoidal trend is formed with the minimum around $\Phi=0^{\circ}$, and the maximum around $\Phi=120^{\circ}$. The maximum efficiency peaks between $10\%$ to $15\%$ with a single case with maximum efficiency around $7\%$, and generally increases as $\alpha_{T/4}$ increases. The experiments and simulations show strong agreement for this regime.

For the third regime in Figure \ref{f:regimeC}, representing the highest $\alpha_{T/4}$ values, the wake phase model also shows a minimum efficiency at approximately $\Phi=0^{\circ}$. The efficiency range is from $7\%$ to $15\%$ on wake phases close to $180^{\circ}$, similar to the LEV regime, but efficiency is from $-5\%$ to $15\%$ on wake phase near $\Phi=0^{\circ}$. The two highest $\alpha_{T/4}$ show an efficiency increase around $\Phi=0^{\circ}$. Compared to Figure \ref{f:regimeC-psi}, the wake phase model in Figure \ref{f:regimeC} demonstrates a partial collapse of the efficiency profiles bringing the minimum efficiency of each case closer to $\Phi=0^{\circ}$.

In contrast, the experimental data in Figure \ref{f:regimeC} still show a similar sinusoidal behavior to the LEV regime in Figure \ref{f:regimeB}, but with a plateau around the maximum efficiency between $\Phi=70^{\circ}-180^{\circ}$. The stark differences between various kinematics, and between the experimental and numerical data, highlight the strong influence of the LEV and TEV within this regime. In particular, the strong TEV formation increases the wake disturbance and vortex dynamics more so than the other two regimes. Through analyzing the flow fields in each regime as exemplified in Figure \ref{f:Qvortinst}, it is found that the numerical data have more concentrated vortices than the experiments of similar angles of attack. This is likely a result of different Reynolds numbers and the three-dimensional effects that are inevitable in experiments and are not captured in the simulations. 

In terms of the primary vortex position when it impinges on the trailing foil in an upstroke motion, Kinsey and Dumas defined four configurations that affect the pressure field around the foil. Two configurations are favorable for power extraction and these occur when the counter-clockwise vortex impinges on the suction side of the foil or a clockwise vortex impinges on the pressure side of the foil. The two unfavorable conditions occur when vortices with signs opposite to each of the above mentioned configuration impinges on the trailing foil.

Kinsey and Dumas also found a strong interaction between the primary vortex and the trailing foil at kinematics of $fc/U_\infty=0.14; h_o=1.00; \theta_o=70^{\circ}$, which has a $\alpha_{T/4}=0.50$, a borderline case between the LEV and LEV+TEV regimes. For this case at limited inter-foil phases, they obtained an optimal global phase of $90^{\circ}$, which is different than the optimal wake phase of $120^{\circ}$ obtained in the LEV regime. However, the foil kinematics and configurations from Kinsey and Dumas only explored two inter-foil phases, $-90^{\circ}, -180^{\circ}$, as well as small variations in the heave and pitch amplitudes compared to the current data.

With a considerably higher parameter variation, the original global phase model is applied to the current data (see Appendix B) to check its performance among regimes in terms of efficiency trends and optimal global phases. The predictions between models is particularly different within the LEV regime, with the global phase model not being able to present as clear of an efficiency trend as the wake phase model. This can be explained by the wake phase model incorporating the mean streamwise wake velocity.

With the sinusoidal efficiency profile in the LEV regime, the wake phase model can be used to predict the efficiency depending only on the leading foil kinematics. For instance, at $\alpha_{T/4}=0.40$ and $\Phi=-60^{\circ}$, the trailing foil efficiency should be within $1\%<\eta<10\%$. This prediction is insightful to obtain optimal configurations for better energy harvesting. For example, a trailing foil efficiency of $15\%$ is obtained for a wake phase of $-180^{\circ}$ and a case with $\alpha_{T/4}=0.49$. Therefore, the same efficiency could be obtained with any set of reduced frequency, heave and pitch amplitude as long as it provides the same $\alpha_{T/4}$ and wake phase. Kinsey and Dumas highlighted the prediction ability of their global phase model but they found it reliable only between cases sharing the same wake profiles or similar $\alpha_{T/4}$. The approach of dividing the wake into regimes makes prediction more feasible across regimes. 

\begin{figure}[htbp]
\centering
	\begin{subfigure}{0.49\textwidth}
	\centering
      	\includegraphics[width=1\linewidth]{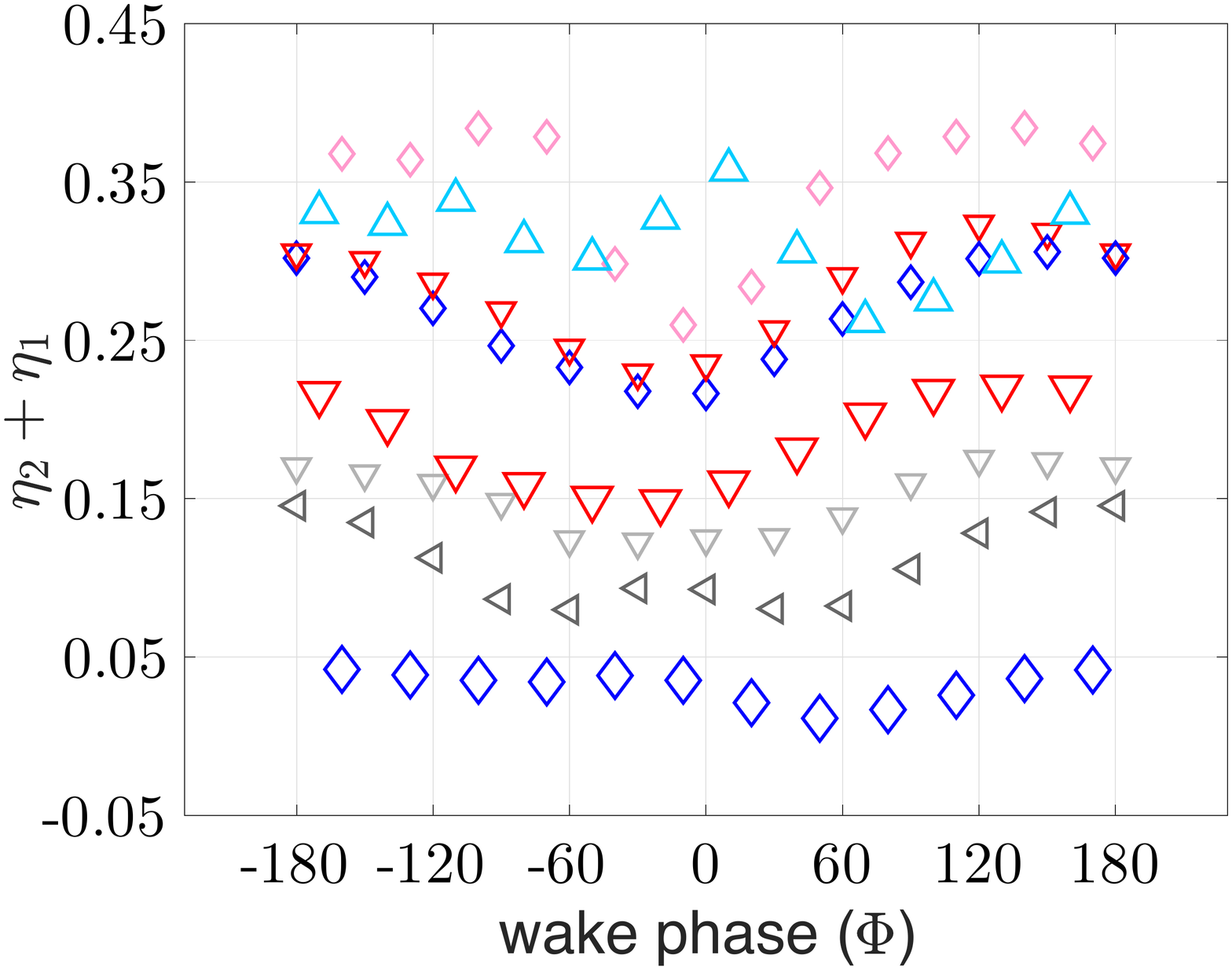}
		\caption{System efficiency, $\eta_1+\eta_2$, for kinematics within all regimes.}
		\label{f:syseta}
	\end{subfigure}
	\begin{subfigure}{0.49\textwidth}
	\centering
		\includegraphics[width=1\linewidth]{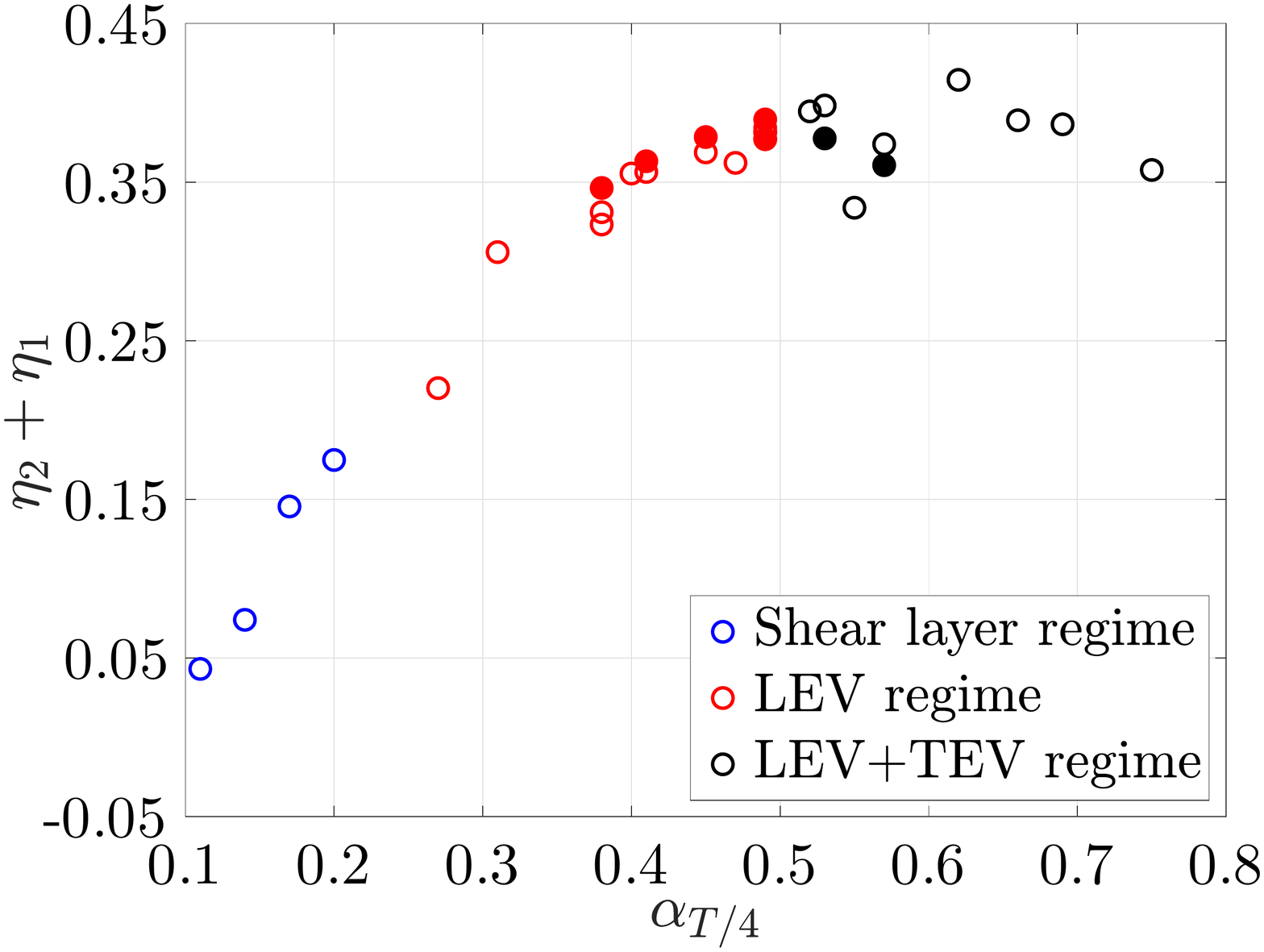}
		\caption{Maximum system efficiency vs $\alpha_{T/4}$. Filled circles correspond to experimental data.}
		\label{f:maxsyseta}
	\end{subfigure}
\caption{System efficiency analysis}
\label{f:sysanalysis}
\end{figure}

Another measure of the tandem array performance, and one that is important in practical configuration,  is the \emph{system} efficiency, defined as the sum of the individual efficiency values ($\eta_1$+$\eta_2$) since the two foils operate over the same swept area. System efficiency is computed with respect to wake phase for a subset of data ranging from $\alpha_{T/4}=0.11-0.75$ in Figure \ref{f:syseta}. Since efficiency increases with increasing $\alpha_{T/4}$ (Figure \ref{f:etaUinf}), system efficiency also increases accordingly. More subtle is the trend with respect to wake phase as $\alpha_{T/4}$ increases. At the lowest $\alpha_{T/4}$ values, the system efficiency is almost constant throughout different wake phases. As $\alpha_{T/4}$ increases, a sinusoidal profile emerges due to the regular formation of a strong LEV and wake-foil interactions become stronger as observed in the instantaneous vorticity field across different $\alpha_{T/4}$ from Figure \ref{f:Qvortinst}. For the highest $\alpha_{T/4}$ value, the sinusoidal trend is disrupted and the maximum and minimum wake phase is shifted. These conclusions are also supported by Figure \ref{f:maxsyseta}, where the maximum system efficiency (at optimal wake phase angle) increases as $\alpha_{T/4}$ increases, reaching a maximum of $\eta_1+\eta_2 \approx 0.40$ at approximately $\alpha_{T/4} \approx 0.50$.

Overall, among all regimes, the wake phase model presents a maximum trailing efficiency at $\Phi\approx120^{\circ}$ and a minimum at around $\Phi=0$. This is obtained for a wide range of kinematics including a variation of each kinematic parameter, inter-foil phase and inter-foil distance. The maximum trailing foil efficiency is similar to the implementation by Xu and Xu \cite{xuWake2017}, where they obtained a maximum efficiency at around $\Phi=160^{\circ}$ and also a minimum efficiency close to $\Phi=0$. However, similar to Xu and Xu \cite{xuWake2017}, the result deviates from the original global phase model from Kinsey and Dumas \cite{kinseytandem2012}. A possible reason for this difference is that Xu and Xu analyzed the global phase model in a single set of kinematics. In the original model, Kinsey and Dumas \cite{kinseytandem2012} considered a variation of the reduced frequency, heave and pitch amplitudes. However, Kinsey and Dumas noticed the optimal global phase of $90^{\circ}$ was found for a specific inter-foil distance, and only tested on a small subset of inter-foil phase angles. 

\subsection{Updating trailing foil efficiency using the wake characteristics}

The wake velocity is not only useful to predict trailing foil efficiency and system efficiency, but can be used to more accurately represent the kinetic energy in the wake. By incorporating energy directly from wake measurements, the trailing foil efficiency is modified. Starting with the foil efficiency defined in Equation \ref{eq:eta}, the mass flow rate per unit span is modified from $\rho U_\infty$ to $\rho \overline{u}_p$ and the energy available in the oncoming flow is modified from $\frac{1}{2} U_\infty^2$ to $\frac{1}{2}\left(\overline{u}_p^2 + 2k_p \right )$, incorporating energy from the mean flow in addition to the turbulent kinetic energy. The addition of the averaged turbulent kinetic energy is correlated with the unsteady fluctuations within the coherent vortices in the wake. Equation \ref{eq:etaupdated} defines the modified trailing foil efficiency, $\eta_2^*$, as
\begin{equation}
\eta_2^* = \frac{\overline{P_2}}{\frac{1}{2}\rho \overline{u}_p \left(\overline{u}_p^2 + 2k_p \right ) Y_p}.
\label{eq:etaupdated}
\end{equation}

\begin{figure}[htbp]
\centering
\includegraphics[width=0.6\textwidth]{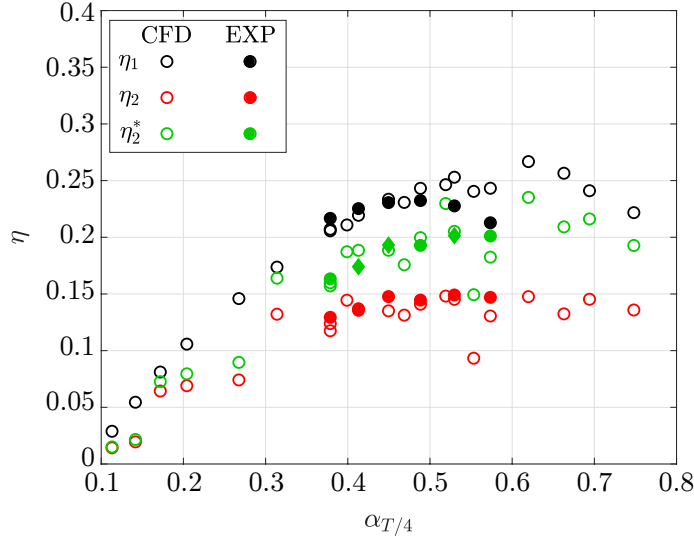}
\caption{Leading and trailing foil efficiencies ($\eta_1$ and $\eta_2$) normalized with freestream velocity and using information from the wake between foils ($\eta_2^*$). Each point for the trailing foil corresponds to the inter-foil phase that provides the highest efficiency. The diamond shape marker corresponds to PIV linear interpolated data. See Appendix A for further details on linear interpolated data.}
\label{f:etaupdated}
\end{figure}

Figure \ref{f:etaupdated} includes numerical and experimental data using the original definition of efficiency (Equation \ref{eq:eta}), in addition to $\eta_2^*$ defined in Equation \ref{eq:etaupdated}. The effect of $\eta_2^*$ is especially prevalent at high $\alpha_{T/4}$ as the turbulent kinetic energy is considerably higher compared to low $\alpha_{T/4}$ (Figure \ref{f:kvsalpha}). For instance, the difference in efficiency values between $\eta_2^*$ and $\eta_2$ at $\alpha_{T/4}=0.66$ is approximately $8\%$, whereas at $\alpha_{T/4}=0.20$ is only $1\%$.

Overall Figure \ref{f:etaupdated} shows that the modified trailing foil efficiency is higher than the efficiency normalized by $U_\infty$ across all $\alpha_{T/4}$, and it is bounded by the efficiency of the leading foil. The efficiency profiles from Figure \ref{f:etaupdated} show that analyzing the wake in terms of its measured mean and unsteady velocities increases the efficiency, as observed by Young et al. \cite{young2020}. Young et al. showed this potential increase in efficiency through an analysis of the additional terms that appear when the instantaneous flow variables are split into their mean and fluctuating components. These terms include a non-uniform and unsteady pressure across the wake between foils, unsteady transport of momentum and energy from freestream into the wake, and viscous flow work and viscous dissipation within two control volumes enclosing each oscillating foil. Whereas they noticed a particularly high contribution for the unsteady pressure term, the net effects from analyzing the unsteady flow components is being considering through Equation \ref{eq:etaupdated} in the current investigation. As the flow becomes more chaotic with increasing $\alpha_{T/4}$, the mean wake velocity decreases and the mean turbulent kinetic energy increases. Thus, the difference between $\eta_2$ and $\eta_2^*$ is more apparent at higher $\alpha_{T/4}$.  

This result may give insight on optimal array configurations since the system efficiency may be higher for one configuration than the other depending on how much energy the trailing foil receives from the wake.
The wake analysis in terms of averaging streamwise velocity and turbulent kinetic energy over a $y-z$ plane can be expanded and applied for multiple foils downstream. In this sense, an efficiency ratio parameter (e.g. $\eta_2^*/\eta_1$) may be useful to predict array performance.

\section{Conclusion}

The wake structure and dynamics on a two-foil array is explored in a wide range of kinematics and configurations, both computationally and experimentally. The goal of this paper is to quantify the effects between foil kinematics and the wake behind leading foil in terms of vortex structure and flow velocity, and use this information to model the trailing foil's performance.

Through an analysis of the primary vortex strength shed from the leading foil, three main regimes are defined and characterized based on their kinematics ($\alpha_{T/4}$), vortex strength, and wake structure. The `shear layer' regime occurs at low $\alpha_{T/4}$ ($\alpha_{T/4} \leq 0.20$) and the wake region is dominated by a shear layer and contains very weak vortex formations. The `LEV' regime ($0.20 < \alpha_{T/4} \leq 0.49$) contains a primary LEV that sheds at each half-stroke forming a path of shed vortices that are distanced from the previous half-stroke vortices of opposite sign. This is more dramatic in the third and final regime, `LEV+TEV' ($0.49 < \alpha_{T/4} \leq 0.75$), in which the development of a large TEV further influences the wake structure. 

The efficiency of the second foil is a function of the inter-foil distance and inter-foil phase angle, as well as the baseline kinematics of the foils. Within the LEV regime, the efficiency is roughly sinusoidal with respect to inter-foil phase angle. This sinusoidal behavior begins to develop in the shear layer regime but occurs to a lesser extent.  At the high $\alpha_{T/4}$ of the `LEV+TEV' regime, the sinusoidal trend is not sustainable for all kinematics due to the more unpredictable wake patterns from the TEV. When selecting the optimal inter-foil phase, the efficiency of the second foil gradually increases with $\alpha_{T/4}$, leveling off at approximately $15\%$ in the LEV regime at approximately $\alpha_{T/4} = 0.40$.

Building off the global phase model proposed by Kinsey and Dumas \cite{kinseytandem2012}, a wake phase model is introduced to better predict the trailing foil efficiency based on kinematics and inter-foil configurations. The model shows good agreement with the data for the LEV regime, when the wake structure is more predictable, and the experimental and numerical data collapse nicely. The wake phase model is less successful in predicting the efficiency for the shear layer regime in which the kinematics do not produce  consistently strong vortices, or the LEV+TEV regime in which the number of vortices increases to the point where the wake is too chaotic. Although the results shown are for the same kinematics applied on both foils, the wake phase model can be used for distinct kinematics between foils since the model is independent of the trailing kinematics.

Finally, the wake is decomposed into its steady and unsteady components to evaluate the contribution of available kinetic energy for the second foil. A modification to the trailing foil efficiency is implemented utilizing measured quantities in the wake. This modification provides a better estimate of the energy available to the trailing foil, improving the nominal efficiency that uses the freestream velocity. These results support the findings by Dabiri \cite{dabiri2020} and Young et al. \cite{young2020} who emphasize the importance of unsteady flow in such configurations.

\section*{Acknowledgments}

This research was funded by the US National Science Foundation, (CBET award 1921594: JF,BR;  CBET award 1921359: KB). KB, YS and QG also benefited from support from the US Air Force Office of Scientific Research (Grant FA9550-18-1-0322).  The authors would like to thank David Burkhart for the support on post-processing the numerical data, Filip Simeski for the preliminary simulations, Nicholas Simone for the efficiency measurements presented in Figure \ref{f:validation}, and Howon Lee for his contributions towards the interpretation of the data.

\section*{Appendix A: Computing blockage effects} \label{appendixA}

Due to the constraints of the flume, blockage effects are accounted for in the experimental data using a method derived by Houlsby et al. \cite{houlsby} and later implemented by Ross et al \cite{ross}.

\begin{figure}[H]
\centering
\includegraphics[width=1\textwidth]{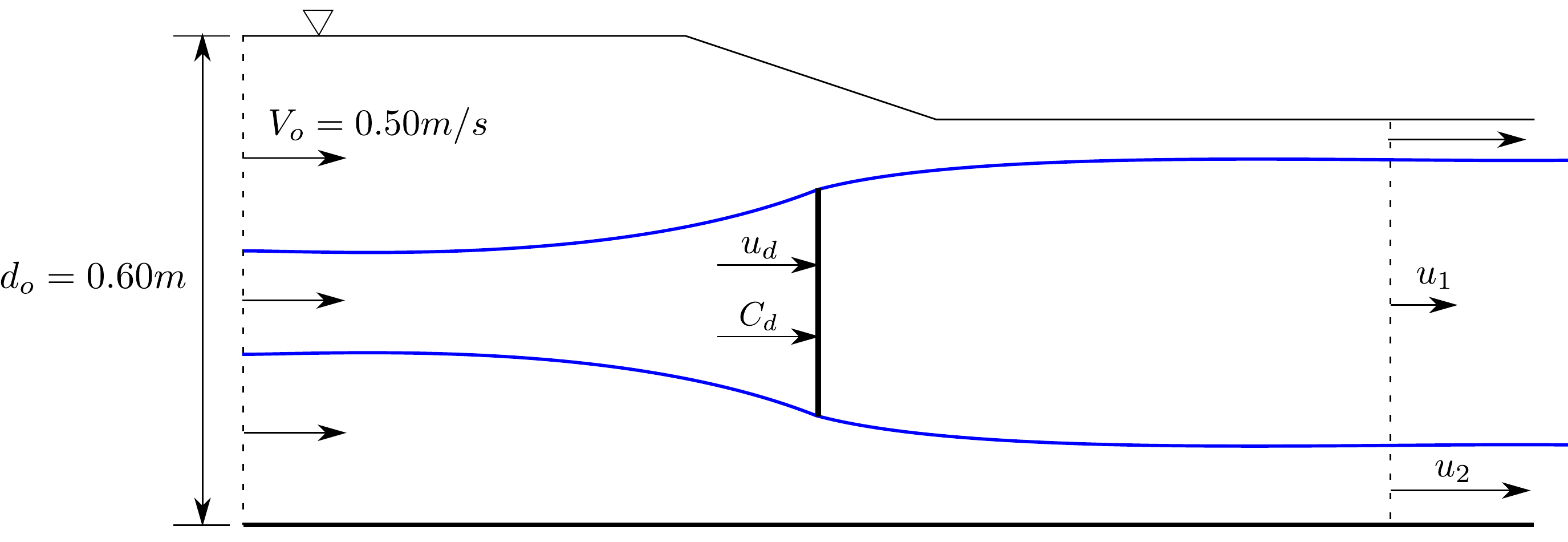}
\caption{Streamtube model of an actuator disc in open channel flow with a deformable free surface.}
\label{f:houlsby}
\end{figure}

A general schematic of an open channel used in Houlsby's method is presented in Figure \ref{f:houlsby}, including values of the current flume's undisturbed upstream water depth, $d_o=0.60$ m, and the freestream velocity, $V_o=0.50$ m/s. The streamwise velocity through the turbine ($u_d$), the velocity of the core flow ($u_1$), and the bypass flow velocity ($u_2$) are iterative variables used in Houlsby's method.

The drag coefficient ($C_d$) is experimentally measured on the leading foil for each set of kinematics. It is also assumed that the drag coefficient is roughly the same on both foils considering the dynamic pressure of the freestream velocity. Additionally, $C_d$ is normalized using the definition implemented by Gauthier et al. \cite{gauthier} to account for the blockage correction on oscillating-foils turbine. The drag coefficient is normalized by the chord length $c$ and heave amplitude $h_o$,

\begin{equation}
\label{e:a1}
C_d^{norm} = C_d \frac{c}{2h_o},
\end{equation}

\noindent where $h_o$ is equal to $1c$ for the experiments performed. The blockage ratio, 

\begin{equation}
\label{e:a2}
\beta = \frac{2h_o \times span}{width \times depth} = 14.6\%,
\end{equation}

\noindent as defined by Gauthier et al. \cite{gauthier}, refers to the ratio of the projected area of the turbine ($chord \times span = 0.10m \times 0.35m$) and the cross-section of the flume ($width \times depth = 0.60m \times 0.80m$). The foil used for the blockage correction has slightly different span and chord lengths, but only a $0.1\%$ difference in blockage ratio.

The implementation of Houlsby's correction method starts with calculating the Froude number,

\begin{equation}
\label{e:h1}
Fr = \frac{V_o}{\sqrt{gd_o}},
\end{equation}

\noindent where $g$ is the gravitational acceleration.  Guessing values for $u_2$, the velocity of the core flow is obtained by iterating the equations

\begin{equation}
\label{e:h3}
u_1 = \frac{Fr^2u^4_2 - (4+2Fr^2)V^2_ou^2_2 + 8V^3_ou_2 - 4V^4_o +4\beta C_d^{norm}V^4_o + Fr^2V^4_o}{-4Fr^2u^3_2 + (4Fr^2+8)V^2_ou_2 - 8V^3_o}
\end{equation}

\noindent and

\begin{equation}
\label{e:h4}
u_1 = \sqrt{u^2_2 - C_d^{norm} V^2_o}
\end{equation}

\noindent until both values for $u_1$ are equal. With $u_1$ and $u_2$ known, $u_d$ is calculated through

\begin{equation}
\label{e:h5}
u_d = \frac{u_1(u_2-V_o)(2gd_o - u^2_2 - u_2V_o)}{2\beta gd_o(u_2-u_1)}.
\end{equation}

Finally, with $u_d$ known, the corrected values for the freestream velocity,

\begin{equation}
\label{e:h6}
V_o'=\frac{V_o((u_d/V_o)^2+C_d^{norm}/4)}{u_d/V_o}, 
\end{equation}

\noindent efficiency,

\begin{equation}
\label{e:h8}
\eta'=\eta \left (\frac{V_o}{V_o'}\right)^3,
\end{equation}

\noindent drag coefficient,

\begin{equation}
\label{e:h7}
C_d^{norm'}=C_d^{norm}\left (\frac{V_o}{V_o'}\right)^2, 
\end{equation}

\noindent and wake velocity,

\begin{equation}
\label{e:h9}
\overline{u}_p'=\overline{u}_p\left (\frac{V_o}{V_o'}\right), 
\end{equation}

\noindent can be determined.

\begin{figure}[H]
\centering
\includegraphics[width=0.6\textwidth]{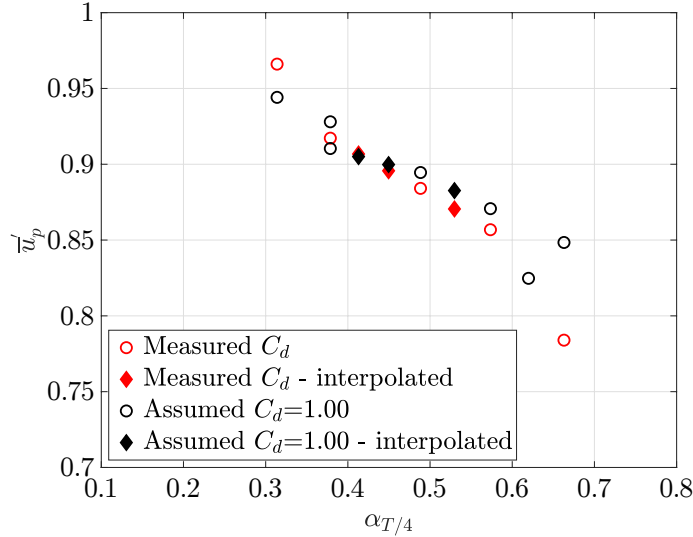}
\caption{Wake velocity comparison between CFD, PIV data corrected with measured $C_d$ and PIV data with the $C_d=1.00$ assumption. Only a PIV data deviation is noticed at $\alpha_{T/4}>0.6$. For these higher values, Houslby's correction method is less accurate.}
\label{f:cdcomp}
\end{figure}

Figure \ref{f:cdcomp} shows the corrected wake velocities for the kinematics analyzed in this paper.

Different values of $C_d$ are evaluated but did not severely impact the corrected data. For instance, even though the measured $C_d$ values are used for the blockage correction, the results are insensitive if $C_d$ is assumed to be $1.00$ (See Figure \ref{f:cdcomp}). Figure \ref{f:cdcomp} also shows the interpolated data taken at few kinematics from PIV. These data are linearly interpolated in kinematics with $fc/U_\infty = 0.11,0.13,0.14$ using $fc/U_\infty = 0.10,0.12,0.15$. All the kinematics used for the PIV interpolation were taken at $h_o/c=1.0$ and $\theta_o=65^{\circ}$.

\section*{Appendix B: Global phase model} \label{appendixB}

\begin{figure}[H]
\centering
	\begin{subfigure}{0.49\textwidth}
	\centering
      	\includegraphics[width=1\linewidth]{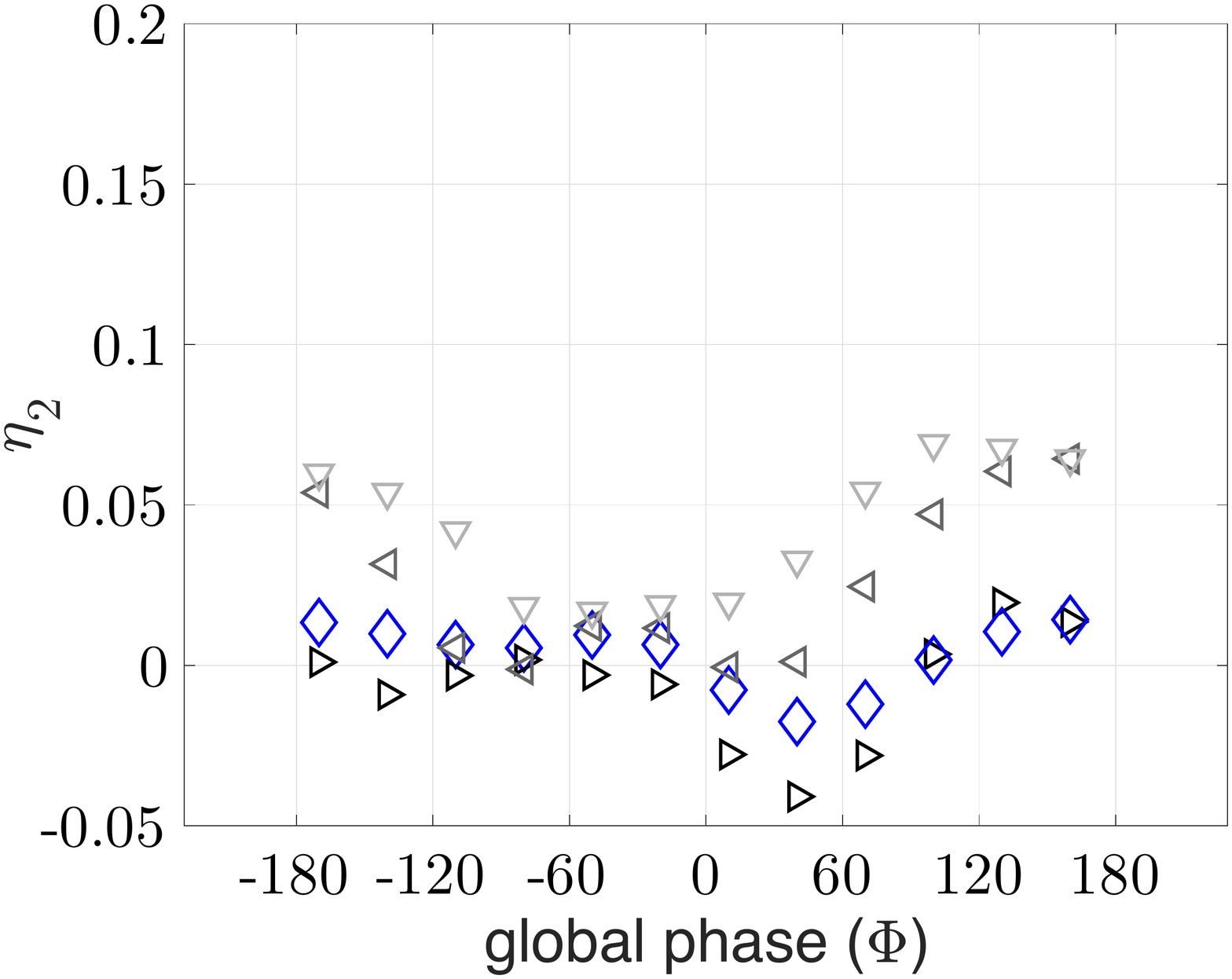}
		\caption{Shear Layer regime}
		\label{f:regimeA_KD}
	\end{subfigure}
	\begin{subfigure}{0.49\textwidth}
	\centering
		\includegraphics[width=1\linewidth]{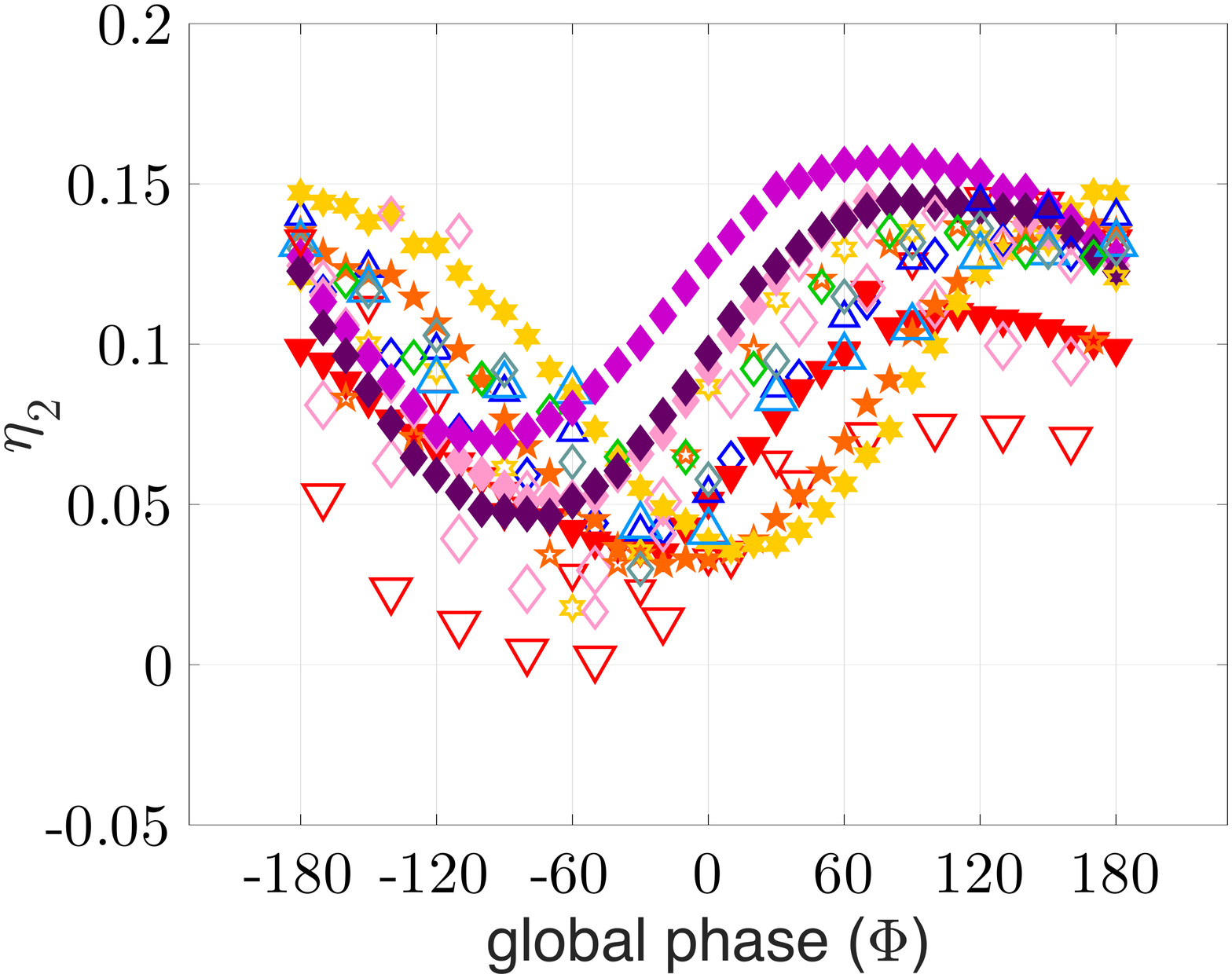}
		\caption{LEV regime}
		\label{f:regimeB_KD}
	\end{subfigure}
	
	\begin{subfigure}{0.49\textwidth}
	\centering
		\includegraphics[width=1\linewidth]{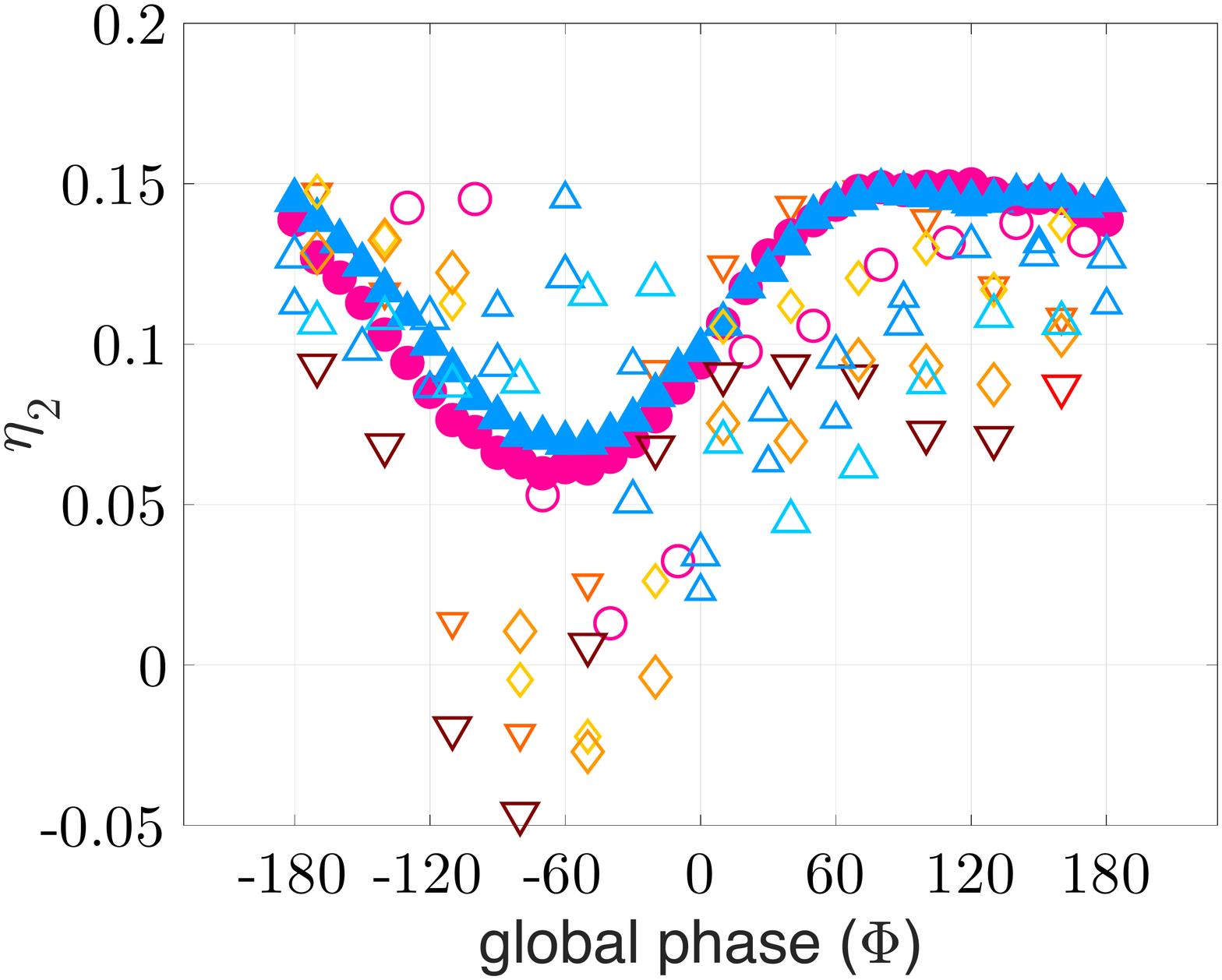}
		\caption{LEV+TEV regime}
		\label{f:regimeC_KD}
	\end{subfigure}
\caption{Trailing foil efficiency ($\eta_2$) regimes using the global phase model from Kinsey and Dumas \cite{kinseytandem2012}. Open symbols are numerical data, whereas filled ones are experimental data. Each curve represents a set of kinematics as defined in Table \ref{t:tablekin}.}
\label{f:regimes_KD}
\end{figure}

Figure \ref{f:regimes_KD} shows trailing foil efficiency profiles computed with the global phase model proposed by Kinsey and Dumas \cite{kinseytandem2012}.

\bibliography{manuscript}

\end{document}